\pdfoutput=1
\documentclass[11pt]{article}

\font\grande=cmr9.5 scaled \magstep4
\font\medio=cmr9.5 scaled \magstep2
\outer\def\beginsection#1\par{\medbreak\bigskip
      \message{#1}\leftline{\bf#1}\nobreak\medskip
\vskip-\parskip
      \noindent}
\usepackage{graphicx} 
\usepackage{mathrsfs}
\usepackage[overload]{textcase}
\begin{document}

\bibliographystyle{unsrt}

\titlepage

\vspace{1cm}
\begin{center}
{\grande Inflationary magnetogenesis in the perturbative regime}\\
\vspace{1cm}
Massimo Giovannini \footnote{e-mail address: massimo.giovannini@cern.ch}\\
\vspace{1cm}
{{\sl Department of Physics, CERN, 1211 Geneva 23, Switzerland }}\\
\vspace{0.5cm}
{{\sl INFN, Section of Milan-Bicocca, 20126 Milan, Italy}}
\vspace*{1cm}
\end{center}
\vskip 0.3cm
\centerline{\medio  Abstract}
\vskip 0.1cm
 While during inflation a phase of increasing gauge coupling allows for a scale-invariant  hyperelectric spectrum, when the  coupling decreases a flat hypermagnetic spectrum can be generated for typical wavelengths larger than the effective horizon. 
 After the gauge coupling flattens out the late-time hypermagnetic power spectra outside the horizon in the radiation epoch are determined by the hyperelectric fields at the end of inflation whereas the opposite is true in the case of decreasing coupling. Instead of imposing an abrupt freeze after inflation, we  consider a smooth evolution of the mode functions by positing that the gauge couplings and their conformal time derivatives are always continuous together with the background extrinsic curvature. The amplified gauge power spectra
are classified according to their transformation properties under the duality symmetry.
After clarifying the role of the comoving and of the physical spectra in the formulation of the relevant magnetogenesis constraints, the parameter space of the scenario is scrutinized. It turns out that a slightly blue hyperelectric spectrum during inflation may lead to a quasi-flat hypermagnetic spectrum prior to matter radiation equality and before the relevant wavelengths reenter the effective horizon. In this framework the gauge coupling is always perturbative but the induced large-scale magnetic fields can be of the order of a few hundredths of a nG and over typical length scales between a fraction of the Mpc and $100$ Mpc prior to the gravitational collapse of the protogalaxy.
\noindent
\vspace{5mm}
\vfill
\newpage
\renewcommand{\theequation}{1.\arabic{equation}}
\setcounter{equation}{0}
\section{Introduction}
\label{sec1}
Besides the invariance under local gauge transformations, the Weyl  \cite{lich} and the duality \cite{duality1,duality2} symmetries are particularly relevant for the dynamics of the gauge fields in general relativity and in scalar-tensor theories of gravity. If the governing equations of a given  field are invariant under Weyl rescaling, the corresponding quantum fluctuations are not parametrically amplified by the evolution of the geometry \cite{parker1,birrell,parker2}. In the absence of sources duality rotates field strengths into their duals (i.e. tensors into pseudotensors) and constrains the gauge power spectra potentially amplified  during conventional or unconventional inflationary phases. 

The symmetries of the evolution equations of the gauge fields in curved space-times
are one of the main handles on the origin of the large-scale magnetism, a perplexing problem 
originally posed by Fermi \cite{ONEa} in connection with the propagation of cosmic rays within the galaxy (see also \cite{ONEb}) and subsequently scrutinized by Hoyle \cite{SIX1} in a cosmological context in view of the comparatively large correlation scales of the fields. Indeed, while the typical diffusion scale in the interstellar medium is of the order of the AU ($1 \, \mathrm{AU} = 1.49 \times 10^{13} \,\, \mathrm{cm}$)  magnetic fields are observed over  larger scales ranging between the $30\, \mathrm{kpc}$ and few $\mathrm{Mpc}$ ($1\, \mathrm{pc}= 3.08\times 10^{18} \, \mathrm{cm}$). 

Three general classes of suggestions have been proposed through the years. Soon after Hoyle's observations \cite{SIX1}  Zeldovich \cite{SIX2} and of Thorne \cite{SIX3} suggested, in the context of anisotropic (but homogeneous) Bianchi-type models \cite{SIX3a}, that the large-scale magnetic fields could be a fossil remnant of a primordial field that originated with the Universe. In a complementary perspective we could imagine 
that the large-scale magnetic fields have been produced at some point during the radiation 
epoch and inside the Hubble radius by the vorticity associated with the turbulent dynamics; the first one suggesting this possibility was probably Harrison \cite{SIX4,SIX5} whose idea found direct applications in the context of phase transitions.
Finally the third class of scenarios implies that a spectrum of gauge fields is produced because of the breaking 
of Weyl  invariance during a standard stage of inflationary expansion; this is the perspective discussed in the present paper.  The collection of themes related to the origin and to the early time effects of large-scale magnetism has been dubbed some time ago magnetogenesis \cite{mgenesis}. While this terminology was so far quite successful, the problem itself has a long history 
as a number of inspiring monographs demonstrates \cite{b1,b2,b3}; see also 
Refs. \cite{rev1,rev2,rev3} for some more recent reviews. 
 
It has been repeatedly argued through the years that magnetic fields with a sufficiently 
large correlation scale could be generated during a phase of accelerated expansion in full analogy 
with what happens for the scalar and tensor modes of the geometry (see e.g. \cite{wein1}). 
The gauge fields parametrically amplified in the early Universe are in fact divergenceless 
vector random fields that  do not  break the spatial isotropy (as it happens instead 
in the case of the fossil remnants discussed in Refs. \cite{SIX2,SIX3}). To avoid the constraints 
imposed by Weyl invariance the gauge fields might directly couple to one or more scalar fields (see \cite{DT1,DT2,DT3,DT4} for an incomplete list of references). The scalar fields may coincide with one or (more inflatons) or even with multiple spectator fields. This class of models is based on the following action:
\begin{equation}
S_{gauge} = - \frac{1}{16 \pi} \int \, d^{4} x\, \sqrt{-G} \biggl[ \lambda(\varphi,\psi) Y_{\alpha\beta} \, Y^{\alpha\beta} + \lambda_{pseudo}(\varphi,\psi) 
Y_{\alpha\beta}\, \widetilde{\,Y\,}^{\alpha\beta}\biggr].
\label{first}
\end{equation}
Within the present notations $Y^{\mu\nu}$ and $\widetilde{\,Y\,}^{\mu\nu}$ are, respectively, the gauge field strength and its dual; $G = \mathrm{det}\,G_{\mu\nu}$ is the determinant of the four-dimensional metric with signature mostly minus\footnote{The Greek indices run over the four space-time dimensions while the Latin (lowercase) indices 
run over the three spatial dimensions. The signature of the four-dimensional metric will be mostly minus i.e. 
$(+, \,, -\,, -\,, -)$. Finally the relation between the Riemann and the Ricci tensors 
will be chosen to be $R_{\mu\nu} = R^{\alpha}_{\,\,\,\,\,\mu\alpha\nu}$.}.  In Eq. (\ref{first}) $\varphi$ denotes the  inflaton field while $\psi$ represents a generic spectator field. During a stage of 
conventional (slow-roll) inflation the variation of $\lambda$ is associated with the variation of the gauge coupling whose evolution may reach into the strong coupling regime (see the third paper of Ref. \cite{DT1}).  The presence of $\lambda(\varphi,\psi)$  in Eq. (\ref{first}) is more relevant than the pseudoscalar  (axion-like \cite{c0,c0a}) coupling which will be ignored even if it has been studied by many authors \cite{c1,c1a,c2} in the context of the magnetic field generation.  For the amplification of the magnetic field itself the pseudoscalar vertex is not essential but it may lead to hypermagnetic flux lines are linked or twisted as originally discussed in \cite{mm1}. The produced Chern-Simons condensates leads to a viable (but unconventional) mechanism for baryogenesis via hypermagnetic knots \cite{mm1,mm2} which are characterized by an average magnetic gyrotropy \cite{mm3}.
These  gyrotropic and helical fields play also a role in anomalous magnetohydrodynamics where the evolution of the magnetic fields at finite conductivity is analyzed in the presence of anomalous charges \cite{cme1}. In the collisions of heavy ions this phenomenon is often dubbed chiral magnetic effect \cite{cme2}.
  
In this paper duality will be explicitly used to deduce the gauge power spectra not only during inflation but 
also during the subsequent decelerated expansion. We will show, in particular, that  after the gauge coupling flattens out the late-time hypermagnetic power spectra outside the horizon in the radiation epoch are determined by the hyperelectric fields at the end of inflation whereas the opposite is true in the case of decreasing coupling. The obtained results suggest that a slightly blue hyperelectric spectrum during inflation may lead to a quasi-flat hypermagnetic spectrum prior to matter radiation equality and before the relevant wavelengths reenter the effective horizon. From the technical viewpoint these
results will arise from the discussion of an appropriate transition matrix whose elements have well defined transformation properties under the duality symmetry and control the form of the late-time spectra. Using 
these results we shall investigate 
the magnetogenesis requirements as well as all other pertinent constraints; 
we shall conclude that large-scale magnetic fields can be generated during a quasi-de Sitter 
stage of expansion while the gauge coupling remains perturbative throughout all the stages 
of the dynamical evolution.

The layout of this paper is in short the following. In section \ref{sec2} after introducing the necessary generalities, we shall discuss the duality symmetry both for the field equations and for the power spectra. 
At the end of the section we shall discuss two complementary (and dual) profiles for the evolution of the gauge couplings.
In section \ref{sec3} the gauge power spectra during inflation will be computed for typical wavelengths 
larger than the effective horizon and related via duality transformations. 
Using the overall continuity of the whole description the electric and the magnetic mode functions 
at the end of inflation will be explicitly related, in section \ref{sec4}, to the gauge spectra in the radiation 
epoch via a transition matrix whose elements transform in a well defined way under duality.
The gauge spectra will then be computed in several approximations schemes with particular 
attention to the relevant phenomenological regimes.
In section \ref{sec5} the magnetogenesis requirements will be examined in conjunction with the physical constraints. Section \ref{sec6} contains our concluding remarks. 
To avoid lengthy digressions various technical results have been relegated to the appendices \ref{APPA}, \ref{APPB} and \ref{APPC}.

\renewcommand{\theequation}{2.\arabic{equation}}
\setcounter{equation}{0}
\section{Equations of motion, duality and dynamical gauge couplings}
\label{sec2}
With the purpose 
of making the whole discussion self-contained, the essential notations will be introduced 
in the first part of this section while in the second part we shall examine the gauge power spectra 
in the light of the duality symmetry. In the last part of the section the physical aspects of the 
evolution of the gauge couplings will be specifically addressed.
\subsection{Notations, conventions and some general considerations}
The field content of the model may be different but it generally follows from a 
total action of the type:
\begin{equation}
S_{tot} = S_{grav} + S_{scalar} + S_{gauge},
\label{ACT1}
\end{equation}
where $S_{gauge}$ has been already introduced in Eq. (\ref{first}) and, as 
already mentioned, $Y_{\mu\nu}$ and $\widetilde{\,Y\,}^{\mu\nu}$ 
denote, respectively, the gauge field strength and its dual:
\begin{equation}
\widetilde{\,Y\,}^{\mu\nu} = \frac{1}{2} \, E^{\mu\nu\rho\sigma} \, Y_{\rho\sigma}, \qquad 
E^{\mu\nu\rho\sigma} = \frac{\epsilon^{\mu\nu\rho\sigma}}{\sqrt{-G}}.
\label{ACT1a}
\end{equation}
In Eq. (\ref{ACT1a}) $\epsilon^{\mu\nu\rho\sigma}$ denotes the totally antisymmetric symbol of Levi-Civita 
in four dimensions.  In more explicit terms the gravitational and the scalar actions of Eq. (\ref{ACT1}) can be expressed as:
\begin{eqnarray}
S_{grav} &=& - \frac{1}{2 \ell_{P}^2} \int d^{4}\, x \,\, \sqrt{- G} \, R,
\nonumber\\
S_{scalar} &=& \int d^{4}\, x \,\, \sqrt{- G} \biggl[ \frac{1}{2} G^{\alpha\beta} \partial_{\alpha} \varphi \partial_{\beta}\varphi + \frac{1}{2} G^{\alpha\beta} \partial_{\alpha} \psi \partial_{\beta} \psi - W(\varphi, \psi)\biggr],
\label{ACT2}
\end{eqnarray}
where $\ell_{P}^2 = 1/\overline{M}_{P}^{2} = 8\pi/M_{P}^2$ and $M_{P}=1.22 \times 10^{19}$ GeV is the 
Planck mass.
The notations of  Eq. (\ref{ACT2}) are purposely schematic and $\varphi$ denotes the inflaton  while $\psi$ is a spectator field. 
Various magnetogenesis models have been discussed in various contexts 
where $W(\varphi,\psi)$ has a well defined 
expression. For instance in the models of Ref. \cite{DT1} $W$ is only function of $\varphi$ 
while an explicit magnetogenesis model based on spectator fields can be found in the 
last paper of Ref. \cite{DT2}. With these specifications, the general equations derived from the actions (\ref{ACT1})--(\ref{ACT2}) 
are:
\begin{eqnarray}
&& R_{\mu}^{\nu} = \ell_{P}^2 \biggl[ \partial_{\mu} \varphi \partial^{\nu} \varphi - W \delta_{\mu}^{\nu} + {\mathcal T}_{\mu}^{\nu}\biggr],
\label{ACT3}\\
&& G^{\alpha\beta} \nabla_{\alpha} \nabla_{\beta} \varphi + \frac{\partial W}{\partial \varphi} +
\frac{1}{16\pi} \frac{\partial \lambda}{\partial \varphi} Y_{\alpha\beta} Y^{\alpha\beta} =0,
\label{ACT4}\\
&& \nabla_{\alpha} \biggl( \lambda Y^{\alpha\beta}\biggr) = 0, \qquad \nabla_{\alpha} \widetilde{Y}^{\alpha\beta}=0.
\label{ACT5}
\end{eqnarray} 
In Eq. (\ref{ACT3}) ${\mathcal T}_{\mu}^{\nu}$ denotes the (traceless) energy-momentum tensor of the gauge fields 
whose explicit form is given by:
\begin{equation}
{\mathcal T}_{\mu}^{\nu} = \frac{\lambda}{4\pi} \biggl[ - Y_{\mu \alpha} \, Y^{\nu\beta} + \frac{1}{4} \delta_{\mu}^{\nu} \, 
Y_{\alpha\beta} \, Y^{\alpha\beta} \biggr]. 
\label{ACT6}
\end{equation}
We shall be mostly concerned with conformally flat background geometries whose associated line element is:
\begin{equation}
d s^2 = G_{\alpha\beta} \,\, d x^{\alpha} \, d x^{\alpha} = a^2(\tau) \bigl[ d\tau^2 - d\vec{\,x\,}^2 \bigr],
\label{ACT7}
\end{equation}
where $a(\tau)$ denotes the scale factor. In the geometry (\ref{ACT7}) the various components of ${\mathcal T}_{\mu}^{\,\,\,\nu}$ 
defined in Eq. (\ref{ACT6}) are:
\begin{eqnarray}
{\mathcal T}_{0}^{\,\,0} &=& \rho_{B} + \rho_{E},\qquad {\mathcal T}_{0}^{\,\,i} = \frac{1}{4 \pi a^4} \bigl( \vec{\,E\,} \times \vec{\,B\,} \bigr)^{i},
\nonumber\\
{\mathcal T}_{i}^{\,\,j} &=& - (p_{E} + p_{B}) \delta_{i}^{j} + \Pi_{E\,\,i}^{\,\,\,j} + \Pi_{B\,\,i}^{\,\,\,j}.
\label{ACT8}
\end{eqnarray}
In Eq. (\ref{ACT8}) we introduced the energy density, the pressure and the anisotropic stresses of the hypermagnetic and hyperelectric fields:
\begin{eqnarray}
\rho_{B} &=& \frac{B^2}{8 \pi a^4} , \qquad \rho_{E} = \frac{E^2}{8 \pi a^4},\qquad p_{B} = \frac{\rho_{B}}{3}, \qquad p_{E} = \frac{\rho_{E}}{3},
\label{ACT9}\\
\Pi_{E\,\,i}^{\,\,\,j} &=& \frac{1}{4 \pi a^4} \biggl( E_{i} \, E^{j} - \frac{E^2}{3} \delta_{i}^{j} \biggr),\qquad 
\Pi_{B\,\,i}^{\,\,\,j} = \frac{1}{4 \pi a^4} \biggl( B_{i} \, B^{j} - \frac{B^2}{3} \delta_{i}^{j} \biggr), 
\label{ACT10}
\end{eqnarray}
where $E^2 = \vec{\,E\,}\cdot \vec{\,E\,}$ and $B^2 = \vec{\,B\,}\cdot \vec{\,B\,}$.
Equations (\ref{ACT8}), (\ref{ACT9}) and (\ref{ACT10}) are expressed in terms of $\vec{\,E\,}$ and $\vec{\,B,}$ i.e.  the  {\em comoving} hyperelectric and hypermagnetic  fields. These rescaled quantities are actually the normal modes of the system and are related to the {\em physical} fields as:
\begin{equation}
\vec{\,E\,}  =  \, a^2 \, \sqrt{\lambda}\, \vec{\,E\,}^{(phys)}, \qquad \vec{\,B\,} = a^2 \, \sqrt{\lambda}\,\vec{\,B\,}^{(phys)}.
\label{ACT13}
\end{equation}
The components of the field strengths can be directly expressed in terms of the components 
of the physical fields; so for instance, in terms of $\epsilon^{i\,j\,k}$ (i.e. the Levi-Civita symbol
in three dimensions) we have
\begin{equation}
Y_{i\, 0} = - a^2 E_{i}^{(phys)}, \qquad Y^{i\, j} = - \epsilon^{i\, j\, k} B_{k}^{(phys)}/a^2,
\label{ACT13a}
\end{equation}
and similarly for the dual strength. The physical fields are essential for the discussion of the actual magnetogenesis constraints but the evolution equations of (\ref{ACT5}) are simpler in terms 
of the comoving fields:
\begin{eqnarray}
&& \vec{\,\nabla\,} \cdot \bigl( \sqrt{\lambda} \, \vec{\,E\,} \bigr) =0, \qquad \vec{\,\nabla\,} \cdot \biggl( \frac{\vec{\,B\,}}{\sqrt{\lambda}} \biggr) =0,
\label{ONE1}\\
&& \partial_{\tau} \biggl( \sqrt{ \lambda} \, \vec{\,E\,} \biggr) = \vec{\nabla} \times \biggl( \sqrt{\lambda}  \, \vec{\,B\,}\biggr),\qquad  \partial_{\tau} \biggl(\frac{\vec{\,B\,}}{\sqrt{\lambda}}\biggr) = - \vec{\nabla} \times \biggl(\frac{\vec{\,E\,}}{\sqrt{\lambda}} \biggr).
\label{ONE2}
\end{eqnarray}
Equations (\ref{ONE1})--(\ref{ONE2}) have been written in the general case where 
$\lambda$ can be inhomogeneous even if, as we shall see in the last part of this section, the gauge coupling will always 
be considered to be time-dependent but homogeneous. Furthermore Eqs. (\ref{ONE1})--(\ref{ONE2}) under the following 
duality tranformation:
\begin{equation}
\sqrt{\lambda} \to \frac{1}{\sqrt{\lambda}}, \qquad \vec{\,B\,} \to \vec{\,E\,}, \qquad \vec{\,E\,} \to - \vec{\,B\,}.
\label{ONE5}
\end{equation}
All the mode functions 
and power spectra obtained by the simultaneous evolution of the geometry 
and of the gauge coupling must be consistent with Eq. (\ref{ONE5}) both during the inflationary 
phase and in the subsequent decelerated stages of expansion before the given scale 
reenters the effective horizon. The gauge coupling $e(\lambda)$ is related to the inverse of $\sqrt{\lambda}$:
\begin{equation}
S_{gauge} = - \frac{1}{4} \int d^{4} x \, \frac{\sqrt{-G} }{e^{2}}\,\, Y_{\mu\nu} \, Y^{\mu\nu}, \qquad e^2 =  \frac{4\pi}{\lambda}.
\label{ONE1a}
\end{equation}
Thus the gauge coupling increases when $\lambda$ decreases and vice versa. 

\subsection{Quantization and canonical form of the power spectra} 
Since we are going to discuss the parametric amplification of the quantum fluctuations of the gauge fields 
the classical evolution  summarized by Eqs. (\ref{ONE1})--(\ref{ONE2}) must be complemented 
by the corresponding quantum treatment. From the semi-classical viewpoint this process can be viewed as the conversion of traveling waves into standing waves;  the same phenomenon occurs for the scalar and tensor modes of the geometry \cite{wein1} and it leads to the so-called Sakharov oscillations \cite{SAK1} (see also \cite{SAK2,SAK3,SAK4}). In the present context the duality symmetry explicitly relates the Sakharov oscillations 
of the hypermagnetic [i.e. $f_{k,\,\alpha}(\tau)$] and hyperelectric mode functions 
[i.e. $g_{k,\,\alpha}(\tau)$] entering the expressions of the corresponding field operators:
\begin{eqnarray}
&& \hat{B}_{i}(\tau, \vec{x}) = - \frac{i\, \epsilon_{m n i}}{(2\pi)^{3/2}}  \sum_{\alpha=\oplus,\,\otimes} \int d^{3} k \,k_{m} \,e^{(\alpha)}_{n}(\hat{k}) \,
\biggl[ f_{k,\,\alpha}(\tau)\, \hat{a}_{\vec{k}, \alpha} e^{- i \vec{k} \cdot \vec{x}}  - f_{k,\,\alpha}^{*}(\tau) \hat{a}^{\dagger}_{\vec{k}, \alpha}
e^{ i \vec{k} \cdot \vec{x}}\biggr], 
\label{ONE6}\\
&& \hat{E}_{i}(\tau,\vec{x}) = - \frac{1}{(2\pi)^{3/2}}  \sum_{\alpha=\oplus,\,\otimes} \int d^{3} k \,e^{(\alpha)}_{i}(\hat{k}) \,
\biggl[ g_{k\,\alpha}(\tau)  \hat{a}_{\vec{k}, \alpha} e^{- i \vec{k} \cdot \vec{x}}  + g_{k,\,\alpha}^{*}(\tau) \hat{a}^{\dagger}_{\vec{k}, \alpha}e^{ i \vec{k} \cdot \vec{x}} \biggr].
\label{ONE7}
\end{eqnarray}
In Eqs. (\ref{ONE6})--(\ref{ONE7})  the two vector polarizations are directed along the orthogonal unit vectors $\hat{e}_{\oplus}$ and $\hat{e}_{\otimes}$ that are also orthogonal to $\hat{k}$ (i.e. $\hat{k} \cdot\hat{e}_{\alpha} =0$).
In Eq. (\ref{ONE7}) $ \hat{a}_{\vec{k}, \alpha} $ and  $ \hat{a}_{\vec{k}, \alpha}^{\dagger} $ are the 
creation and annihilation operators obeying, within the present notations,  $[\hat{a}_{\vec{q}, \alpha}, \, \hat{a}_{\vec{p}, \beta}^{\dagger}] = \delta^{(3)}(\vec{q} - \vec{p})\, \delta_{\alpha\beta}$.
In Eqs. (\ref{ONE6}) and (\ref{ONE7}) the sum is performed over the physical polarizations $e^{(\alpha)}_{i}(\hat{k})$ while the mode functions 
$f_{k,\alpha}$ and $g_{k\,\alpha}$ obey, in the absence of conductivity, the following pair of  equations:
\begin{equation}
f_{k,\,\alpha}^{\,\prime} = g_{k,\,\alpha} + {\mathcal F} f_{k,\,\alpha},\qquad  g_{k,\,\alpha}^{\, \prime}= -  k^2 f_{k,\,\alpha} - {\mathcal F} g_{k,\,\alpha}, \qquad {\mathcal F} = \frac{\sqrt{\lambda}^{\,\,\prime}}{\sqrt{\lambda}},
\label{ONE8}
\end{equation}
where the prime denotes a derivation with respect to the conformal time coordinate $\tau$.
The mode functions in Eq. (\ref{ONE8}) must also be correctly normalized so that the Wronskian 
of any solution must satisfy for each polarization:
\begin{equation}
f_{k,\,\alpha}(\tau) \, g_{k, \alpha}^{*}(\tau) - f_{k,\,\alpha}^{*}(\tau) \, g_{k, \alpha}(\tau) = i.
\label{ONE8w}
\end{equation}
The field operators of Eqs. (\ref{ONE6})--(\ref{ONE7}) can be represented in Fourier space as\footnote{Equation (\ref{ONE8a}) follows from Eqs. (\ref{ONE6}) and (\ref{ONE7}) by recalling that
$\hat{B}_{j}(\vec{p},\tau)= \int d^{3} \, x \, \hat{B}_{j}(\vec{x},\tau) e^{i \vec{p}\cdot\vec{x}}/(2\pi)^{3/2}$ 
and that $\hat{E}_{i}(\vec{q},\tau)= \int d^{3} \, x \, \hat{E}_{i}(\vec{x},\tau) e^{i \vec{q}\cdot\vec{x}}/(2\pi)^{3/2}$.}
\begin{eqnarray}
&&\hat{B}_{j}(\vec{p},\tau)= - i \, p_{m} \, \epsilon_{m n j}\,  \sum_{\beta} \biggl[ e^{\beta}_{n}(\hat{p}) \, \hat{a}_{\vec{p},\, \beta} \, f_{q,\beta}(\tau) + e^{\alpha}_{n}(-\hat{p}) \, \hat{a}_{-\vec{p},\, \beta}^{\dagger} \, f_{p,\beta} ^{*}(\tau) \biggr],
\nonumber\\
&&\hat{E}_{i}(\vec{q},\tau) = \sum_{\alpha} \biggl[ e^{\alpha}_{i}(\hat{q}) \, \hat{a}_{\vec{q},\, \alpha} \, g_{q,\alpha}(\tau) + e^{\alpha}_{i}(-\hat{q}) \, \hat{a}_{-\vec{q},\, \alpha}^{\dagger} \, g_{q,\alpha} ^{*}(\tau) \biggr].
\label{ONE8a}
\end{eqnarray}
Therefore the corresponding two-point functions in Fourier space become: 
\begin{eqnarray}
&& \langle \hat{B}_{i}(\vec{k}, \tau)\, \hat{B}_{j}(\vec{p},\tau) \rangle = \frac{2\pi^2}{k^3}\, P_{B}(k,\tau)\, p_{ij}(\hat{k})  \,\delta^{(3)}(\vec{k} + \vec{p}),
\label{ONE13}\\
&& \langle \hat{E}_{i}(\vec{k},\tau)\, \hat{E}_{j}(\vec{p},\tau) \rangle = \frac{2\pi^2}{k^3}\, P_{E}(k,\tau)\, p_{ij}(\hat{k})  \,\delta^{(3)}(\vec{k} + \vec{p}),
\label{ONE14}
\end{eqnarray}
where $p(\hat{k}) = (\delta_{i j} - \hat{k}_{i} \, \hat{k}_{j} )$;  $P_{B}(k,\tau)$ and $P_{E}(k,\tau)$ are 
the {\em comoving magnetic and electric power spectra}, respectively\footnote{If the mode functions for the two polarizations coincide the sums appearing in 
Eqs. (\ref{ONE14a})--(\ref{ONE14b}) can be 
 performed trivially since $f_{k\, \oplus} = f_{k\, \otimes} =f_{k}$  and similarly 
 for the hyperelectric mode function.}: 
\begin{eqnarray}
 P_{B}(k,\tau) &=& \frac{k^{5}}{4\, \pi^2\,} \, \sum_{\alpha= \oplus,\,\otimes} |f_{k,\,\,\alpha}(\tau)|^2 \equiv \frac{k^{5}}{2\, \pi^2\,} \,  |f_{k}(\tau)|^2,
\label{ONE14a}\\ 
 P_{E}(k,\tau) &=& \frac{k^{3}}{4\, \pi^2\,} \, \sum_{\alpha= \oplus,\,\otimes} |g_{k,\,\,\alpha}(\tau)|^2 \equiv \frac{k^{3}}{2\, \pi^2\,} \,  |g_{k}(\tau)|^2.
\label{ONE14b}
\end{eqnarray}
 For the present purposes it will be  important to distinguish 
the comoving from the physical power spectra: while the comoving power spectra are obtained 
from the comoving field operators $\hat{E}_{i}$ and $\hat{B}_{i}$, the {\em physical power spectra} 
follow from the the corresponding physical fields  $\hat{E}^{(phys)}_{i}$ and $\hat{B}^{(phys)}_{i}$, 
defined in Eq. (\ref{ACT13}); the relation between the physical and the comoving power 
spectra is therefore given by:
\begin{equation}
P^{(phys)}_{B}(k,\tau) =\frac{P_{B}(k,\tau)}{\lambda(\tau) a^{4}(\tau)}, \qquad  P^{(phys)}_{E}(k,\tau) =\frac{P_{E}(k,\tau)}{\lambda(\tau) a^{4}(\tau)}.
\label{ONE14c}
\end{equation}
The phenomenological requirements (e.g. the magnetogenesis constraints to be 
discussed in section \ref{sec5}) must be typically expressed in terms of the physical spectra. Occasionally this distinction has not been clearly spelled out and the consequences have been confusing, as we shall remark 
later on. Let us conclude this discussion by recalling that the two coupled first-order equations given in Eq. (\ref{ONE8})  can be 
transformed in two (decoupled) second-order differential equations: 
\begin{eqnarray}
f_{k,\,\alpha}^{\prime\prime} + \biggl[ k^2 - \frac{\sqrt{\lambda}^{\prime\prime}}{\sqrt{\lambda}}\biggr] f_{k,\,\alpha} =0,\qquad 
g_{k,\,\alpha}^{\prime\prime} + \biggr[ k^2 - \sqrt{\lambda}\biggl(\frac{1}{\sqrt{\lambda}}\biggr)^{\prime\prime}\biggr] g_{k,\,\alpha} =0,
\label{ONE10}
\end{eqnarray}
which have the same content of Eq. (\ref{ONE8}) provided the initial conditions are correctly 
imposed by taking into account that  $g_{k,\,\alpha}$ is ultimately determined from $f_{k,\alpha}$ and its derivative
as $g_{k,\alpha} = f_{k\,\alpha}^{\prime} - {\mathcal F} f_{k\,\alpha}$.  Equations (\ref{ONE8})--(\ref{ONE10}) are invariant under the following duality transformations\footnote{In connection with Eqs. (\ref{ONE8}) and (\ref{ONE10a}) we recall that ${\mathcal F}= \sqrt{\lambda}^{\,\prime}/\sqrt{\lambda}$; this means that for $\sqrt{\lambda} \to 1/\sqrt{\lambda}$, ${\mathcal F} \to - {\mathcal F}$.}:
\begin{equation}
\sqrt{\lambda} \to 1/\sqrt{\lambda}, \qquad f_{k,\,\alpha} \to g_{k,\,\alpha}/k,\qquad g_{k,\,\alpha} \to - k f_{k,\,\alpha}.
\label{ONE10a}
\end{equation}

\subsection{Spectral energy density ad backreaction constraint} 
Since the amplification of the gauge fields takes place in a 
homogeneous and isotropic background geometry (see Eq. (\ref{ACT7})) 
the corresponding energy density must not exceed the critical energy density $\rho_{crit} = 3 \, \overline{M}_{P}^2 H^2$ where 
$H$ is the Hubble rate. The averaged energy density of the gauge fields follows from Eqs. 
(\ref{ONE13}) and (\ref{ONE14})  and from the definitions of $\rho_{B}$ and $\rho_{E}$ 
appearing in Eqs. (\ref{ACT8}), (\ref{ACT9}) and (\ref{ACT10}). Thanks to 
Eqs.  (\ref{ONE14a})--(\ref{ONE14b}) the final result is:
\begin{equation}
\langle \hat{\rho}_{Y} \rangle = \langle \hat{\rho}_{B} \rangle + \langle \hat{\rho}_{E} \rangle=  \frac{1}{4 \pi a^4(\tau)} \int \frac{d k}{k} \biggl[ P_{E}(k,\tau) + P_{B}(k,\tau)\biggr].
\label{ONE11}
\end{equation}
In terms of the physical power spectra of Eq.  (\ref{ONE14c}) the result (\ref{ONE11}) becomes:
\begin{equation}
\langle \hat{\rho}_{Y} \rangle =  \frac{\lambda(\tau)}{4 \pi} \int \frac{d k}{k} \biggl[ P^{(phys)}_{E}(k,\tau) + P_{B}^{(phys)}(k,\tau)\biggr].
\label{ONE11a}
\end{equation}
To compare energy density of the parametrically amplified gauge fields
with the energy density of the background geometry we introduce the spectral energy density in critical units:
\begin{equation}
\Omega_{Y} = \frac{1}{\rho_{crit}} \frac{ d \langle \hat{\rho} \rangle}{d \ln{k}} = \frac{2}{3 H^2 \, M_{P}^2 a^4}  \biggl[ P_{E}(k,\tau) + P_{B}(k,\tau)\biggr].
\label{ONE15}
\end{equation}
 To guarantee the absence of dangerous backreaction effects $\Omega_{Y}(k,\tau)$ 
must always be much smaller $1$ throughout all the stages of the evolution and for all relevant scales; this requirement must be separately verified both during and after inflation.

\subsection{Gauge couplings and their continuity}
\subsubsection{Increasing gauge coupling}
\label{subs241}
\begin{figure}[!ht]
\centering
\includegraphics[height=7cm]{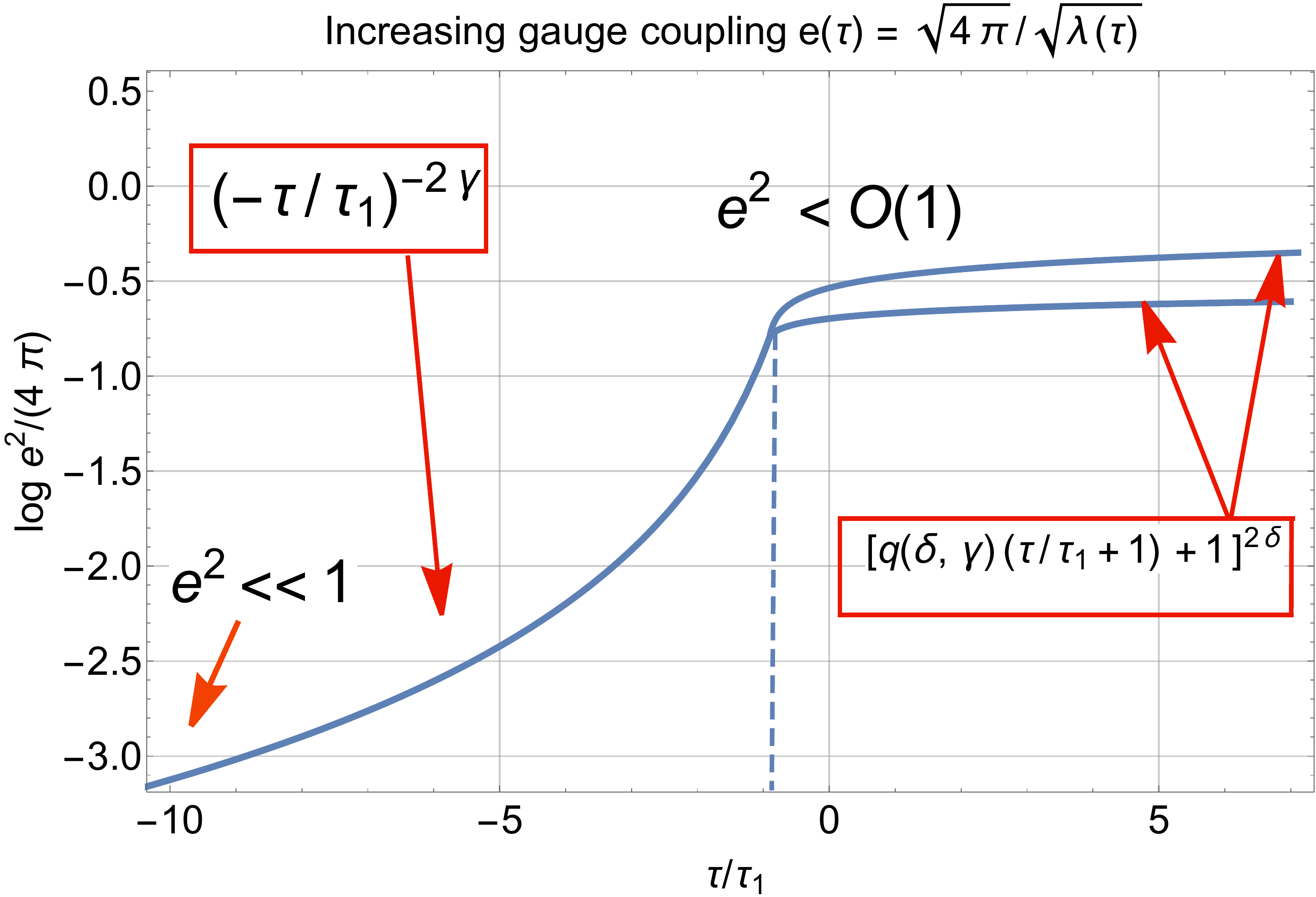}
\caption[a]{The logic and the main notations employed for the dynamical description of an increasing 
gauge coupling. The two different curves 
for $\tau \geq - \tau_{1}$ correspond to different values of $\delta \ll 1$. Note that $q(\delta,\gamma) = \delta/\gamma$ [see also Eq. (\ref{FIVE6})].}
\label{FFF0a}      
\end{figure}

In Fig. \ref{FFF0a} the profile describing the evolution of the gauge 
coupling is illustrated together with the main 
notations employed throughout the discussion.  
During the inflationary phase (i.e. for $\tau \leq - \tau_{1}$ in Fig. \ref{FFF0a}) $\gamma$ describes the rate of increase of the gauge coupling in the conformal time parametrization
 while $\delta$ controls the evolution during the post-inflationary stage 
of expansion when the gauge coupling flattens out:
\begin{eqnarray}
\sqrt{\lambda} &=& \sqrt{\lambda_{1}} \biggl(-\frac{\tau}{\tau_{1}}\biggr)^{\gamma}, \qquad \tau \leq - \tau_{1}, 
\label{TWO1}\\
\sqrt{\lambda} &=& \sqrt{\lambda_{1}} \biggl[\frac{\gamma}{\delta} \biggl( \frac{\tau}{\tau_{1}} + 1\biggr) +1 \biggr]^{- \delta}, \qquad \tau \geq - \tau_{1}.
\label{FIVE2}
\end{eqnarray}
The explicit form of Eqs. (\ref{TWO1}) and (\ref{FIVE2}) is dictated by the continuity of $\sqrt{\lambda}$ and 
of $\sqrt{\lambda}^{\,\prime}$; furthermore the physical range of $\gamma$ and $\delta$ is given by: 
\begin{equation} 
\gamma > 0, \qquad \mathrm{and}\qquad 0 \leq  \delta \ll  \gamma.
\label{FIVE3}
\end{equation}
For short the limit $\delta \to 0$ shall be referred to as the sudden approximation while 
the smooth approximation corresponds to the case of Eq. (\ref{FIVE3}) where $ 0 \leq  \delta \ll  \gamma$.
As we shall see it will always be possible to derive the 
results of the sudden approximation by taking the limit $\delta \to 0$ from the formulas 
valid in the case of the smooth approximation. The continuity of the magnetic and electric mode functions introduced in Eqs. (\ref{ONE6})--(\ref{ONE7}) and obeying Eqs. (\ref{ONE8})--(\ref{ONE8w}) 
rely however on the continuity of $\sqrt{\lambda}$ and its derivative. If we would simply assume 
that $\sqrt{\lambda}$ is constant the conformal time derivative will have a jump discontinuity 
and Eq. (\ref{ONE10}) will develop a singularity either in $\sqrt{\lambda}^{\, \prime\prime}/\sqrt{\lambda}$
or in $(1/\sqrt{\lambda})^{\,\prime\prime}\,\sqrt{\lambda}$.  Since the present approach 
guarantees the continuity of gauge mode functions the related power spectra 
will also be continuous. Note that the numerical value of $\lambda_{1}$ may well coincide with $1$ but it could also be slightly larger than $1$ since this range is compatible with a gauge coupling that is perturbative for $\tau = -\tau_{1}$, as illustrated 
in Fig. \ref{FFF0a}. 

In the geometry of Eq. (\ref{ACT7}) the background equations derived from Eqs. (\ref{ACT3})--(\ref{ACT4}) 
are:
\begin{equation}
3 \overline{M}_{P}^2\,{\mathcal H}^2 = \frac{1}{2}{\varphi'}^2 +a^2\, W(\varphi), \qquad\qquad 
2 \overline{M}_{P}^2\, ({\mathcal H}^2 - {\mathcal H}' )= {\varphi'}^2,\qquad\qquad 
\varphi'' + 2 {\mathcal H} \varphi' + \frac{\partial W}{\partial \varphi} a^2 =0,
\label{FL3a}
\end{equation}
where, as already mentioned, the prime 
denotes a derivation with respect to the conformal time coordinate $\tau$; 
furthermore  ${\mathcal H}= (\ln{a})^{\prime} = a H$ where $H=\dot{a}/a$ is the conventional Hubble rate
and the overdot denotes a derivation with respect to the cosmic time coordinate $t$.
The slow roll approximation specifies the evolution during the inflationary phase where the parameters
$\epsilon$, $\eta$ and $\overline{\eta}$ are all much smaller than $1$ and eventually get to $1$ when inflation ends.
The definitions of the slow roll parameters within the notations of this paper are as follows:
\begin{equation}
 \epsilon = - \frac{\dot{H}}{H^2} = \frac{\overline{M}_{P}^2}{2} \biggl(\frac{W_{,\,\varphi}}{W}\biggr)^2, \qquad
\eta = \frac{\ddot{\varphi}}{H \dot{\varphi}}, \qquad \overline{\eta} = \overline{M}_{P}^2 \biggl(\frac{W_{,\,\varphi\varphi}}{W}\biggr),
\label{sr2}
\end{equation}
note that $W_{,\,\varphi}$ and 
$W_{,\,\varphi\varphi}$ are shorthand notations for the first and second derivatives of the potential 
$W(\varphi)$ with respect to $\varphi$.  The slow roll parameters $\eta$, $\overline{\eta}$ and $\epsilon$ are not 
independent and their mutual relation, i.e. $\eta=\epsilon-\overline{\eta}$,  follows 
from the slow roll version of Eqs. (\ref{FL3a}) written in the cosmic time coordinate $t$:
\begin{equation}
3 H \dot{\varphi} + \frac{\partial W}{\partial\varphi} =0, \qquad\qquad 
3 \overline{M}_{P}^2 H^2 = W,\qquad\qquad 2 \overline{M}_{P}^2 \dot{H} = - \dot{\varphi}^2.
\label{sr4}
\end{equation}
We shall be assuming that the inflationary stage of expansion  takes place for $\tau \leq - \tau_{1}$: 
\begin{equation}
{\mathcal H}=  a H = - \frac{1}{(1- \epsilon) \tau}, \qquad \epsilon = - \frac{\dot{H}}{H^2} \ll 1,
\label{inf1}
\end{equation}
whereas after inflation (i.e. for $\tau \geq -\tau_{1}$) the background will be decelerated and
dominated by radiation. The continuity of the scale factor demands:
\begin{eqnarray}
a_{inf}(\tau) &=& \biggl(-\frac{\tau}{\tau_1}\biggr)^{- \beta}, \qquad   \tau \leq - \tau_{1},
\nonumber\\
a_{rad}(\tau) &=& \frac{\tau + (\beta+1) \tau_{1}}{\tau_{1}}, \qquad \tau \geq - \tau_{1}.
\label{FIVE1} 
\end{eqnarray}
with  $\beta \simeq 1/(1 - \epsilon)$. Both Eqs. (\ref{inf1}) and (\ref{FIVE1}) 
assume that $\epsilon$ changes very slowly during inflation. The continuity of the scale 
factor and of its conformal time derivative also imply the continuity of the extrinsic 
curvature whose background value is given by $\overline{K}_{i\,j} = - a\, {\mathcal H} \, \delta_{i\, j}$.

For a correct treatment of the problem the continuity of  the extrinsic curvature is essential. 
However various other effects could be always added with the purpose of providing a more
accurate description of the  transition regime. For instance it could be observed that the slow-roll approximation breaks down at the end of inflation 
so that $\epsilon$ is not truly constant across the transition. Another 
possibility is represented by the dynamics of reheating. These situations 
will affect the maximal frequency of the spectrum which is therefore 
rather difficult to predict with great accuracy. This situation is actually 
analog to what happens in the case of the spectrum of relic gravitons \cite{GW1}.
In analogy with the graviton spectrum we shall therefore parametrize the maximal wavenumber as
$k_{max} = \zeta a_{1} H_{1}$ where $\zeta$ is a numerical factor which may even be, in some cases, 
dependent upon the wavenumber. The impact of the breakdown of the slow-roll
approximation on the maximal frequency of the spectrum has been discussed in \cite{GW2}. 
 In the present case these considerations are anyway immaterial since the typical wavenumbers 
 relevant for magnetogenesis are typically much smaller i.e. ${\mathcal O}(\mathrm{Mpc}^{-1})$ 
 as we shall see. 
 
In a complementary perspective it should also be mentioned that the maximal frequency of the spectrum could be 
affected by the details of the reheating dynamics. As already suggested in the previous paragraph 
the continuity of the extrinsic curvature is an excellent starting point also in this case.
More detailed models of reheating could be certainly useful and might affect the magnetogenesis scenarios 
(as suggested in \cite{GW3}).  In this paper we adopted the approximation of the sudden reheating which
 is also the one often employed in order to set limits on the tensor to scalar ratio
and on the total number of inflationary $e$-folds. We shall however relax this approximation in phenomenological 
discussion of section \ref{sec5}.

\subsubsection{Decreasing gauge coupling}
\label{subs242}
\begin{figure}[!ht]
\centering
\includegraphics[height=7cm]{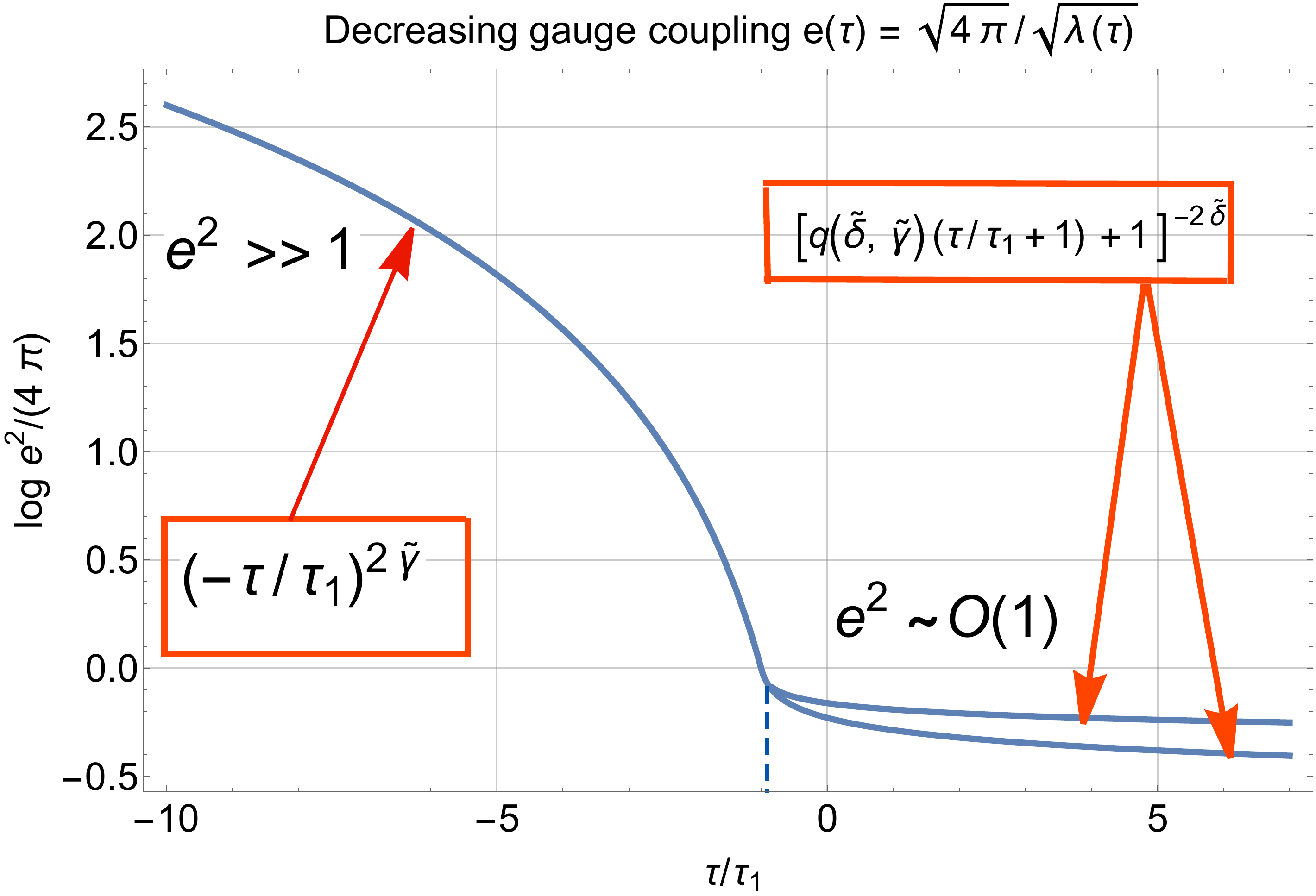}
\caption[a]{We illustrate the main notations employed when discussing the case of the decreasing gauge coupling. As in the case 
of Fig. \ref{FFF0a} the two different curves 
for $\tau \geq - \tau_{1}$ correspond to different values of $\widetilde{\,\delta\,} \ll 1$. Note that $q(\widetilde{\,\delta\,}, \widetilde{\,\gamma\,}) = \widetilde{\,\delta\,}/\widetilde{\,\gamma\,}$ [see also Eq. (\ref{SIX7})].}
\label{FFF0b}      
\end{figure}

For the purposes of the present analysis it is useful to complement the timeline of Fig. \ref{FFF0a} 
with Fig. \ref{FFF0b} describing the dual situation where the gauge coupling is initially 
much larger than $1$.
In the situation illustrated by Fig. \ref{FFF0b} the continuity of $\sqrt{\lambda}$ and $\sqrt{\lambda}^{\,\,\prime}$ 
imply the following parametrization
\begin{eqnarray}
\sqrt{\lambda} &=& \sqrt{\lambda_{1}} \biggl( - \frac{\tau}{\tau_{1}} \biggr)^{-\widetilde{\,\gamma\,}} , \qquad \tau < - \tau_{1},
\label{THREE1}\\
\sqrt{\lambda} &=& \sqrt{\lambda_{1}} \biggl[\frac{\widetilde{\,\gamma\,}}{\widetilde{\,\delta\,}} \biggl( \frac{\tau}{\tau_{1}} + 1\biggr) +1 \biggr]^{\widetilde{\,\delta\,}}, \qquad \tau \geq - \tau_{1},
\label{SIX2} 
\end{eqnarray}
The physical region of the parameters is given by:
\begin{equation} 
\widetilde{\,\gamma\,} > 0, \qquad  \widetilde{\,\delta\,} \geq 0, \qquad \mathrm{and}\qquad 0< \widetilde{\,\delta\,} \ll  \widetilde{\,\gamma\,}.
\label{SIX3}
\end{equation}
The limit $\widetilde{\,\delta\,} \to 0$
denotes again the sudden approximation while the smooth approximation holds for $\widetilde{\,\delta\,} \ll 1$
or, which is the same, $\widetilde{\,\delta\,} \ll \widetilde{\,\gamma\,}$ since we shall assume throughout that $\widetilde{\,\gamma\,}$ is of order $1$.  As in the case of increasing gauge coupling (see Eqs. (\ref{TWO1})--(\ref{FIVE2})), Eqs. (\ref{THREE1})--(\ref{SIX2})
will be complemented by the smooth evolution of the geometry illustrated in Eq. (\ref{FIVE1}). 
The dual profiles of Figs. \ref{FFF0a} and \ref{FFF0b} are not physically equivalent. If we consider a certain reference time $\tau= - \tau_{i}$ close to the onset of the inflationary phase, we 
will have, according to Eq. (\ref{THREE1}) that 
\begin{equation}
\sqrt{\lambda_{i}} = \sqrt{\lambda_{1}} \biggl(\frac{a_{i}}{a_{f}}\biggr)^{\widetilde{\,\gamma\,}} \ll 1 \Rightarrow
e(\tau_{i}) = \frac{\sqrt{4\pi}}{\sqrt{\lambda_{i}}} \gg 1,
\label{COMP1}
\end{equation}
where, by definition, $\lambda_{i} = \lambda(- \tau_{i})$; in Eq. (\ref{COMP1}) we traded the 
conformal time for the scale factors by using Eq. (\ref{FIVE1}) in the limit $\epsilon \ll 1$.
Note in fact that $(a_{i}/a_{f})^{\widetilde{\,\gamma\,}} = e^{ - N \,\widetilde{\,\gamma\,}}\ll 1$ 
where $N$ is the total number of inflationary $e$-folds. Since $N = {\mathcal O}(60)$ (or larger)
we have that  $\sqrt{\lambda_{i}}$ will be ${\mathcal O}(10^{-60})$ (or smaller). 
Equation (\ref{COMP1}) implies that the evolution of the gauge coupling starts from a non-perturbative regime unless $\sqrt{\lambda_{1}}$ is 
extremely large: only in this way we would have $\sqrt{\lambda_{i}} = {\mathcal O}(1)$. Whenever 
$\sqrt{\lambda_{1}} \gg 1$ the gauge coupling will be extremely minute at the end of inflation 
and this is at odds with the fact that during the decelerated stage of expansion 
we would like to have $e^2 = {\mathcal O}(10^{-2})$ but not much smaller. Furthermore, as we shall 
see in section \ref{sec5}, the physical power spectra are suppressed as $\lambda_{1}^{-1}$ and this 
will make their contribution marginal for the phenomenological implications. A possibility 
suggested in Ref. \cite{poss} has been that the $\sqrt{\lambda}$ increases during inflation, decreases 
sharply during reheating, and then flattens out again. This suggestion 
is often assumed by various authors but rarely justified. 

\renewcommand{\theequation}{3.\arabic{equation}}
\setcounter{equation}{0}
\section{Inflationary gauge spectra and their constraints}
\label{sec3}

The qualitative description of large-scale cosmological perturbations 
suggests that a given wavelength exits the Hubble radius at some typical conformal time 
during an inflationary stage of expansion and approximately reenters 
at $\tau_{k} \sim 1/k$, when the Universe still expands but in a decelerated manner. 
By a mode being beyond the horizon we only mean that the physical wavenumber 
is much less than the expansion rate:  this does not necessarily 
have anything to do with causality \cite{wein1,wein2}. Similarly the physical wavenumbers of the hyperelectric and hypermagnetic fields 
can be much smaller than the rate of variation of the gauge coupling
which now plays the role of the effective horizon. 
During inflation the relevant regime will be the one where the wavelengths of the gauge 
fluctuations are inside the effective horizon which means that  $k /{\mathcal F} \simeq k \tau  \ll 1$.
As long as $k \tau <1$ duality will be a valid symmetry but as soon 
as $k \tau_{k} \sim 1$ the electric spectra will be suppressed by conductivity and the gauge spectra 
will not be related by duality (see in this respect section \ref{sec5}).
Not all the wavelengths that are larger than  the effective horizon will 
will reenter at the end of inflation (i.e. for $\tau = -\tau_{1}$) but throughout the whole 
radiation phase. Thus the gauge spectra for wavelengths larger than the effective 
horizon during inflation do not necessarily coincide with the gauge spectra when the given mode reenters
(see, in this respect, section \ref{sec4}).

The logic of this section will be to compute the gauge spectra 
during inflation for the profiles schematically introduced in 
Figs. \ref{FFF0a} and \ref{FFF0b}.  We shall then
demonstrate that the gauge power spectra in the two cases
 are explicitly related by duality. This will mean, for instance, that during inflation the generation 
 of potentially scale-invariant hyperelectric spectrum is only compatible with a phase where the gauge coupling decreases 
  while a flat hypermagnetic spectrum may only arise when the gauge coupling decreases. 

\subsection{Increasing gauge coupling}
When $\sqrt{\lambda}$ is given by (\ref{TWO1}) the solution of Eq. (\ref{ONE8})
compatible with the Wronskian normalization dictated by Eq. (\ref{ONE8w}) is given by\footnote{For $\gamma \to 1/2$, $\mu\to 0$ and, in this limit, Eqs. (\ref{TWO3}) and (\ref{TWO4}) coincide exactly since $H_{-1}^{(1)}(z)= e^{i\,\pi} H_{1}^{(1)}(z)$. See also Eq. (\ref{TWO4b}).}:
\begin{eqnarray}
f_{k}(\tau) &=& \frac{N_{\mu}}{\sqrt{2 k}} \, \sqrt{- k\tau} \, H_{\mu}^{(1)}(-k\tau), \qquad \mu= \biggl| \gamma - \frac{1}{2} \biggr|,
\label{TWO2}\\
g_{k}(\tau) &=& N_{\mu} \,\sqrt{\frac{k}{2}} \,  \sqrt{- k\tau} \, H_{\mu+1}^{(1)}(-k\tau),\qquad \gamma > \frac{1}{2},
\label{TWO3}\\
g_{k}(\tau)  &=& - N_{\mu} \,\sqrt{\frac{k}{2}} \,  \sqrt{- k\tau} \, H_{\mu-1}^{(1)}(-k\tau),\qquad 0< \gamma < \frac{1}{2}. 
\label{TWO4}
\end{eqnarray}
In general $H_{\nu}^{(1)}(z)$ will denote the Hankel functions first kind \cite{abr1,abr2} with argument $z$ and index $\nu$;  $N_{\nu}$ is a complex number whose phase is required 
for a correct asymptotic normalization of the mode functions:
\begin{equation}
H_{\nu}^{(1)}(z) = J_{\nu}(z) + i Y_{\nu}(z), \qquad N_{\nu} = \sqrt{\frac{\pi}{2}} e^{i \pi ( 2 \nu + 1)/4}.
\label{TWO4a}
\end{equation}
Throughout the whole discussion we shall assume, as in the standard theory of Hankel functions 
\cite{abr1,abr2} that $\nu$ is real and positive semi-definite:
\begin{equation}
\mathrm{Re}\,\nu \geq 0, \qquad \mathrm{Im}\,\nu =0, \qquad H_{-\nu}^{(1)}(z)= e^{i\,\pi\, \nu} H_{\nu}^{(1)}(z),\qquad 
  H_{-\nu}^{(2)}(z)= e^{-i\,\pi\, \nu} H_{\nu}^{(2)}(z).
\label{TWO4b}
\end{equation}
Inserting the correctly normalized mode functions of Eqs. (\ref{TWO2}), (\ref{TWO3}) and 
(\ref{TWO4}) into Eqs. (\ref{ONE14a})--(\ref{ONE14b}) the gauge spectra
for $\tau \leq - \tau_{1}$ turn out to be\footnote{A term $|1 -\epsilon|^4$ has been omitted in the prefactors since it coincides 
with $1$ for $\epsilon \ll 1$. }:
\begin{eqnarray}
P_{B}(k,\tau) &=& \frac{a^4 H^4}{8\pi} (- k\tau)^5 \bigl|H_{\mu}^{(1)}(-k\tau)\bigr|^2 \to a^4 \,\, H^4\,\, D(\mu)\,\, |k \tau|^{ 5 - 2 \mu}
\label{TWO5}\\
P_{E}(k,\tau) &=& \frac{a^4 H^4}{8\pi} (- k\tau)^5 \bigl|H_{\mu+1}^{(1)}(-k\tau)\bigr|^2 \to a^4 \,\, H^4\,\, D(\mu+1) \,\,|k \tau|^{ 5 - 2 (\mu+1)}, \quad \gamma > 1/2,
\label{TWO6}\\
P_{E}(k,\tau) &=& \frac{a^4 H^4}{8\pi} (- k\tau)^5 \bigl|H_{\mu-1}^{(1)}(-k\tau)\bigr|^2 \to a^4 \,\, H^4\,\, D(|\mu-1|) \,\,|k \tau|^{ 5 - 2 |\mu-1|}, \quad 0< \gamma < 1/2,
\label{TWO7}
\end{eqnarray}
where we introduced the function:
\begin{equation}
D(x) = 2^{2 x- 3} \frac{\Gamma^2(x)}{\pi^3},
\label{TWO7a}
\end{equation}
that will be used throughout the whole paper exactly with the same meaning; the function $D(x)$
arises when the gauge spectra are evaluated outside the effective horizon (i.e. for $|k\tau| < 1$)  and the corresponding Hankel 
functions are estimated using their limit for small arguments \cite{abr1,abr2}.
 Note, in this respect, that $(-k\tau)$ can also be expressed as\footnote{$N$ denotes the total number of inflationary $e$-folds and has nothing to do with the normalization $N_{\mu}$ of the mode functions.} 
\begin{equation}
(- k \tau) = \frac{k}{(1 - \epsilon) \, a\, H} \simeq \frac{k}{\, a\, H} =\frac{k}{\, a_{1}\, H} e^{- N}, \qquad \epsilon \ll 1, \qquad 
N= \ln{\biggl(\frac{a}{a_{1}}\biggr)}.
\label{TWO7b}
\end{equation}
Since, by definition, $\mu = |\gamma-1/2|$ in terms of $\gamma$ the comoving power spectra of Eqs. (\ref{TWO5}), (\ref{TWO6}) and (\ref{TWO7}) become:
\begin{eqnarray} 
P_{B}(k,\tau) &=&  a^4 \, H^4\, D(|\gamma -1/2|) |k \tau|^{ 5 - |2 \gamma-1|}, 
\label{TWO9}\\
P_{E}(k,\tau) &=&  a^4 \, H^4\, D(\gamma +1/2) |k \tau|^{4 - 2 \gamma}.
\label{TWO10}
\end{eqnarray}
The derivation of Eq. (\ref{TWO9}) from Eq. (\ref{TWO5}) is immediate from the definition of $\mu$ in terms of $\gamma$ (i.e. 
 $\mu = |\gamma-1/2|$). On the contrary Eq. (\ref{TWO10}) is the common expression of Eqs. (\ref{TWO6}) and (\ref{TWO7}).
In fact, in Eq. (\ref{TWO6}) $\mu = \gamma -1/2$ (and therefore $\mu +1 = \gamma+1/2$), conversely 
in Eq. (\ref{TWO7}) $\mu = 1/2 -\gamma$ and $|\mu -1|$ always equals $\gamma+1/2$. After direct insertion of Eqs. 
 (\ref{TWO9})--(\ref{TWO10})  into  Eq. (\ref{ONE15}) the spectral energy density  for $\tau \leq - \tau_{1}$ is
\begin{eqnarray}
\Omega_{Y}(k,\tau) &=& \frac{2}{3} \biggl(\frac{H}{M_{P}}\biggr)^2 \biggl[ D_{B}(|\gamma -1/2|) |k\tau|^{5 - |2 \gamma-1|} + D_{E}(\gamma +1/2) |k\tau|^{4 - 2 \gamma} \biggr].
\label{TWO11}
\end{eqnarray}
The spectral energy density of Eq. (\ref{TWO11}) must be always 
subcritical (i.e. $\Omega_{Y}(k,\tau) \ll 1$)  for $\tau \leq -\tau_{1}$ and $|k \tau| \leq 1$; this requirement is not always 
satisfied even if, during the inflationary phase, $H \ll M_{P}$. 
The first term inside the square bracket at the right hand side of Eq. (\ref{TWO11}) denotes the magnetic contribution while the second term is the electric result\footnote{To stress this 
we just added a subscript to the function $D(x)$ (by writing $D_{B}(x)$ and $D_{E}(x)$) even if the definition of $D(x)$ (see after Eq. (\ref{TWO7})) is the same in both cases.}. By looking together at Eqs. (\ref{TWO9}), (\ref{TWO10}) 
and (\ref{TWO11}) the following conclusions naturally emerge:
\begin{itemize}
\item{} if $\gamma= 2$ the hyperelectric spectrum is exactly scale-invariant while 
the magnetic spectrum is steeply increasing [i.e. $P_{B}(k,\tau) \propto |k \tau|^2$ for $k \tau \ll 1$]
so that  the condition $\Omega_{Y}(k,\tau) \ll 1$ is safely satisfied; 
\item{} if  $1/2 < \gamma \leq 2$ the requirement $\Omega_{Y}(k,\tau) \ll 1$ always holds for $| k \tau \ll 1$ but, 
according to the general wisdom, the 
magnetogenesis constraints cannot be satisfied since the hypermagnetic power spectrum is still too violet;
\item{} if $\gamma > 2$ the hypermagnetic spectrum becomes even steeper while the hyperelectric spectrum diverges
in the limit $k\tau \ll 1$: in this case the bound $\Omega_{Y} \ll 1$ is not satisfied so that the whole class of models 
$\gamma > 2$ must be excluded.
\end{itemize}
The only case not explicitly covered is  $\gamma \to 1/2$ but in this case both the electric and the magnetic spectra are steeply increasing while the magnetic spectrum inherits a logarithmic correction\footnote{When $\gamma \to 1/2$ the mode functions become
$ f_{k}(\tau) = N_{0} \sqrt{- k\tau} \, H_{0}^{(1)}(-k\tau)/\sqrt{2 k}$ and
$ g_{k}(\tau) = N_{0} \, \sqrt{k/2} \, \sqrt{- k\tau} \, H_{1}^{(1)}(- k \tau)$,
where $N_{0} = \sqrt{\pi/2} \,e^{i\pi/4}$.}:
\begin{eqnarray}
P_{B}(k,\tau) &=& \frac{a^4 H^4}{8 \pi} (-k\tau)^5 \, \bigl| H_{0}^{(1)}(- k\tau)\bigr|^2 \to \frac{a^4 H^4}{2 \pi^3} |k\tau|^{5} \ln^2|k\tau|,
\label{TWO15}\\
P_{E}(k,\tau) &=& \frac{a^4 H^4}{8 \pi} (-k\tau)^5 \, \bigl| H_{1}^{(1)}(- k\tau)\bigr|^2 \to \frac{a^4 H^4}{2 \pi^3} |k\tau|^{3},
\label{TWO16}\\
\Omega_{Y}(k,\tau) &=& \frac{1}{3 \pi^3} \biggl(\frac{H}{M_{P}}\biggr)^2 |k\tau|^3 \biggl[ 1 + |k\tau|^2 \ln^2{|k\tau|} \biggr].
\label{TWO17}
\end{eqnarray}
Therefore, also in the case $\gamma \to 1/2$ the spectral energy density is subcritical when the typical wavelengths are larger than the effective horizon.

If we consider Eq. (\ref{TWO11}) at face value we must admit that for $\gamma \leq 2$ the critical density 
constraint is always satisfied since both the hyperelectric and the hypermagnetic 
power spectra are extremely minute; the relation among them 
can be easily deduced from Eqs. (\ref{TWO9})--(\ref{TWO10})  and it is given by:
\begin{eqnarray}
P_{B}(k,\tau) &=& (- k \tau)^2 P_{E}(k, \tau), \qquad\qquad \mathrm{for}\qquad \gamma > 1/2,
\label{TWO17a}\\
P_{B}(k,\tau) &=& (- k \tau)^{4 \gamma}\, P_{E}(k, \tau), \qquad\qquad \mathrm{for} \qquad 0< \gamma < 1/2.
\label{TWO17b}
\end{eqnarray}
In both cases the hypermagnetic field spectrum is smaller than its hyperelectric counterpart.
This feature may also arise in other models but it is not problematic 
as long as the power spectra are both very small.  There have been recently various suggestions  attempting 
to include spectator (electric) fields in curved backgrounds \cite{SCH1,SCH2} with the purpose 
of estimating the Schwinger effect in de Sitter space-time. These estimates are typically
done by assuming a constant and uniform electric field in de Sitter space and by also postulating a vector-like coupling to fermions. 
These three assumptions do not specifically hold in the present case since the fields obtained here are non-uniform and non-homogeneous, 
they are not constant in time and they have, in general, a chiral coupling to fermions.  

It is interesting to note that the estimates of the Schwinger-like effects are highly non-generic in the sense 
that to have a sizeable effect we need to concoct specific currents forbidding the dilution of the electric energy density 
by the expansion of the universe. Indeed, the constancy and uniformity of the field configurations entering the estimate 
of the above effect is achieved by considering a class of appropriate currents and this physical situation is markedly different  from the flat space-time 
case where the constancy and uniformity of the field imposes 
the absence of time-dependent currents. The idea of Refs. \cite{SCH1} (somehow questioned in Ref. \cite{SCH2})
is that to achieve a physical electric field that is constant\footnote{It turns out to be useful, in this respect, the distinction between 
physical and comoving fields given in Eqs. (\ref{ACT13}). The only caveat is that, in standard estimates of the Schwinger effect, the 
gauge coupling does not evolve. } one must actually require the presence of a comoving current so that the explicit form of
the corresponding evolution equations will be:
\begin{equation}
\nabla\times \vec{E} =0,\qquad \nabla\times \vec{B} =0, \qquad \vec{J} + \vec{E}^{\,\,\prime }=0, \qquad \vec{B}^{\,\,\prime} =0.
\label{TWO17b1}
\end{equation}
If one would simply choose $\vec{E} = E_{0} \hat{n}$ (where $\hat{n}$ is a certain unit vector), $\vec{B}=0$ and $\vec{J}=0$ this would be the standard Schwinger ansatz 
but the rate of pair production in de Sitter space would be negligible since the physical fields will be suppressed as $\vec{E}^{(phys)} = E_{0} \hat{n}/a^2$. 
To have a non-negligible effect the common wisdom is therefore to postulate an {\em ad hoc} current constructed in such a way that the 
resulting {\em physical field} is a space-time constant. This can be achieved by postulating a comoving field in the form 
$\vec{E} = a^{\lambda} E_{0} \hat{n}$ \cite{SCH2} implying that the corresponding current becomes, from Eq. (\ref{TWO17b1}),
$\vec{J} = - \lambda \, a \, H\, \vec{E}$. If $\lambda \leq 2$ the corresponding physical field is less suppressed and it becomes 
constant in the limit $\lambda\to 2$. These spectator electric fields can be understood as the result 
of an effective conductivity $\sigma= - \lambda H a $ which is negative if the universe expands (as we are discussing here).
It has been argued that this approach may cause a violation of the second law of thermodynamics \cite{SCH2}.
It is debatable if this kind of estimates are at all relevant for the situation described in this paper.
Even assuming that the coupling of hypercharge fields is vector-like (which is not the case as discussed above), 
 the effects of the current $\vec{J} = - \lambda \, a \, H\, \vec{E}$ are physically quite different from the 
 ones associated with the spectra (\ref{TWO17a})--(\ref{TWO17b}) which are fully inhomogeneous, time-dependent and 
 obtained in the absence of {\em any} supporting current.

Having specified what is the real physical situation we can always try to see under which conditions 
 the probability of pair creation  per unit volume per unit time is under control in some averaged sense. 
 Taking into account all the preceding caveats, and denoting with $\Gamma$ the rate of pair production 
 per unit volume and per unit time a reasonable condition to consider seems to be:
\begin{equation}
\frac{\Gamma}{H^4} = \frac{\alpha}{\pi^2} \langle E^2(\tau,\vec{x}) \rangle < 1,\qquad \Rightarrow 
\frac{2}{\pi^2 \, \lambda} \int_{0}^{k_{max}}\frac{d\, k}{k}\, P_{E}(k,\tau) < H^4,
\label{TWO17c}
\end{equation}
where we recalled that, within the conventions spelled out in Eq. (\ref{ONE1a}), $\alpha = e^2/(4\pi) = 1/\lambda$; we also 
assumed that the flat-space result holds (locally) over typical scales ${\mathcal O}(H^{-1})$ during the quasi-de Sitter 
stage.  Let us now consider, for simplicity, the case of Eq. (\ref{TWO17a}) and see if and how the condition 
(\ref{TWO17c}) is satisfied. In Eq. (\ref{TWO17c}) $k_{max}$ denotes the maximally amplified 
wavenumber which is given, in the present context $k_{max} = {\mathcal O}(1/\tau_{1})$. 
By using the explicit form of the power spectrum given in Eq. (\ref{TWO10}) 
the condition (\ref{TWO17c}) implies 
\begin{equation}
\frac{e_{1}^2 \, 2^{2 \gamma - 3}\Gamma^2(\gamma+1/2)\, \zeta^{4 - 2 \gamma}}{(4 - 2\gamma)\pi^{5} }\biggl(\frac{a_1}{a} \biggr)^{- 4 \gamma}\ll 1, \qquad \qquad 0< \gamma <2,
\label{TWO17d}
\end{equation}
where $\zeta$ is a numerical factor ${\mathcal O}(1)$ [we parametrized $k_{max} = \zeta/\tau_{1}$ by following 
the observations discussed after Eq. (\ref{FIVE1})]. The condition Eq. (\ref{TWO17d}) is always verified 
for $\tau < -\tau_{1}$ since in the wanted range of $\gamma$ we  have that $(a_{1}/a)^{- 4 \gamma} = (-\tau/\tau_{1})^{-4 \gamma} \ll 1$.
This suggests that the condition (\ref{TWO17d}) is always satisfied except, probably, at the very end of inflation. This is,
however, not a problem of the physics but just a problem of the acuracy in the estimate of the inflationary 
spectrum around $k_{max}$. Even in the worse situation the ideas conveyed in this paper come do not contradict the 
spirit of the bound (\ref{TWO17d}): since the gauge coupling at the end 
of inflation is effectively a tuneable parameter, it is sufficient to reduce $e_{1}^2$ by one order of magnitude to safely satisfy the conditions
(\ref{TWO17c}) and (\ref{TWO17d}) also at the end of inflation. All in all Eqs. (\ref{TWO17b1})--(\ref{TWO17c}) and (\ref{TWO17d}) 
suggest that the constraints coming from pair production are therefore not essential in the present context even if an improved 
understanding of the whole problem would be desirable (see e.g. \cite{SCH2}) and discussions therein).

In summary when the gauge coupling increases during a quasi-de Sitter stage of expansion the spectral energy density is subcritical for $0 < \gamma \leq 2$ and overcritical for $\gamma >2$ so that the latter range is excluded while the former is still viable. Since the hypermagnetic spectrum is rather steep (i.e. violet) when $\gamma=2$ the conventional wisdom is that it will also be minute at the galactic scale after the gauge coupling flattens out. This swift conclusion is only  true provided the hypermagnetic magnetic power spectrum at the end of inflation is not modified for $\tau \geq -\tau_{1}$. In section \ref{sec4} 
the gauge power spectra will be explicitly computed in the regime where the gauge coupling flattens out (i.e. for $\tau \geq - \tau_{1}$) and it will be shown that the late-time hypermagnetic spectrum does not coincide with the hypermagnetic spectrum at the end of inflation when the gauge coupling increases.

\subsection{Decreasing gauge coupling}

When $\sqrt{\lambda}$ evolves as in Eq. (\ref{THREE1}) (see also Fig. \ref{FFF0b}) the
correctly normalized solution of Eqs. (\ref{ONE8}) and (\ref{ONE8w}) are now given by
\begin{eqnarray}
f_{k}(\tau) = \frac{N_{\widetilde{\mu}}}{\sqrt{2 k}} \, \sqrt{- k\tau} \, H_{\widetilde{\mu}}^{(1)}(-k\tau),\qquad 
g_{k}(\tau) = N_{\widetilde{\mu}} \,\sqrt{\frac{k}{2}} \,  \sqrt{- k\tau} \, H_{\widetilde{\mu}-1}^{(1)}(-k\tau), 
\label{THREE2}
\end{eqnarray}
where 
\begin{equation}
\widetilde{\mu}= \widetilde{\,\gamma\,} + \frac{1}{2},\qquad N_{\widetilde{\mu}} = \sqrt{\frac{\pi}{2}} e^{i \pi ( 2 \widetilde{\mu} + 1)/4}.
\label{THREE3}
\end{equation}
Inserting Eqs. (\ref{THREE2}) and (\ref{THREE3}) into Eqs. (\ref{ONE14a}) and (\ref{ONE14b}) the comoving 
 power spectra are: 
 \begin{eqnarray}
\widetilde{\,P\,}_{B}(k,\tau) &=& \frac{a^4 H^4}{8\pi} (- k\tau)^5 \bigl|H_{\widetilde{\mu}}^{(1)}(-k\tau)\bigr|^2 \to a^4 \, H^4\, D(\widetilde{\mu})\, |k \tau|^{ 5 - 2 \widetilde{\mu}},
\label{THREE4}\\
\widetilde{\,P\,}_{E}(k,\tau) &=& \frac{a^4 H^4}{8\pi} (- k\tau)^5 \bigl|H_{\widetilde{\mu}-1}^{(1)}(-k\tau)\bigr|^2 \to a^4 \, H^4\, D(|\widetilde{\mu}-1|)\, |k \tau|^{ 5 - 2 \,| \widetilde{\mu} -1|},
\label{THREE5}
\end{eqnarray}
where the function $D(x)$ has been already defined in Eq. (\ref{TWO7a}) for a generic argument $x$.  To stress that the power spectra (\ref{THREE4})--(\ref{THREE5}) correspond to the case of decreasing gauge coupling a tilde has been added on top of each expression.  Since, according to Eq. (\ref{THREE3}), $\widetilde{\mu} = \widetilde{\,\gamma\,} +1/2$,  Eqs. (\ref{THREE4}) and (\ref{THREE5}) are directly expressible in terms of $\widetilde{\,\gamma\,}$ and the result is:
\begin{eqnarray} 
\widetilde{\,P\,}_{B}(k,\tau) &=&  a^4 \, H^4\, D(\widetilde{\,\gamma\,} + 1/2) |k \tau|^{ 4 - 2 \widetilde{\,\gamma\,}}, 
\label{THREE6}\\
\widetilde{\,P\,}_{E}(k,\tau) &=& a^4 \, H^4\, D(|\widetilde{\,\gamma\,} - 1/2|) |k \tau|^{ 5 - | 2 \widetilde{\,\gamma\,}- 1|}.
\label{THREE7}
\end{eqnarray}
Equations (\ref{THREE6})--(\ref{THREE7}) will now be inserted into Eq. (\ref{ONE15}); the explicit 
expression of the spectral energy density is:
\begin{eqnarray}
\widetilde{\,\Omega\,}_{Y}(k,\tau) = \frac{2}{3} \biggl(\frac{H}{M_{P}}\biggr)^2 \biggl[ D_{B}(\widetilde{\,\gamma\,} +1/2) \, |k\tau|^{4 - 2 \widetilde{\,\gamma\,}} + D_{E}(|\widetilde{\,\gamma\,} -1/2|) \,|k\tau|^{5 - |2 \widetilde{\,\gamma\,} -1|} \biggr],
\label{THREE9}
\end{eqnarray}
where in analogy with Eq. (\ref{TWO11})  the subscripts $E$ and $B$ have been added to the function $D(x)$ with the purpose of reminding that the origin of the corresponding terms.
As in the case of Eq. (\ref{TWO11}) the spectral energy density of Eq. (\ref{THREE9}) should always be subcritical (i.e. $\widetilde{\Omega}_{Y}(k,\tau) \ll 1$ for $\tau\leq -\tau_{1}$ and $|k \tau|\ll 1$). In this respect  the following three observations are in order: 
\begin{itemize}
\item{}  when $\widetilde{\,\gamma\,} =2 $ the hypermagnetic spectrum is flat while the hyperelectric spectrum is violet and it goes as
$\widetilde{\,P\,}_{E}(k,\tau) \propto |k\tau|^2$; the spectral energy density is always subcritical for $|k\tau| \ll 1$;
\item{} if $\widetilde{\,\gamma\,} =3$ the hyperelectric spectrum is flat however the hypermagnetic power spectrum diverges as $|k\tau|^{-2}$ in the limit $|k \tau| \ll1$ but, in this case, the spectral energy density constraint is not satisfied;
\item{} in the $0<\widetilde{\,\gamma\,} < 1/2$ the power spectra are both violet;  all in all we conclude 
that the bound (\ref{THREE9}) is satisfied for $0 < \widetilde{\,\gamma\,} \leq 2$.
\end{itemize}
The case $\widetilde{\,\gamma\,}\to 1/2$ is not explicitly covered by Eq. (\ref{THREE9}) so that, in this limit, we have\footnote{When $\widetilde{\,\gamma\,} =1/2$ 
the solution of Eqs. (\ref{ONE8})--(\ref{ONE8w}) is given by 
$f_{k}(\tau) = N_{1}\sqrt{- k\tau} \, H_{1}^{(1)}(-k\tau)/\sqrt{2k}$ and by
$g_{k}(\tau) = - N_{1} \sqrt{\frac{k}{2}} \sqrt{- k\tau} \, H_{0}^{(1)}(-k\tau)$.}
\begin{eqnarray}
\widetilde{\,P\,}_{B}(k,\tau) &=& \frac{a^{4}\, H^{4}}{8 \pi}  (-k\tau)^5 \bigl| H_{1}^{(1)}(-k\tau)\bigr|^2 \to \frac{a^{4}\, H^{4}}{2\pi^3}  |k\tau |^{3},
\label{THREE13}\\
\widetilde{\,P\,}_{E}(k,\tau) &=& \frac{a^{4}\, H^{4}}{8 \pi}  (-k\tau)^5 \bigl| H_{0}^{(1)}(-k\tau)\bigr|^2 \to \frac{a^{4}\, H^{4}}{2\pi^3} |k\tau|^{5} \ln^2{|k\tau|},
\label{THREE14}\\
\widetilde{\, \Omega\,}_{Y}(k,\tau) &=& \frac{1}{3 \pi^3} \biggl(\frac{H}{M_{P}}\biggr)^2 |k\tau|^3 \biggl[ 1 + |k\tau|^2 \ln^2{|k\tau|} \biggr].
\label{THREE15}
\end{eqnarray}

\subsection{Duality and gauge power spectra during inflation}
A flat hypermagnetic spectrum is only consistent with the critical density bound provided  $\widetilde{\,\gamma\,} =2$ (i.e. only when the gauge coupling is decreasing). Conversely a flat hyperelectric spectrum is 
only viable when the gauge coupling increases and $\gamma =2$. Overall the only 
intervals where the critical density constraint is satisfied are given by $0\leq \gamma \leq 2$ 
and $0\leq \widetilde{\,\gamma\,} \leq 2$. These results follow in fact from the duality 
symmetry of Eqs. (\ref{ONE5}) and (\ref{ONE10a}). For the gauge spectra during inflation 
the duality symmetry implies, in general:
\begin{equation}
\sqrt{\lambda} \to \frac{1}{\sqrt{\lambda}}, \qquad P_{B}(k,\tau) \to \widetilde{\,P\,}_{E}(k,\tau), \qquad  P_{E}(k,\tau) \to \widetilde{\,P\,}_{B}(k,\tau)
\label{THREE16}
\end{equation}
If now  Eqs.  (\ref{TWO9})--(\ref{TWO10}) 
are compared with Eqs. (\ref{THREE6})--(\ref{THREE7}) the spectra with increasing and decreasing gauge coupling during inflation are explicitly related by the following transformation:
\begin{equation}
 \gamma \to \widetilde{\,\gamma\,},\qquad P_{B}(k,\tau) \to \widetilde{\,P\,}_{E}(k,\tau), \qquad P_{E}(k,\tau) \to \widetilde{\,P\,}_{B}(k,\tau).
\label{FOUR1}
\end{equation}
Equation (\ref{FOUR1}) has actually the same content of Eq. (\ref{THREE16}) since, according to the explicit 
parametrizations of Eqs. (\ref{TWO1}) and (\ref{THREE1}), the expression
of $\sqrt{\lambda}$ is actually inverted when $ \gamma \to \widetilde{\,\gamma\,}$. 
Thanks Eq. (\ref{FOUR1}) the spectral energy density is left invariant [i.e. $\Omega_{Y}(k,\tau) \to \widetilde{\Omega}_{Y}(k,\tau)$] since the hyperelectric and the hypermagnetic power spectra 
are interchanged.  All in all the standard lore stipulates that the only phenomenologically relevant case 
is the one $\widetilde{\,\gamma\,} =2$ since the other cases lead to an hypermagnetic spectrum 
that is either too steep or anyway inconsistent with the critical density bound. As already mentioned, this 
statement assumes that the gauge spectra are unmodified  when 
 coupling flattens out and the relevant wavelengths are still larger than the effective horizon.
 \renewcommand{\theequation}{4.\arabic{equation}}
\setcounter{equation}{0}
\section{Post-inflationary gauge spectra and continuity}
\label{sec4}
When the gauge coupling flattens out as 
illustrated in Figs. \ref{FFF0a} and \ref{FFF0b} the background geometry enters a stage of decelerated 
expansion after the end of the inflationary phase. The post-inflationary gauge fields are 
determined by the continuous evolution of the corresponding mode functions whose late-time behaviour 
follows from the elements of an appropriate transition matrix mixing together
 the hyperelectric and  the hypermagnetic  mode functions at the end of inflation 
and leading to specific standing oscillations. These Sakharov phases
are different for the $(\gamma,\delta)$  transition 
illustrated in Fig. \ref{FFF0a} and in the case of the 
$(\widetilde{\,\gamma\,}, \widetilde{\,\delta\,})$ profile of Fig. \ref{FFF0b}.

\subsection{Increasing gauge coupling}
\subsubsection{Mixing matrix for the $(\gamma, \, \delta)$ transition}
The continuous parametrization of $\sqrt{\lambda}$ given in
Eqs. (\ref{TWO1}) and (\ref{FIVE2}) implies that the late-time values values of $f_{k}(\tau)$ 
and $g_{k}(\tau)$ for $\tau \geq - \tau_{1}$ are given by:
\begin{equation}
\left(\matrix{ f_{k}(\tau) &\cr
g_{k}(\tau)/k&\cr}\right) = \left(\matrix{ A_{f\, f}(k, \tau, \tau_{1})
& A_{f\,g}(k,\tau, \tau_{1})&\cr
A_{g\,f}(k,\tau, \tau_{1}) &A_{g\,g}(k,\tau, \tau_{1})&\cr}\right) \left(\matrix{ \overline{f}_{k} &\cr
\overline{g}_{k}/k&\cr}\right),
\label{FIVE11}
\end{equation}
where $\overline{f}_{k}= f_{k}(-\tau_{1})$ and $\overline{g}_{k} = g_{k}(-\tau_1)$ 
denote the values of the mode functions at end of the inflationary phase and the matrix elements 
at the right hand side of Eq. (\ref{FIVE11}) are determined from the continuity of the mode functions as described in appendix \ref{APPA}:
\begin{eqnarray}
A_{f\, f}(k,\tau, \tau_{1}) &=& \frac{\pi}{2} \sqrt{q x_{1}} \sqrt{ k y} \biggl[ Y_{\nu -1}( q x_{1}) J_{\nu}(k y) - J_{\nu-1}(q x_{1}) Y_{\nu}(k y) \biggr],
\nonumber\\
A_{f\, g}(k, \tau, \tau_{1}) &=& \frac{\pi}{2} \sqrt{q x_{1}} \sqrt{ k y} \biggl[ J_{\nu}( q x_{1}) Y_{\nu}(k y) - Y_{\nu}(q x_{1}) J_{\nu}(k y) \biggr],
\nonumber\\
A_{g\, f}(k, \tau, \tau_{1}) &=& \frac{\pi}{2} \sqrt{q x_{1}} \sqrt{ k y} \biggl[ Y_{\nu - 1}( q x_{1}) J_{\nu -1}(k y) - J_{\nu-1}(q x_{1}) Y_{\nu-1}(k y) \biggr],
\nonumber\\
A_{g\, g}(k, \tau, \tau_{1}) &=& \frac{\pi}{2} \sqrt{q x_{1}} \sqrt{ k y} \biggl[ J_{\nu}( q x_{1}) Y_{\nu-1}(k y) - Y_{\nu}(q x_{1}) J_{\nu-1}(k y) \biggr].
\label{FIVE13}
\end{eqnarray}
Since the inflationary mode functions $\overline{f}_{k}$ and $\overline{g}_{k}$ obey the Wronskian 
normalization of Eq. (\ref{ONE8w}), also $f_{k}(\tau)$ and $g_{k}(\tau)$ must obey the same 
condition for $\tau \geq \tau_{1}$ and this happens provided\footnote{The validity of this condition can also be verified by plugging the explicit matrix elements of Eq. (\ref{FIVE13}) into Eq. (\ref{FIVE12a})
and by using the standard recurrence relations involving the Bessel functions and their Wronskians \cite{abr1,abr2}.}: 
\begin{equation}
A_{f\, f}(k, \tau, \tau_{1}) A_{g\, g}(k, \tau, \tau_{1}) - A_{f\, g}(k, \tau, \tau_{1}) A_{g\, f}(k, \tau, \tau_{1}) =1.
\label{FIVE12a}
\end{equation}
The arguments of the Bessel functions appearing in (\ref{FIVE13}) depend on $q x_{1}$ and $k \, y$ while the corresponding indices depend on $\delta$; the explicit expressions of $q$, $y$ and $\nu$ are: 
\begin{equation}
q(\delta, \gamma) = \frac{\delta}{\gamma}, \qquad y(\tau,\delta,\gamma) = \tau + \tau_{1}[1 + q(\delta,\gamma)], \qquad \nu(\delta)  = \delta +1/2.
\label{FIVE13def1}
\end{equation}
Equation (\ref{FIVE13def1}) shows that, within the present notations,  $y(-\tau_{1}) = q \tau_{1}$ which also 
implies (by definition of $x_{1}$) that  $k y(-\tau_{1}) = q \, k\, \tau_{1} = q x_{1}$. Consequently, as expected from the continuity of the mode functions, 
\begin{equation}
A_{f\,g}(k, - \tau_{1}, \tau_{1}) =A_{g\,f}(k, - \tau_{1}, \tau_{1})=0, \qquad A_{f\,f}(k, - \tau_{1}, \tau_{1}) =A_{g\,g}(k, - \tau_{1}, \tau_{1})=1.
\end{equation}
Since all the expression entering Eq. (\ref{FIVE13})  ultimately depend on
the dimensionless variables $x = k \tau$, $x_{1} = k \tau_{1}$ and $\nu$,  the matrix appearing in Eq. (\ref{FIVE11}) is in fact a function of $\delta$, $x$ and $x_{1}$ for any fixed value of $\gamma$:
\begin{equation}
{\mathcal M}(\delta, x,\, x_{1}) =  \left(\matrix{ A_{f\, f}(\delta, \,x,\, x_{1})
& A_{f\,g}(\delta,\, x,\, x_{1})&\cr
A_{g\,f}(\delta,\, x,\,x_{1}) &A_{g\,g}(\delta,\, x,\,x_{1})&\cr}\right).
\label{FIVE13def2}
\end{equation} 
We stress that the variable $x_{1} = k \tau_{1} \leq 1$ measures $k$ in units of the maximal wavenumber of the spectrum (i.e. $1/\tau_{1} = a_{1} H_{1}$) and this is why it cannot be larger than ${\mathcal O}(1)$. 
  
\subsubsection{General form of the power spectra}
When the mode functions $f_{k}(\tau)$ and $g_{k}(\tau)$ are deduced from Eq. (\ref{FIVE11}) the gauge power spectra of Eqs. (\ref{ONE14a})--(\ref{ONE14b})  become:
\begin{eqnarray}
P_{B}(k,\tau) &=& \frac{k^5}{2\pi^2} \biggl| A_{f\, f} \, \overline{f}_{k} + A_{f\, g} \,\frac{\overline{g}_{k}}{k}\biggr|^2,
\label{FIVE15}\\
P_{E}(k,\tau) &=& \frac{k^3}{2\pi^2} \biggl| k\, A_{g\, f} \, \overline{f}_{k} + A_{g\, g} \,\overline{g}_{k}\biggr|^2.
\label{FIVE15aa}
\end{eqnarray}
One of the two terms 
inside each of the squared moduli appearing in Eqs. (\ref{FIVE15}) and (\ref{FIVE15aa}) will be alternatively dominant. 
For the sake of concreteness  we shall now verify that the first term 
inside the squared modulus of Eq. (\ref{FIVE15}) dominates against the second:
\begin{equation}
\biggl| A_{f\, g}(\delta, x, x_{1})\, \frac{\overline{g}_{k}}{k} \biggr| \gg \biggl| A_{f\, f}(\delta, x, x_{1}) \, \overline{f}_{k} \biggr|.
\label{DEM1}
\end{equation}
Since all the variables have been explicitly defined, the validity of the condition (\ref{DEM1}) could 
be investigated numerically\footnote{This analysis has been performed by fixing $\gamma$ to a reference 
value and by scanning the relative weight of the two terms of Eq. (\ref{DEM1}) for different values 
of $\delta \ll \gamma$. For the sake of conciseness this numerical analysis will not be explicitly 
discussed.}. For a more general proof  
of Eq. (\ref{DEM1}) it is sufficient to consider Eq. (\ref{DEM1}) in the limit $x_{1} < 1$ (which is 
always verified for all the amplified modes of the spectrum) together with the subsidiary condition that  $ 0 \leq \delta \ll \gamma$. An equivalent form of Eq. (\ref{DEM1}) and the result is:
 \begin{equation}
 \biggl|J_{\nu}(q x_{1}) Y_{\nu}(k y) - Y_{\nu}(q x_{1}) J_{\nu}(k y) \biggr|\gg \biggl|\frac{H_{|\gamma -1/2|}^{(1)}(x_{1})}{H_{\gamma +1/2}^{(1)}(x_{1})} \biggr|\, \biggl| Y_{\nu-1 }(q x_{1}) J_{\nu}(k y) - J_{\nu-1}(q x_{1}) Y_{\nu}(k y) \biggr|.
 \label{DEM3}
 \end{equation}
If $x_{1} \ll x\ll 1$ the arguments of the Bessel and Hankel 
functions appearing in Eq. (\ref{DEM3}) are all smaller than $1$. It then follows that, in this limit  (i.e. 
$x_{1}/x= \tau_{1}/\tau \ll 1$), the inequalities (\ref{DEM1})--(\ref{DEM3}) 
are verified provided:
\begin{equation}
\biggl(\frac{q x_{1}}{2}\biggr)^{ - 2 \delta} \gg \frac{\Gamma(1/2 -\delta) \Gamma(|\gamma -1/2|)}{\Gamma(1/2 +\delta) \Gamma(\gamma +1/2)}\, \biggl(\frac{x_{1}}{2}\biggr)^{\gamma +1/2 - |\gamma -1/2|},
\label{DEM4}
\end{equation}
where we used the explicit result of Eq. (\ref{FIVE12aa}). 
When $\gamma>1/2$ the condition (\ref{DEM4}) implies, up 
to irrelevant numerical factors, that $(q x_{1}/2)^{-2 \delta} > (x_{1}/2)$ which is always verified\footnote{In the limit $\delta \to 0$, since $q = \delta/\gamma$, the condition  $(q x_{1}/2)^{-2 \delta} \gg (x_{1}/2)$ is also 
verified as long as $x_{1} < 1$. Since the gauge coupling freezes (either partially 
or totally) for $\tau > -\tau_{1}$ the physical situation discussed here 
corresponds to a range of parameters where $ 0 \leq \delta \ll 1/2$
which also implies $ 0 \leq \delta \ll \gamma$; these two conditions 
will be used interchangeably.}
as long as $x_{1} < 1$ and $\delta \geq 0$. Similarly, if $0< \gamma < 1/2$, Eq. (\ref{DEM4}) implies 
$(q x_{1}/2)^{-2 \delta} > (x_{1}/2)^{\gamma}$ which is also verified in the physical range of the 
parameters. 

So far we demonstrated  that Eq. (\ref{DEM1}) holds in the range $x_{1} \ll x \ll 1$.
When $x_{1} \ll 1$ and $x\gg 1$  the functions whose argument  coincides with $k y \simeq x \gg 1$ can be always represented as\footnote{Recall, in this respect, Eq. (\ref{FIVE13def1}). Even if the value of $x$ can be either smaller or larger 
than $1$, as soon as $ x = k \tau = {\mathcal O}(1)$ the conductivity cannot be neglected and this situation will be more specifically discussed in section \ref{sec5}. For the moment we shall just consider 
the case $x\gg 1$ as a mere mathematical limit.} :
\begin{equation}
J_{\nu}(k y) = M_{\nu} \cos{\theta_{\nu}}, \qquad Y_{\nu}(k y) = M_{\nu} \sin{\theta_{\nu}},
\label{DEM5}
\end{equation}
where, for $x\gg 1$, $\theta_{\nu}(x) \to  x $ while $M_{\nu}(x) \to \sqrt{2/\pi} x^{-1/2}[ 1 + {\mathcal O}(x^{-2})]$; this is the so-called modulus-phase approximation for the Bessel functions \cite{abr1,abr2}.
Inserting Eq. (\ref{DEM5}) into Eq. (\ref{DEM3}) and expanding the remaining functions for $x_{1}\ll1$ 
it then follows that the inequality (\ref{DEM1}) is satisfied. Consequently,  thanks to Eq. (\ref{DEM1})
 the comoving spectrum of Eq. (\ref{FIVE15}) is:
\begin{equation}
P_{B}(k,\tau) = \frac{k^5}{2\pi^2} \biggl| A_{f\, g}(\delta,\, x_{1}, \,x)\, \frac{\overline{g}_{k}}{k}\biggr|^2.
\label{DEM6}
\end{equation}
The results of 
Eqs. (\ref{DEM1})--(\ref{DEM6}) show that the 
hyperelectric field at the end of inflation determines the late-time hypermagnetic field 
for $\tau \gg - \tau_{1}$. This is of course not a general truism but it 
happens provided the gauge coupling first increases during inflation and then flattens out 
in the radiation-dominated epoch. For 
the  hyperelectric spectrum of Eq. (\ref{FIVE15aa}) the inequality of Eq. (\ref{DEM1}) is in fact replaced by 
the following condition 
\begin{equation}
\biggl| A_{g\, g}(\delta, x, x_{1})\,  \overline{g}_{k}\biggr| \gg \biggl| A_{g\, f}(\delta, x, x_{1}) \,k\, \overline{f}_{k} \biggr|,
\label{DEM2}
\end{equation}
which can be verified explicitly by using the same strategy illustrated in the case of Eq. (\ref{DEM1}); for the 
sake of conciseness these details will not be explicitly discussed.
Therefore, thanks to Eq. (\ref{DEM2}), the late-time expression of the comoving hyperelectric spectrum  is:
\begin{eqnarray}
P_{E}(k,\tau) = \frac{k^3}{2\pi^2} \biggl|  A_{g\, f}(\delta, \, x_{1}\, x) \, \overline{g}_{k} \biggr|^2.
\label{DEM7}
\end{eqnarray}
Equation (\ref{DEM7}) mirrors the result of Eq. (\ref{DEM6}) and it shows that the the hyperelectric power spectrum for $\tau \gg - \tau_{1}$ is determined by the hyperelectric power spectrum at $\tau = - \tau_{1}$. As we shall see
in a moment when the gauge coupling decreases the dual result will hold.

\subsubsection{Explicit expressions of the power spectra in various regimes}
To obtain a more explicit form of the gauge power spectra in the decelerated stage of expansion the 
matrix elements of Eq. (\ref{FIVE13}) can be expanded in powers of  $x_{1} \ll1$. The strategy will be 
to fix $k y$ and expand the various terms in powers of  
of $x_{1}$  with the subsidiary conditions $ 0\leq \delta \ll 1/2$. With these 
specifications Eq. (\ref{FIVE13}) implies\footnote{According to Eq. (\ref{FIVE13def1}),  
$k y= x + x_{1} (q + 1)$ and, to lowest order in $x_{1}$, we have that 
$k y = x + {\mathcal O}(x_{1})$.}:
\begin{eqnarray}
A_{f\, f}(\delta,\,x_{1},\,x) &=& \biggl( \frac{q x_{1}}{2}\biggr)^{\delta} \,\,\biggl[ \sqrt{\frac{x}{2}} \Gamma(1/2 -\delta) J_{- \delta -1/2}(x) + 
{\mathcal O}(x_{1}) \biggr] + {\mathcal O}\biggl[(q\,x_{1})^{1 - \delta}\biggr],
\label{EXPA1}\\
A_{f\, g}(\delta,\,x_{1},\,x) &=& \biggl( \frac{q x_{1}}{2}\biggr)^{-\delta} \,\,\biggl[ \sqrt{\frac{x}{2}} \Gamma(1/2 +\delta) J_{\delta +1/2}(x) + {\mathcal O}(x_{1}) \biggr] + {\mathcal O}\biggl[ (q\,x_{1})^{1 + \delta}\biggr],
\label{EXPA2}\\
A_{g\, f}(\delta,\,x_{1},\,x) &=& \biggl( \frac{q x_{1}}{2}\biggr)^{\delta} \,\,\biggl[- \sqrt{\frac{x}{2}} \Gamma(1/2 -\delta) J_{1/2- \delta}(x) + 
{\mathcal O}(x_{1}) \biggr] + {\mathcal O}\biggl[ (q\,x_{1})^{1 - \delta}\biggr],
\label{EXPA3}\\
A_{g\, g}(\delta,\,x_{1},\,x) &=& \biggl( \frac{q x_{1}}{2}\biggr)^{-\delta} \,\,\biggl[\sqrt{\frac{x}{2}} \Gamma(1/2 +\delta) J_{\delta -1/2}(x) + 
{\mathcal O}(x_{1}) \biggr] + {\mathcal O}\biggl[(q\,x_{1})^{1 + \delta}\biggr].
\label{EXPA4}
\end{eqnarray}
Thanks to the results of Eqs. (\ref{EXPA1})--(\ref{EXPA2}) the inequality Eq. (\ref{DEM1}) can be again verified in a different manner. Furthermore, inserting Eq. (\ref{EXPA2}) into Eq. (\ref{DEM6}) and recalling the expressions for $\overline{f}_{k}$ and $\overline{g}_{k}$ in the case 
of increasing coupling (see e.g. Eq. (\ref{FIVE12})) the hypermagnetic power spectrum becomes:
\begin{eqnarray}
P_{B}(k,\tau) &=& a_{1}^{4} \, H_{1}^4 \, D(\gamma + 1/2) \, \biggl(\frac{k}{a_{1} \, H_{1}}\biggr)^{4 - 2 \gamma - 2 \delta} \, F_{B}(k \tau, \delta),
\nonumber\\
F_{B}(x,\delta) &=& \biggl(\frac{q}{2}\biggr)^{-\,2 \delta} \, \biggl(\frac{x}{2} \biggr) \, \Gamma^2(\delta+1/2) \, J_{\delta+1/2}^2(x). 
\label{EXPA5}
\end{eqnarray}
Similarly, from Eq. (\ref{DEM7}) and (\ref{EXPA4}) the hyperelectric spectrum turns out to be
\begin{eqnarray}
P_{E}(k,\tau) &=& a_{1}^{4} \, H_{1}^4 \, D(\gamma + 1/2) \, \biggl(\frac{k}{a_{1} \, H_{1}}\biggr)^{4 - 2 \gamma - 2 \delta} \, F_{E}(k \tau, \delta),
\nonumber\\
F_{E}(x,\delta) &=& \biggl(\frac{q}{2}\biggr)^{-\,2 \delta} \, \biggl(\frac{x}{2} \biggr) \, \Gamma^2(\delta+1/2) \, J_{\delta-1/2}^2(x). 
\label{EXPA6}
\end{eqnarray}
The results of Eqs. (\ref{EXPA5})--(\ref{EXPA6}) only assume $x_{1} <1 $ and $0\leq \delta \ll \gamma$ and can be evaluated either for $k\tau \ll 1$ or for $k\tau \gg 1$. As long as  $ k \tau \ll 1$  it is enough to recall that $J_{\alpha}(z) \simeq (z/2)^{\alpha}/\Gamma(\alpha +1)$ \cite{abr1,abr2}; 
in the opposite regime (i.e. $k\tau \gg 1$) Eq. (\ref{DEM5}) provides instead the valid approximation scheme\footnote{ Equations (\ref{EXPA5}) and (\ref{EXPA6}) hold for any value of $k \tau$; however, as we shall argue in section \ref{sec5}, for $\tau > \tau_{k} \sim 1/k$ the power spectra will be modified  
by the finite value of the conductivity.}.

Another interesting limit is the sudden approximation which is not well defined a priori but only as the   
$\delta \to 0$ limit; in this case $x$ and $x_{1}$ are kept fixed and the matrix elements of Eq. (\ref{FIVE13}) assume a rather simple form implying:
\begin{eqnarray}
P_{B}(k,\tau) &=& \frac{k^5}{2\pi^2} \biggl| \cos{(x+ x_{1})} \overline{f}_{k} + \sin{(x+ x_{1})} \frac{\overline{g}_{k}}{k}\biggr|^2,
\label{S15a}\\
P_{E}(k,\tau) &=& \frac{k^3}{2\pi^2} \biggl| - k\, \sin{(x+ x_{1})} \, \overline{f}_{k} + \cos{(x+ x_{1})}\overline{g}_{k}\biggr|^2.
\label{S15aa}
\end{eqnarray}
Using Eq. (\ref{FIVE12})  the gauge power spectra will be:
\begin{eqnarray}
P_{B}(k,\tau) &=& a_{1}^{4} \, H_{1}^4 \, D(\gamma + 1/2) \, \biggl(\frac{k}{a_{1} \, H_{1}}\biggr)^{4 - 2 \gamma} \, \sin^2{k\tau},
\label{EXPA7a}\\
P_{E}(k,\tau) &=& a_{1}^{4} \, H_{1}^4 \, D(\gamma + 1/2) \, \biggl(\frac{k}{a_{1} \, H_{1}}\biggr)^{4 - 2 \gamma} \, \cos^2{k\tau}.
\label{EXPA7b}
\end{eqnarray}
The same results of Eqs. (\ref{EXPA7a})--(\ref{EXPA7b}) follow immediately from  Eqs. (\ref{EXPA5})--(\ref{EXPA6})
 by recalling that $q^{-\delta} = (\delta/\gamma)^{-\delta} \to 1$  in the limit $\delta \to 0$.
 All in all, in the sudden approximation $x_{1}$ and $x$ are kept fixed while $\delta\to 0$; in the smooth limit  
$\delta$ may be very small (i.e.  $\delta \ll 1$) but it is always different 
from zero. These two complementary approximations commute since it can be shown that:
\begin{equation}
\lim_{\delta \to 0} \,\, \biggl[ \lim_{x_{1} \ll x \ll 1} {\mathcal M}(\delta, \, x_{1}, \, x) \biggr] = 
\lim_{x_{1} \ll x \ll 1}\,\, \biggl[\lim_{\delta \to 0} {\mathcal M}(\delta, \, x_{1}, \, x)\biggr] = \left(\matrix{1
& x &\cr
-x &1 &\cr}\right).
\label{COMM0}
\end{equation}

\subsection{Decreasing gauge coupling}

\subsubsection{Matrix for the $(\widetilde{\,\gamma\,}, \widetilde{\,\delta\,})$ transition} 

 When the gauge coupling decreases and the dynamics of the transition follows 
the timeline of Fig. \ref{FFF0b} the analog of Eq. (\ref{FIVE11}) 
can be written as:
 \begin{equation}
\left(\matrix{ f_{k}(\tau) &\cr
g_{k}(\tau)/k&\cr}\right) = \left(\matrix{ \widetilde{\,A\,}_{f\, f}(k, \tau, \tau_{1})
& \widetilde{\,A\,}_{f\,g}(k,\tau, \tau_{1})&\cr
\widetilde{\,A\,}_{g\,f}(k,\tau, \tau_{1}) & \widetilde{\,A\,}_{g\,g}(k,\tau, \tau_{1})&\cr}\right) \left(\matrix{ \overline{f}_{k} &\cr
\overline{g}_{k}/k&\cr}\right),
\label{SIX7a}
\end{equation}
where, as usual,  $\overline{f}_{k}= f_{k}(-\tau_{1})$ and $\overline{g}_{k} = g_{k}(-\tau_{1})$ 
follow from Eq. (\ref{THREE2}) and their explicit expression can be found in Eq. (\ref{SIX14b}).
As usual we also added a tilde on top of the various matrix elements to stress 
that they are computed in the framework of the $(\widetilde{\,\gamma\,}, \widetilde{\,\delta\,})$ transition.
Thanks to the Wronskian normalization (\ref{ONE8w}) the analog of Eq. (\ref{FIVE12a}) is now:
\begin{equation}
\widetilde{\,A\,}_{f\, f}(k, \tau, \tau_{1}) \widetilde{\,A\,}_{g\, g}(k, \tau, \tau_{1}) - \widetilde{\,A\,}_{f\, g}(k, \tau, \tau_{1}) \widetilde{\,A\,}_{g\, f}(k, \tau, \tau_{1}) =1.
\label{SIX16a}
\end{equation}
The various entries of the transition matrix appearing in Eq. (\ref{SIX7a}) are:
\begin{eqnarray}
\widetilde{\,A\,}_{f\, f}(k, \tau, \tau_{1}) &=& \frac{\pi}{2} \sqrt{q x_{1}} \sqrt{ k y} \biggl[ Y_{\widetilde{\,\nu\,} -1}( q x_{1}) J_{\widetilde{\,\nu\,}}(k y) - J_{\widetilde{\,\nu\,}-1}(q x_{1}) Y_{\widetilde{\,\nu\,}}(k y) \biggr],
\nonumber\\
\widetilde{\,A\,}_{f\, g}(k, \tau, \tau_{1}) &=& \frac{\pi}{2} \sqrt{q x_{1}} \sqrt{ k y} \biggl[ J_{\widetilde{\,\nu\,}}( q x_{1}) Y_{\widetilde{\,\nu\,}}(k y) - Y_{\widetilde{\,\nu\,}}(q x_{1}) J_{\widetilde{\,\nu\,}}(k y) \biggr],
\nonumber\\
\widetilde{\,A\,}_{g\, f}(k, \tau, \tau_{1}) &=& \frac{\pi}{2} \sqrt{q x_{1}} \sqrt{ k y} \biggl[ Y_{\widetilde{\,\nu\,} - 1}( q x_{1}) J_{\widetilde{\,\nu\,} -1}(k y) - J_{\widetilde{\,\nu\,}-1}(q x_{1}) Y_{\widetilde{\,\nu\,}-1}(k y) \biggr],
\nonumber\\
\widetilde{\,A\,}_{g\, g}(k, \tau, \tau_{1}) &=& \frac{\pi}{2} \sqrt{q x_{1}} \sqrt{ k y} \biggl[ J_{\widetilde{\,\nu\,}}( q x_{1}) Y_{\widetilde{\,\nu\,}-1}(k y) - Y_{\widetilde{\,\nu\,}}(q x_{1}) J_{\widetilde{\,\nu\,}-1}(k y) \biggr].
\label{SIX11a}
\end{eqnarray}
The matrix elements of Eq. (\ref{SIX11a}) and the values of $\overline{f}_{k}$ and $\overline{g}_{k}$ 
are determined from the continuity of the mode functions as described in appendix 
\ref{APPB}. The variables $y$, $q$ and $\widetilde{\,\nu\,}$ are now functions of $\widetilde{\,\delta\,}$ and 
$\widetilde{\,\gamma\, }$ in full analogy with
Eq. (\ref{FIVE13def1}):
\begin{equation}
q(\widetilde{\,\delta\,}, \widetilde{\,\gamma\,}) = \frac{\widetilde{\,\delta\,}}{\widetilde{\,\gamma\,}}, \qquad y(\tau,\widetilde{\,\delta\,},\widetilde{\,\gamma\,}) = \tau + \tau_{1}\biggl[1 + q(\widetilde{\,\delta\,},\widetilde{\,\gamma\,})\biggr], \qquad \widetilde{\,\nu\,}(\widetilde{\,\delta\,})  = |\widetilde{\,\delta\,} -1/2|.
\label{SIX11b}
\end{equation}
Note that Eq. (\ref{SIX11a}) has the same form of Eq. (\ref{FIVE13}); formally Eq. (\ref{SIX11a}) can be 
obtained from Eq. (\ref{FIVE13}) by replacing $\nu$ with $\widetilde{\,\nu\,}$.
With the same notations of Eq(\ref{FIVE13def2}) we can define the matrix 
\begin{equation}
\widetilde{\,{\mathcal M}\,}(\widetilde{\,\delta\,}, x,\, x_{1}) =  \left(\matrix{ \widetilde{\,A\,}_{f\, f}(\widetilde{\,\delta\,}, \,x,\, x_{1})
& \widetilde{\,A\,}_{f\,g}(\widetilde{\,\delta\,},\, x,\, x_{1})&\cr
\widetilde{\,A\,}_{g\,f}(\widetilde{\,\delta\,},\, x,\,x_{1}) & \widetilde{\,A\,}_{g\,g}(\widetilde{\,\delta\,},\, x,\,x_{1})&\cr}\right).
\label{SIX13def2}
\end{equation} 
Inserting the expressions of Eqs. (\ref{SIX7a}) into  Eqs. (\ref{ONE14a})--(\ref{ONE14b})  the comoving power spectra 
for the $(\widetilde{\,\gamma\,}, \widetilde{\,\delta\,})$ transition are:
\begin{eqnarray}
\widetilde{\,P\,}_{B}(k,\tau) &=& \frac{k^5}{2\pi^2} \biggl| \widetilde{\,A\,}_{f\, f} \, \overline{f}_{k} + \widetilde{\,A\,}_{f\, g} \frac{\overline{g}_{k}}{k}\biggr|^2,
\label{D15}\\
\widetilde{\,P\,}_{E}(k,\tau) &=& \frac{k^3}{2\pi^2} \biggl| k\, \widetilde{\,A\,}_{g\, f} \, \overline{f}_{k} + \widetilde{\,A\,}_{g\, g} \overline{g}_{k}\biggr|^2.
\label{D15aa}
\end{eqnarray}
As already argued for the $(\gamma,\delta)$ transition, the terms appearing inside the 
squared moduli at the right hand side of Eqs. (\ref{D15}) and (\ref{D15aa}) 
are not of the same order but one of the two terms will be alternatively dominant. 
Owing to the specific form of Eqs. (\ref{SIX13def2}) and (\ref{SIX11a})--(\ref{SIX14b}) the hierarchies between the different terms contributing to Eqs. (\ref{D15})--(\ref{D15aa}) turn out to be:
\begin{eqnarray}
| \widetilde{\,A\,}_{f\, f}(\widetilde{\,\delta\,}, x, x_{1}) \, \overline{f}_{k} | &\gg& \biggl| \widetilde{\,A\,}_{f\, g}(\widetilde{\,\delta\,}, x, x_{1})\, \frac{\overline{g}_{k}}{k} \biggr|,
\label{TEM1}\\
| \widetilde{\,A\,}_{g\, f}(\widetilde{\,\delta\,}, x, x_{1}) \, \overline{f}_{k} | &\gg&  \biggl| \widetilde{\,A\,}_{g\, g}(\widetilde{\,\delta\,}, x, x_{1})\, \frac{\overline{g}_{k}}{k} \biggr|.
\label{TEM2}
\end{eqnarray}
The inequalities Eqs. (\ref{TEM1}) and (\ref{TEM2}) imply that the late-time hypermagnetic power spectra 
are determined by the hypermagnetic power spectra at $\tau_{1}$ and they can be explicitly verified by following the same strategy illustrated in the case of the $(\gamma, \, \delta)$ transition. So for instance we can rewrite Eq. (\ref{TEM1}) as:
\begin{equation}
 \biggl| Y_{\widetilde{\,\nu\,}-1 }(q x_{1}) J_{\widetilde{\,\nu\,}}(k y) - J_{\widetilde{\,\nu\,}-1}(q x_{1}) Y_{\widetilde{\,\nu\,}}(k y) \biggr| 
\gg \biggl|\frac{H_{\widetilde{\mu} -1}^{(1)}(x_{1})}{H_{\widetilde{\mu}}^{(1)}(x_{1})}\biggr|\, \biggl|J_{\widetilde{\,\nu\,}}(q x_{1}) Y_{\widetilde{\,\nu\,}}(k y) - Y_{\widetilde{\,\nu\,}}(q x_{1}) J_{\widetilde{\,\nu\,}}(k y) \biggr|.
\label{TEM3}
\end{equation}
If both sides of Eq. (\ref{TEM3}) are expanded for $x_{1} \ll x < 1 $ and with the subsidiary condition $0\leq \widetilde{\,\delta\,} \ll \widetilde{\,\gamma\,}$ we obtain the following condition:
\begin{equation}
\biggl(\frac{q x_{1}}{2} \biggr)^{ 2 \widetilde{\,\delta\,}} \gg \frac{\Gamma(3/2+ \widetilde{\,\delta\,})\,\Gamma(|\widetilde{\,\gamma\,} -1/2|)}{\Gamma(1/2 - \widetilde{\,\delta\,})\, \Gamma(\widetilde{\,\gamma\,} +1/2)} \, \biggl(\frac{x_{1}}{2}\biggr)^{\widetilde{\,\gamma\,} +1/2 - |\widetilde{\,\gamma\,} -1/2|}.
\label{TEM4}
\end{equation}
When $\widetilde{\,\gamma\,} > 1/2$ Eq. (\ref{TEM4})  implies $x_{1}^{2 \widetilde{\,\delta\,}} > x_{1}$: this requirement is always verified for $x_{1} < 1$ since, by assumption, $0 \leq \widetilde{\,\delta\,} \ll 1/2$. Similarly, in the range $\widetilde{\,\gamma\,} < 1/2$, Eq. (\ref{TEM4}) demands $x_{1}^{2 \widetilde{\,\delta\,}} > x_{1}^{2 \widetilde{\,\gamma\,}}$ which is also true since $\widetilde{\,\delta\,} \ll \widetilde{\,\gamma\,}$. The same logic can be applied to Eq. (\ref{TEM2}) but this discussion will be omitted for the sake of conciseness. All in all Eqs. (\ref{TEM1}) and (\ref{TEM2}) imply that the comoving power spectra
become
\begin{eqnarray}
\widetilde{P}_{B}(k,\tau) = \frac{k^5}{2\pi^2} \biggl| \widetilde{\,A\,}_{f\, f}(\widetilde{\delta},\,x_{1},\,x)\,\overline{f}_{k}\biggr|^2,\qquad 
\widetilde{P}_{E}(k,\tau) = \frac{k^3}{2\pi^2} \biggl| \widetilde{\,A\,}_{g\, f} (\widetilde{\delta},\,x_{1},\,x)\, k\,\overline{f}_{k}\biggr|^2.
\label{TEM6}
\end{eqnarray}

\subsubsection{Explicit forms of the power spectra in various limits}
The spectra of Eq. (\ref{TEM6}) can be evaluated in various limits and, for this 
purpose, it is useful to expand  
Eq. (\ref{SIX11a}) for $x_{1}$ with the subsidiary conditions $ 0\leq \delta < 1/2$. The result 
of this stem can be expressed as:
\begin{eqnarray}
\widetilde{\,A\,}_{f\, f}(\widetilde{\,\delta\,},\,x_{1},\,x) &=& \biggl( \frac{q x_{1}}{2}\biggr)^{-\widetilde{\,\delta\,}} \,\,\biggl[ \sqrt{\frac{x}{2}} \Gamma(1/2 +\widetilde{\,\delta\,}) J_{\widetilde{\,\delta\,} -1/2}(x) + 
{\mathcal O}(x_{1}) \biggr] + {\mathcal O}\biggl[ (q\,x_{1})^{1 +\widetilde{\,\delta\,}}\biggr],
\label{EXPADD1}\\
\widetilde{\,A\,}_{f\, g}(\widetilde{\,\delta\,},\,x_{1},\,x) &=& \biggl( \frac{q x_{1}}{2}\biggr)^{\widetilde{\,\delta\,}} \,\,\biggl[ \sqrt{\frac{x}{2}} \Gamma(1/2 - \widetilde{\,\delta\,}) J_{-\widetilde{\,\delta\,} +1/2}(x) + {\mathcal O}(x_{1}) \biggr] + {\mathcal O}\biggl[ (q\,x_{1})^{1 - \widetilde{\,\delta\,}}\biggr],
\label{EXPADD2}\\
\widetilde{\,A\,}_{g\, f}(\widetilde{\,\delta\,},\,x_{1},\,x) &=& \biggl( \frac{q x_{1}}{2}\biggr)^{-\widetilde{\,\delta\,}} \,\,\biggl[- \sqrt{\frac{x}{2}} \Gamma(1/2 +\widetilde{\,\delta\,}) J_{1/2+ \widetilde{\,\delta\,}}(x) + 
{\mathcal O}(x_{1}) \biggr] + {\mathcal O}\biggl[ (q\,x_{1})^{1+\widetilde{\,\delta\,}}\biggr],
\label{EXPADD3}\\
\widetilde{\,A\,}_{g\, g}(\widetilde{\,\delta\,},\,x_{1},\,x) &=& \biggl( \frac{q x_{1}}{2}\biggr)^{\widetilde{\,\delta\,}} \,\,\biggl[\sqrt{\frac{x}{2}} \Gamma(1/2- \widetilde{\,\delta\,}) J_{- \widetilde{\,\delta\,} -1/2}(x) + 
{\mathcal O}(x_{1}) \biggr] + {\mathcal O}\biggl[ (q\,x_{1})^{1 -\widetilde{\,\delta\,}}\biggr].
\label{EXPADD4}
\end{eqnarray}
Again using Eqs. (\ref{EXPADD1})--(\ref{EXPADD2}) and (\ref{EXPADD3})--(\ref{EXPADD4}) 
the inequalities (\ref{TEM1})--(\ref{TEM2}) are verified  and the corresponding power spectra 
turn out to be:
\begin{eqnarray}
\widetilde{\,P\,}_{B}(k,\tau) &=& a_{1}^{4} \, H_{1}^4 \, D(\widetilde{\,\gamma\,} + 1/2) \, \biggl(\frac{k}{a_{1} \, H_{1}}\biggr)^{4 - 2 \widetilde{\,\gamma\,} - 2 \widetilde{\,\delta\,}} \, \widetilde{\,F\,}_{B}(k\,\tau, \widetilde{\,\delta\,}),
\nonumber\\
\widetilde{\,F\,}_{B}(x,\widetilde{\,\delta\,}) &=& \biggl(\frac{q}{2}\biggr)^{-\,2 \widetilde{\,\delta\,}} \, \biggl(\frac{x}{2} \biggr) \, \Gamma^2(\delta+1/2) \, J_{\widetilde{\,\delta\,} - 1/2}^2(x). 
\label{EXPADD5}
\end{eqnarray}
From Eq. (\ref{DEM7}) and (\ref{EXPA4}) the hyperelectric spectrum is
\begin{eqnarray}
\widetilde{\,P\,}_{E}(k,\tau) &=& a_{1}^{4} \, H_{1}^4 \, D(\widetilde{\,\gamma\,} + 1/2) \, \biggl(\frac{k}{a_{1} \, H_{1}}\biggr)^{4 - 2 \widetilde{\,\gamma\,} - 2 \widetilde{\,\delta\,}} \, \widetilde{\,F\,}_{E}(k \,\tau, \widetilde{\,\delta\,}),
\nonumber\\
\widetilde{\,F\,}_{E}(x,\widetilde{\,\delta\,}) &=& \biggl(\frac{q}{2}\biggr)^{-\,2 \widetilde{\,\delta\,}} \, \biggl(\frac{x}{2} \biggr) \, \Gamma^2(\widetilde{\,\delta\,}+1/2) \, J_{\widetilde{\,\delta\,}+1/2}^2(x). 
\label{EXPADD6}
\end{eqnarray}
Equations (\ref{EXPADD5}) and (\ref{EXPADD6}) can be evaluated in the sudden approximation 
by taking the limit $\widetilde{\,\delta\,} \to 0$ while $x$ and $x_{1}$ are kept fixed; the gauge power spectra are, in this case\footnote{Note that the same results of Eqs. (\ref{EXPADD7a})--(\ref{EXPADD7b}) follow immediately from  Eqs. (\ref{EXPADD5})--(\ref{EXPADD6})
 by recalling that $q^{-\widetilde{\,\delta\,}} = (\widetilde{\,\delta\,}/\widetilde{\,\gamma\,})^{-\widetilde{\,\delta\,}} \to 1$  in the limit $\widetilde{\,\delta\,} \to 0$.}
\begin{eqnarray}
\widetilde{\,P\,}_{B}(k,\tau) &=& a_{1}^{4} \, H_{1}^4 \, D(\widetilde{\,\gamma\,} + 1/2) \, \biggl(\frac{k}{a_{1} \, H_{1}}\biggr)^{4 - 2 \widetilde{\,\gamma\,}} \, \cos^2{k\tau},
\label{EXPADD7a}\\
\widetilde{\,P\,}_{E}(k,\tau) &=& a_{1}^{4} \, H_{1}^4 \, D(\widetilde{\,\gamma\,} + 1/2) \, \biggl(\frac{k}{a_{1} \, H_{1}}\biggr)^{4 - 2 \widetilde{\,\gamma\,}} \, \sin^2{k\tau}.
\label{EXPADD7b}
\end{eqnarray}
By comparing Eqs. (\ref{EXPADD7a})--(\ref{EXPADD7b}) with the analog results 
 given in Eqs. (\ref{EXPA7a})--(\ref{EXPA7b}) we note that 
the phases of the Sakharov oscillations are exchanged: while $P_{B}(k,\tau)$ oscillates like 
$\sin^2{k\tau}$, the oscillations of $\widetilde{\,P\,}_{B}(k,\tau)$ go like $\cos^2{k\tau}$. The opposite 
is true for the hyperelectric power spectra.
The differences in the phases of oscillation are a particular consequence of duality
which will now be analyzed in general terms.

\subsection{Duality and continuity of the power spectra after inflation}
During inflation duality relates the gauge  power spectra 
in different dynamical situations (see Eqs. (\ref{THREE16})--(\ref{FOUR1})). 
When the gauge coupling flattens out the gauge power spectra are also related  by duality which 
is equivalent to the following transformation:
\begin{equation}
(\gamma, \delta)\,\,\to\,\,(\widetilde{\gamma}, \, \widetilde{\delta}). 
\label{DUAL1}
\end{equation}
The action of Eq. (\ref{DUAL1}) transforms the elements of the transition matrix according to 
\begin{equation}
 A_{f\,f} \to \widetilde{\,A\,}_{g\,g}, \qquad A_{g\,g} \to \widetilde{\,A\,}_{f\,f},\qquad 
  A_{f\,g} \to -  \widetilde{\,A\,}_{g\,f}, \qquad \, A_{g\,f} \to -  \widetilde{\,A\,}_{f\,g}.
\label{DUAL2}
\end{equation}
Equation (\ref{DUAL2}) follows from the expressions of the matrix elements reported in Eqs. (\ref{FIVE13}) and (\ref{SIX11a}). For the sake of accuracy the results of Eq. (\ref{DUAL2}) have been explicitly derived in appendix \ref{APPC}. Moreover 
$\overline{f}_{k}$ and $\overline{g}_{k}$ will transform under duality as $\overline{f}_{k} \to \overline{g}_{k}/k$ and $\overline{g}_{k} \to - k \, \overline{f}_{k}$. Thanks to Eqs. (\ref{ONE10a}) and (\ref{DUAL2}) the action of Eq. (\ref{DUAL1})  transforms the gauge power spectra as:
\begin{equation}
P_{B}(k,\tau) \to \widetilde{\,P\,}_{E}(k,\tau), \qquad P_{E}(k,\tau) \to \widetilde{\,P\,}_{B}(k,\tau).
\label{APC11}
\end{equation}
Also the approximate expressions of the power spectra, if they are correct, must transform according to Eq. (\ref{APC11})
since they must be consistent with duality; in particular:
\begin{itemize}
\item{}  the approximate power spectrum $P_{B}(k,\tau)$ 
of Eq. (\ref{EXPA5}) gives the approximate form of $\widetilde{\,P\,}_{E}$ (i.e. Eq. (\ref{EXPADD6}));
\item{} the transformation (\ref{DUAL1}) applied to Eq. (\ref{EXPA6}) gives the approximate 
form of the hyeprmagnetic power spectrum in the dual description (i.e. $P_{E}(k,\tau) \to \widetilde{\,P\,}_{B}(k,\tau)$);
\item{} in the sudden approximation (when $\delta\to 0$ and $\widetilde{\,\delta\,} \to 0$) the gauge power  
spectra of Eqs. (\ref{FIVE15})--(\ref{FIVE15aa}) and (\ref{D15})--(\ref{D15aa}) are  related by duality
\end{itemize}

All in all during inflation duality implies that when the gauge coupling increases the magnetic spectrum is never flat and if the gauge coupling is instead decreasing the electric spectrum is never  flat. After inflation duality constrains the explicit forms of the gauge power 
spectra and the phases of the Sakharov oscillations. If the gauge coupling flattens out  
after a phase of increasing coupling the late-time gauge spectra are determined 
by the  hyperelctric spectrum at the end of inflation. In the dual situation the gauge coupling 
freezes after a stage of decreasing coupling. Barring for the possible physical justifications 
of a strongly coupled phase at the beginning of inflation, the late-time gauge power spectra 
follow from the hypermagnetic spectrum at the end of inflation.

\renewcommand{\theequation}{5.\arabic{equation}}
\setcounter{equation}{0}
\section{Phenomenological considerations}
\label{sec5}
The hypercharge
field projects on the electromagnetic fields through the cosine of the Weinberg angle (i.e. $\cos{\theta_{W}}$) so that the late-time gauge power spectra could be compared  both with the magnetogenesis requirements and with other phenomenological constraints. In this section we adopt the standard notations commonly employed in the context of the concordance paradigm.
So for instance  $\Omega_{R0}$ denotes present critical fraction of relativistic particles, $\Omega_{M0}$  is total critical fraction of massive species, $r_{T}$ defines the ratio between the tensor and scalar power spectra at the pivot scale $k_{p} =0.002\, \mathrm{Mpc}^{-1}$ and so on and so forth. The present value of the scale  factor will be normalized to $1$ (i.e. $a_{0} =1$) and, thanks to this widely used convention, the physical and the comoving frequencies coincide at the present time. The present section 
is devoted to phenomenological implications; in particular the following topics will be specifically treated;
\begin{itemize}
\item{} in subsection \ref{subs1} a brief summary of the typical scales of the 
problem has been provided in a phenomenological perspective;
\item{}  in subsection \ref{subs2} 
some simplified estimates of the physical spectra before reentry have been discussed;
\item{} subsection \ref{subs3} is devoted to the physical spectra after reentry;
\item{} in subsection \ref{subs4} we give a specific discussion of the magnetogenesis constraints 
in a self-contained perspective;
\item{}  the dependence the parameter space of the model has been 
charted in the subsection \ref{subs5};
\item{} the subsection \ref{subs6} has been devoted to the effects of a post-inflationary
phase preceding the radiation epoch;
\item{} a final assessment of the obtained results has been
presented in the subsection \ref{subs7}.
\end{itemize}
 
\subsection{Scales of the problem}
\label{subs1}
Defining with $\tau_{k}=1/k$ the reentry time of a generic wavelength, the ratio between $\tau_{k}$ and the time of matter-radiation equality $\tau_{eq}$ is:
\begin{equation}
\frac{\tau_{k}}{\tau_{eq}} = 1.06 \times 10^{-2} \biggl(\frac{h_{0}^2 \Omega_{M0}}{0.1386} \biggr) 
\biggl(\frac{h_{0}^2 \Omega_{R 0}}{4.15\times 10^{-5}}\biggr)^{-1/2} \, \biggl( \frac{k}{\mathrm{Mpc}^{-1}}\biggr)^{-1}, 
\label{PH1}
\end{equation}
since $(\tau_{k}/\tau_{eq}) = \sqrt{2} (H_{0}/k) \Omega_{M0}/\sqrt{\Omega_{R0}}$ and $H_{0} 
= 100 \, h_{0} \, (\mathrm{km}/\mathrm{Mpc}) \, \mathrm{Hz}$ denotes the present value of the Hubble rate. Consequently the relevant wavenumbers for magnetogenesis considerations [i.e. $k = {\mathcal O}(\mathrm{Mpc}^{-1})$] reentered prior to matter radiation equality.

After the gauge coupling flattens out the amplitude of the comoving power spectra is ${\mathcal O}(H_{1}^4)$
 (see, for instance, Eqs. (\ref{EXPA5})--(\ref{EXPA6}) and (\ref{EXPADD5})--(\ref{EXPADD6})). The inflationary 
 Hubble rate $H_{1}$ is  related to the Planck scale as\footnote{In Eq. (\ref{PH2}) we traded $\epsilon$ for the tensor scalar ratio $r_{T}$ by using the consistency relation  $r_{T} = 16 \,\epsilon$.}:
  \begin{equation}
\frac{H_1}{M_{P}} = \sqrt{ \pi \, \epsilon\, {\mathcal A}_{{\mathcal R}}} = 2.17 \times 10^{-6} \biggl(\frac{r_{T}}{0.01}\biggr)^{1/2} \biggl(\frac{{\mathcal A}_{{\mathcal R}}}{2.41\times 10^{-9} }\biggr)^{1/2}, 
\label{PH2}
\end{equation}
where ${\mathcal A}_{{\mathcal R}}$ denotes the amplitude of the scalar power spectrum at the pivot scale $k_{p}=0.002\,\mathrm{Mpc}^{-1}$.  Thanks to Eq. (\ref{PH2}) it is then possible to express $H_{1}^4$ in units of $\mathrm{nG}^2$ by recalling that $M_{P}^4 = 4.62 \times 10^{132}\, \mathrm{nG}^2$. The scale $H_{1}$ also enters the explicit expression of  $x_{1}= k\tau_{1} = k/(a_{1}\, H_{1})$:
\begin{eqnarray}
\frac{k}{a_{1}\, H_{1}} &=& \frac{k}{a_{0}\, H_{0}} \biggl(\frac{a_{0} \, H_{0}}{a_{1} \, H_{1}}\biggr) = \frac{k}{ H_{0}}\, \zeta_{r}^{1/2 - \alpha} \biggl(2 \pi \epsilon \Omega_{R0} 
{\mathcal A}_{{\mathcal R}}\biggr)^{-1/4}\, \sqrt{\frac{H_{0}}{M_{P}}},  
\label{PH3}
\end{eqnarray}
where the first equality is just true by definition while the second one is the explicit estimate. 
As already mentioned, in Eq. (\ref{PH3}) $\Omega_{R0}$ 
is the (present) critical fraction of massless species (in the concordance paradigm $h_{0}^2 \Omega_{R0} = 4.15 \times 
10^{-5}$); note that $H_{0}$ is usefully expressed in Planck units as $(H_{0}/M_{P}) = 1.22 \times 
10^{-61} (h_0/0.7)$.  In terms of the typical values of the various parameters, Eq. (\ref{PH3}) can also be written in the following revealing form: 
\begin{equation}
\frac{k}{a_{1}\, H_{1}} = 10^{-23.05}\,\, \biggl(\frac{k}{\mathrm{Mpc}^{-1}}\biggr)\, \biggl(\frac{r_{T}}{0.01}\biggr)^{-1/4}\,\,\biggl(\frac{h_{0}^2 \Omega_{R0}}{4.15\times 10^{-5}}\biggr)^{-1/4} \,\,\biggl(\frac{{\mathcal A}_{{\mathcal R}}}{2.41\times10^{-9}}\biggr)^{-1/4} \,\, \zeta_{r}^{1/2-\alpha}.
\label{PH4}
\end{equation}
In Eqs. (\ref{PH3})--(\ref{PH4}) $\zeta_{r} = H_{r}/H_{1}$ and $H_{r}$ is the value of the Hubble rate at the moment when the radiation starts dominating; in the concordance paradigm $\zeta_{r}=1$ (and this is the value we shall 
assume most of the time). However it is sometimes useful to check for the stability 
of the obtained results with respect to a change of the reference scenario. This is why Eqs. 
(\ref{PH3})--(\ref{PH4}) have been derived by considering a generalized post-inflationary evolution where the radiation epoch is preceded by a phase expanding at a rate parametrized\footnote{If this intermediate stage of expansion 
is dominated by a perfect  fluid with barotropic index $w$, we have that $\alpha= 2/[3(w+1)]$.}
by $\alpha$.  Finally, throughout the explicit numerical evaluation it will be useful to bear in mind 
the following two general expressions:
\begin{equation}
\biggl(\frac{a_{1}}{a_{0}}\biggr) = \zeta_{r}^{\alpha -1/2} \, \biggl(\frac{2 \, \Omega_{R0}}{\pi \epsilon {\mathcal A}_{{\mathcal R}}}\biggr)^{1/4} \, \sqrt{\frac{H_{0}}{M_{P}}},\qquad \qquad
\biggl(\frac{a_{0} \, H_{0}}{a_{1} \, H_{1}} \biggr) = \zeta_{r}^{1/2 - \alpha} \biggl(2 \pi \epsilon \Omega_{R0} 
{\mathcal A}_{{\mathcal R}}\biggr)^{-1/4}\, \sqrt{\frac{H_{0}}{M_{P}}}.
\label{PH6}
\end{equation}

\subsection{Physical spectra prior to reentry}
\label{subs2}
In the previous sections we discussed mainly the comoving spectra of the hyperelectric and hypermagnetic fields 
but in various situations the distinction between the physical and the comoving fields 
has played a crucial role. The distinction between physical and comoving power spectra 
has been given in Eq. (\ref{ONE14c}) and here we just want to illustrate, in a preliminary 
respective, the evolution of the comoving fields as  a function of $k\tau$.
\begin{figure}[!ht]
\centering
\includegraphics[height=5cm]{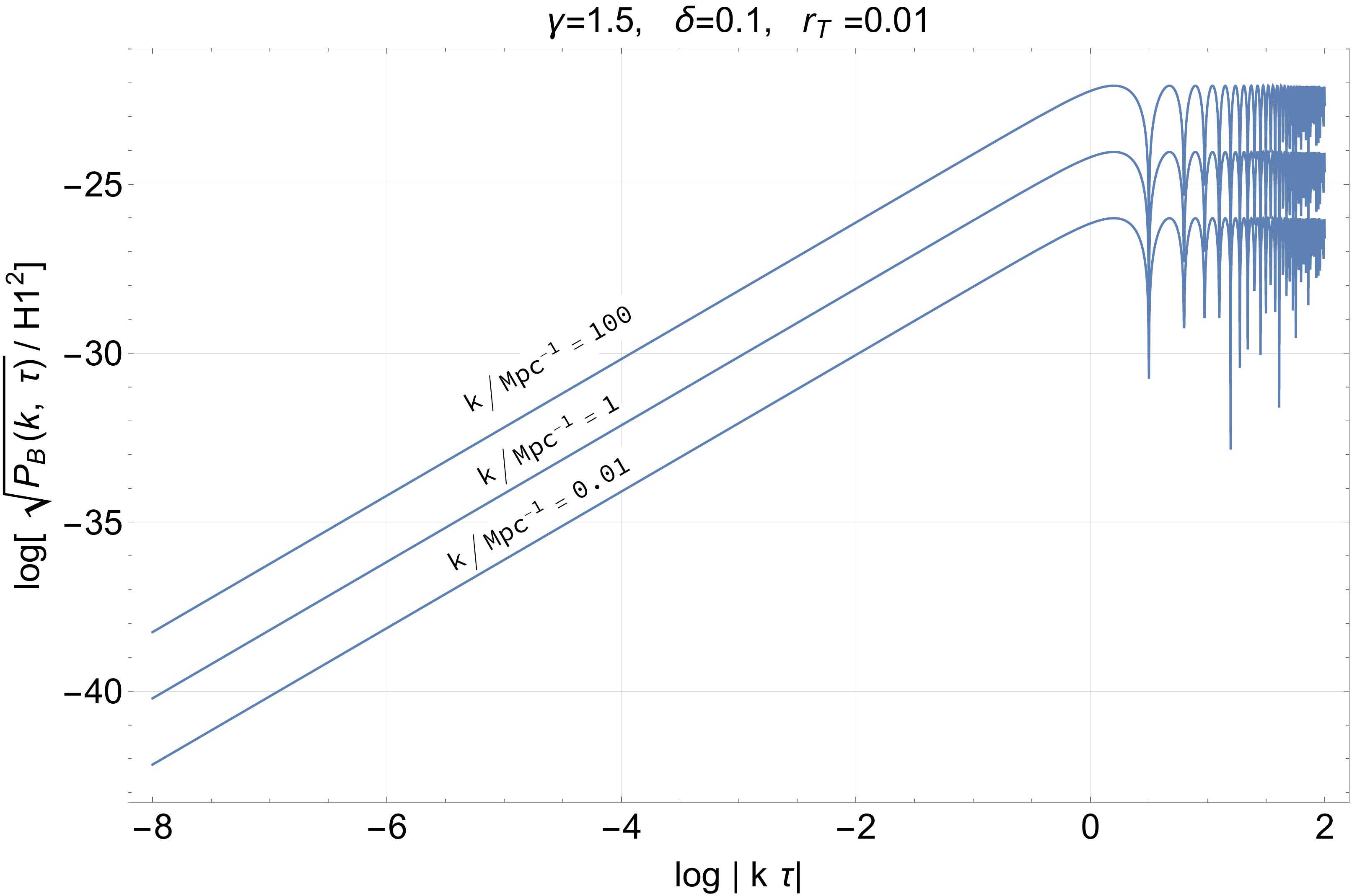}
\includegraphics[height=5cm]{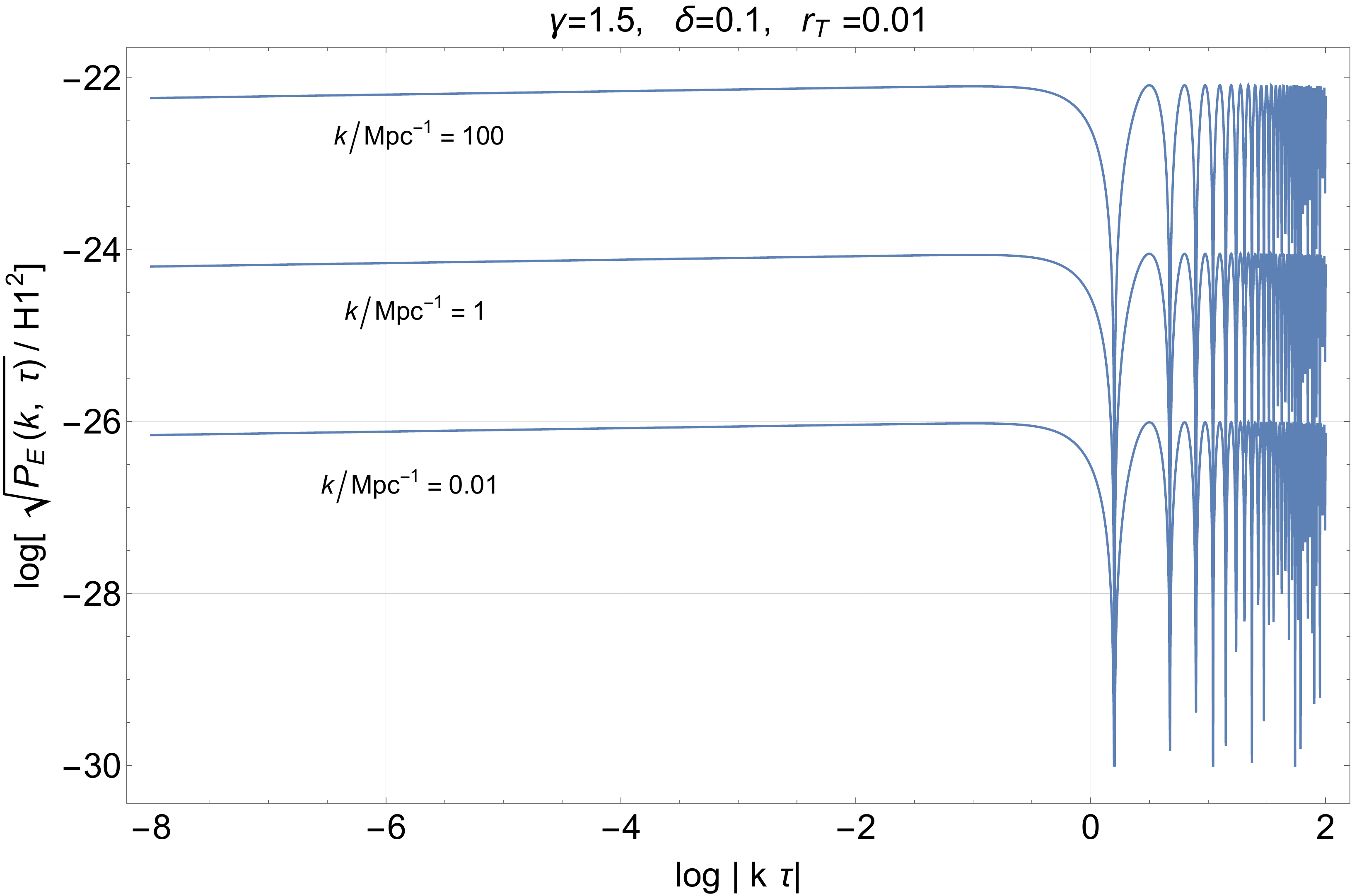}
\caption[a]{We illustrate the comoving power spectra prior to reentry for a particular choice of the parameters and for three 
different comoving scales. }
\label{FFFSP}      
\end{figure}
In Fig. \ref{FFFSP} we illustrate the evolution of the common logarithm of the 
square roots of the hyperelectric and hyperamagnetic power spectra\footnote{The electromagnetic projection of the hypercharge through the Weinberg angle has not been included but it will be carefully taken into account when charting the parameter space in subsection \ref{subs5}.} as a function of the 
common logarithm of $| k \tau|$. The values of $\gamma$ and $\delta$ have been randomly chosen 
since the parameter space will be more carefully scrutinised in subsection \ref{subs5}.
In Fig.  \ref{FFFSP}  three different scales have been included. The most relevant one is the 
scale of the primordial gravitational collapse (i.e. $k = {\mathcal O}(\mathrm{Mpc}^{-1})$).
The other two scales are just illustrative but the same bunch of wavenumbers will be illustrated in 
the contour plots of subsection \ref{subs5}.

For $\tau \ll \tau_{k}$ the 
power spectra follow from Eqs.   (\ref{EXPA5})--(\ref{EXPA6}) and (\ref{EXPADD5})--(\ref{EXPADD6})
evaluated in the limit $k \tau \ll 1$. When the gauge coupling increases the physical spectra (\ref{ONE14c}) associated with the comoving expressions of Eqs. (\ref{EXPA5})--(\ref{EXPA6}) are\footnote{Concerning Eqs. (\ref{PH7}) and (\ref{PH8}) we note that $\lambda_{1}= \lambda(-\tau_{1})$. For the sake of simplicity 
we shall assume $\lambda_{1} = 1$ even if $\lambda_{1}$ could also be slightly larger 
than $1$.}:
\begin{eqnarray}
P_{B}^{(phys)}(k,\tau) &=& \frac{H_{1}^4}{\lambda_{1}} \, \biggl(\frac{a_{1}}{a}\biggr)^4 \, \frac{D(\gamma+1/2)}{(2 \delta +1)^2 q^{4 \delta}} \biggl(\frac{k}{a_{1} \, H_{1}}\biggr)^{6 - 2 \gamma} \, \biggl(\frac{a_{1} \, H_{1}}{a\, H}\biggr)^{ 4 \delta + 2},
\label{PH7}\\
P_{E}^{(phys)}(k,\tau) &=& \frac{H_{1}^4}{\lambda_{1}} \, \biggl(\frac{a_{1}}{a}\biggr)^4 \, 
\frac{D(\gamma+1/2)}{ q^{\,4 \delta}}\, \biggl(\frac{k}{a_{1} \, H_{1}}\biggr)^{4 - 2 \gamma} \, 
\biggl(\frac{a_{1} \, H_{1}}{a\, H}\biggr)^{ 4 \delta},\qquad \tau < \tau_{k}.
\label{PH8}
\end{eqnarray}
When the gauge coupling decreases the comoving spectra  of Eqs. (\ref{EXPADD5})--(\ref{EXPADD6}) must be inserted into Eq. (\ref{ONE14c}) and the result is:
\begin{eqnarray}
\widetilde{\,P\,}_{B}^{(phys)}(k,\tau) &=& \frac{H_{1}^4}{\lambda_{1}} \, \biggl(\frac{a_{1}}{a}\biggr)^4 \, 
\frac{D(\widetilde{\,\gamma\,}+1/2)}{ q^{\,4 \widetilde{\,\delta\,}}}\, \biggl(\frac{k}{a_{1} \, H_{1}}\biggr)^{4 - 2 \widetilde{\,\gamma\,}} \, 
\biggl(\frac{a_{1} \, H_{1}}{a\, H}\biggr)^{ 4 \widetilde{\,\delta\,}},
\label{PH9}\\
\widetilde{\,P\,}_{E}^{(phys)}(k,\tau) &=& \frac{H_{1}^4}{\lambda_{1}} \, \biggl(\frac{a_{1}}{a}\biggr)^4 \, \frac{D(\widetilde{\,\gamma\,}+1/2)}{(2 \widetilde{\,\delta\,} +1)^2 q^{4 \widetilde{\,\delta\,}}} \biggl(\frac{k}{a_{1} \, H_{1}}\biggr)^{6 - 2 \widetilde{\,\gamma\,}} \, \biggl(\frac{a_{1} \, H_{1}}{a\, H}\biggr)^{ 4 \widetilde{\,\delta\,} + 2},\qquad \tau < \tau_{k}.
\label{PH10}
\end{eqnarray} 
The explicit formulas of Eqs. (\ref{PH2})--(\ref{PH3}) and (\ref{PH6}) 
allow for an explicit evaluation of Eqs. (\ref{PH7})--(\ref{PH8}) and (\ref{PH9})--(\ref{PH10}).
If $\lambda_{1} = {\mathcal O}(1)$ the only spectra consistent with the perturbative evolution 
of the gauge coupling are the ones of Eqs. (\ref{PH7})--(\ref{PH8}). In the case of decreasing gauge coupling
$\lambda_{1}$ must be instead extremely large (see also Eq. (\ref{COMP1}) and discussion therein). 
This means, however, that the magnetic spectrum (\ref{PH9}) will be suppressed as 
$\lambda_{1}^{-1} \simeq e^{- 2\widetilde{\,\gamma\,} N}$ where $N$ now denotes the total 
number of inflationary $e$-folds. 
Let us finally remark that in the sudden limit (i.e. $\delta \to 0$ and $\widetilde{\,\delta\,} \to 0$) the 
physical power spectra will always follow from Eq. (\ref{ONE14c}) but the related comoving spectra 
will be given, respectively, by Eqs. (\ref{EXPA7a})--(\ref{EXPA7b}) and (\ref{EXPADD7a})--(\ref{EXPADD7b}).
For instance, in the case of increasing gauge coupling the result will be:
\begin{eqnarray}
P_{B}^{(phys)}(k,\tau) &=& \frac{H_{1}^4}{\lambda_{1}} \, \biggl(\frac{a_{1}}{a}\biggr)^4 \, D(\gamma+1/2)
\biggl(\frac{k}{a_{1} \, H_{1}}\biggr)^{6 - 2 \gamma} \, \biggl(\frac{a_{1} \, H_{1}}{a\, H}\biggr)^{  2}, 
\label{PH11}\\
P_{E}^{(phys)}(k,\tau) &=& \frac{H_{1}^4}{\lambda_{1}} \, \biggl(\frac{a_{1}}{a}\biggr)^4 \, 
D(\gamma+1/2)\, \biggl(\frac{k}{a_{1} \, H_{1}}\biggr)^{4 - 2 \gamma}, \,\qquad \tau < \tau_{k}.
\label{PH12}
\end{eqnarray}

\subsection{Physical spectra after reentry}
\label{subs3}
For $\tau \geq \tau_{k}$  the evolution of the mode functions is modified by 
the presence of the conductivity and it is approximately given by\footnote{Note that in a different system of units 
(where the gauge coupling is defined without the $\sqrt{4 \pi}$ factor) we would have $4 \pi \sigma$ 
(and not simply $\sigma$) in Eq. (\ref{PH13}). We are assuming here that the gauge coupling 
does not evolve anymore and it is frozen to its constant value.}:
\begin{equation}
g_{k}^{\prime} = - k^2 f_{k} - \sigma \, g_{k}, \qquad f_{k}^{\prime} = g_{k}.
\label{PH13}
\end{equation}
These equations can be systematically solved as an expansion in $(k/\sigma)$ by setting initial conditions 
at $\tau =\tau_{k}$. To lowest 
order the solution of Eq. (\ref{PH13}) for $\tau \geq \tau_{k}$ is:
\begin{equation}
f_{k}(\tau) = A_{g\,f}(k, \tau_{1}, \tau_{k}) \frac{\overline{g}_{k}}{k} e^{- \frac{k^2}{k_{\sigma}^2}}, 
\qquad\qquad 
g_{k}(\tau) = \biggl(\frac{k}{\sigma}\biggr) A_{g\,g}(k, \tau_{1}, \tau_{k}) \overline{g}_{k} e^{- \frac{k^2}{k_{\sigma}^2}},
\label{PH17}
\end{equation}
where  the magnetic diffusivity scale $k_{\sigma}$ has been defined as
 $k_{\sigma}^{-2}  = \int_{\tau_{k}}^{\tau} \, d z /\sigma(z)$. While the estimate of $k_{\sigma}$ 
can be made accurate by computing the transport coefficients of the plasma in different regimes, 
for the present purposes this is not necessary since the ratio $(k/k_{\sigma})^2$  is so small, for the 
phenomenologically interesting scales, that the negative exponentials  in Eq. (\ref{PH17}) 
evaluate to $1$. In fact by taking $\tau = \tau_{\mathrm{eq}}$ we have that $k_{\sigma}$ can be estimated as: 
\begin{equation}
\biggl(\frac{k}{k_{\sigma}}\biggr)^2 = \frac{4.75 \times 10^{-26}}{ \sqrt{2 \, h_{0}^2 \Omega_{M0} (z_{\mathrm{eq}}+1)}} \, \biggl(\frac{k}{\mathrm{Mpc}^{-1}} \biggr)^2,
\label{PH18}
\end{equation}
where $\Omega_{M0}$ is the present critical fraction in matter and $z_{\mathrm{eq}} + 1 = a_{0}/a_{\mathrm{eq}}\simeq {\mathcal O}(3200)$ is the redshift of matter-radiation equality. The physical power spectra after for  $\tau \gg \tau_{k}$ will then be given by:
\begin{eqnarray}
P_{B}^{(phys)}(k,\tau) &=& H_{1}^4 \biggl(\frac{a_{1}}{a}\biggr)^4 D(\gamma +1/2) \, \biggl(\frac{k}{a_{1} H_{1}} \biggr)^{4 - 2 \gamma - 2 \delta} \, F_{B}(k \tau_{k}, \delta)\,e^{-2 \frac{k^2}{k_{\sigma}^2}},
\label{PH19}\\
P_{E}^{(phys)}(k,\tau) &=& \biggl(\frac{k}{\sigma}\biggr)^2 \, H_{1}^4 \biggl(\frac{a_{1}}{a}\biggr)^4 D(\gamma +1/2) \, \biggl(\frac{k}{a_{1} H_{1}} \biggr)^{4 - 2 \gamma - 2 \delta} \,F_{E}(k\tau_{k}, \delta) \,e^{-2 \frac{k^2}{k_{\sigma}^2}}.
\label{PH20}
\end{eqnarray}
where $F_{B}(x,\delta)$ and $F_{E}(x,\delta)$ coincide with the ones defined in Eqs. (\ref{EXPA5})--(\ref{EXPA6}).
\begin{figure}[!ht]
\centering
\includegraphics[height=7.3cm]{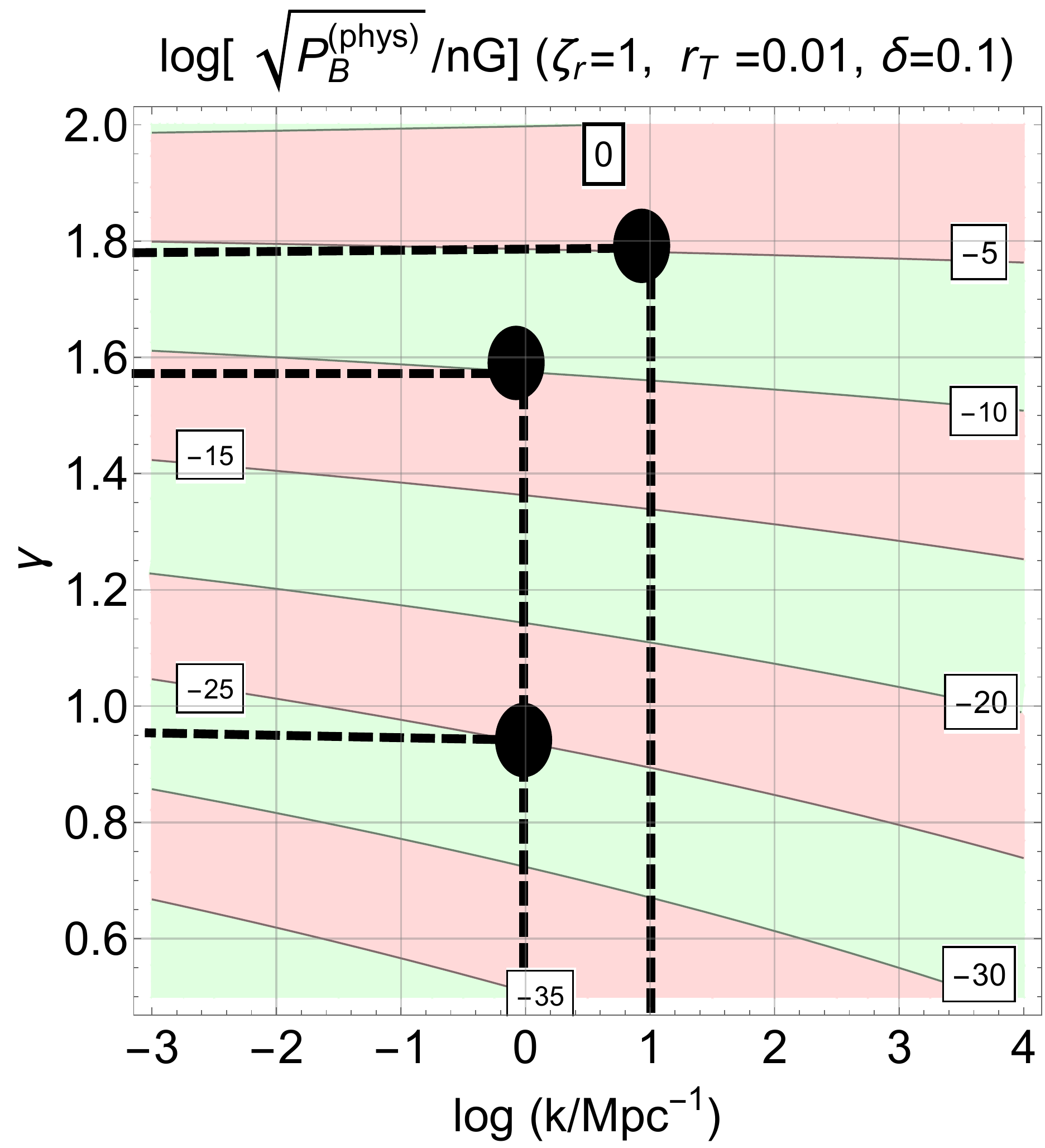}
\includegraphics[height=7.3cm]{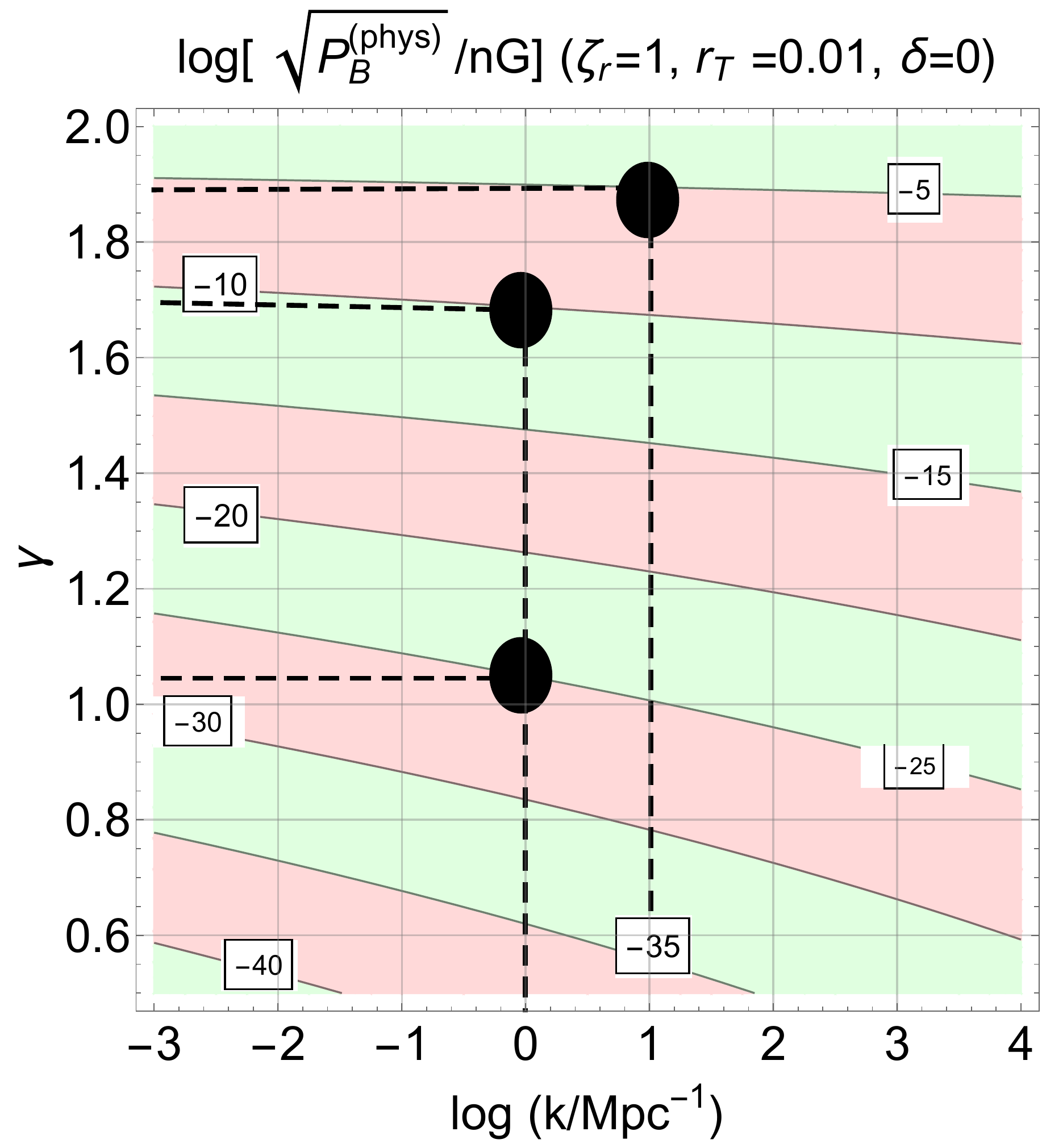}
\caption[a]{The parameter space is illustrated in the $(\gamma, \, k)$ plane. The numbers appearing on the various contours correspond to the common logarithm of the physical power 
spectrum. The common logarithm of $k$ in units of $\mathrm{Mpc}^{-1}$ is reported 
on the horizontal axis.}
\label{FFF1}      
\end{figure}

According to Eqs. (\ref{PH19})--(\ref{PH20})  the magnetic spectrum for 
$k < k_{\sigma}$ is practically not affected by the conductivity while the electric power spectrum is
suppressed by $k^2/\sigma^2\ll 1$. Since prior to decoupling the  electron-photon and electron-proton interactions tie the temperatures close together, the conductivity scales as $\sigma \sim 
\sqrt{T/m_{e}} T/\alpha_{em}$ where $m_{e}$ is the electromagnetic mass, $T$ is the temperature and $\alpha_{em}$ is the fine structure constant. For  instance, if we take $T= {\mathcal O}(\mathrm{eV})$ we get $k/\sigma = {\mathcal O}(10^{-30})$ for $k= {\mathcal O}(\mathrm{Mpc}^{-1})$. The electric power spectrum of Eq. (\ref{PH20}) will then be suppressed, in comparison with the magnetic spectrum, by a factor ranging between $40$ and $60$ orders of magnitude. This also demonstrates, in practical terms, that the duality symmetry is broken for $\tau > \tau_{k}$. 

\subsection{Magnetogenesis constraints}
\label{subs4}

According to the standard lore the observed large-scale fields in galaxies (and to some extent in clusters) 
should have been much smaller before the gravitational collapse of the protogalaxy. 
Compressional amplification typically increases the initial values of the magnetic seeds 
by $4$ or even $5$ orders of magnitude and the logic underlying this statement is 
in short the following (see e. g. \cite{rev1,rev2,rev3}). When the protogalactic matter 
collapsed by gravitational instability over a typical scale ${\mathcal O}(\mathrm{Mpc})$
the mean matter density before collapse was of the order of $\rho_{crit}$. 
whereas right after the collapse the mean matter density 
became, approximately, six orders of magnitude larger than the critical density.
Since the physical size of the patch decreases from $1$ Mpc to 
$30$ kpc the magnetic field increases, because of flux conservation, 
of a factor $(\rho_{a}/\rho_{b})^{2/3} \sim 10^{4}$ 
where $\rho_{a}$ and $\rho_{b}$ are, respectively the energy densities 
right after and right before gravitational collapse. 
\begin{figure}[!ht]
\centering
\includegraphics[height=7.3cm]{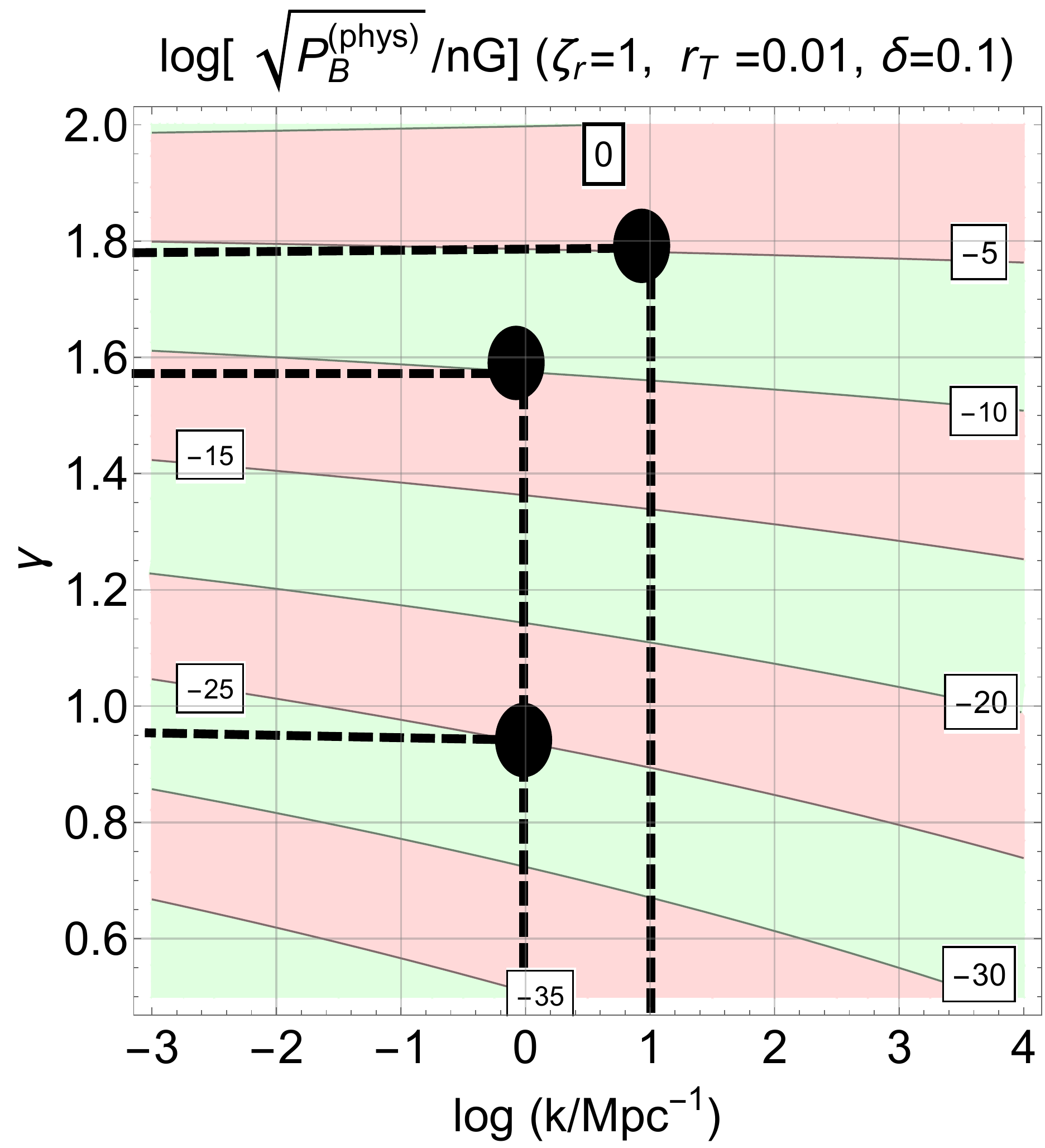}
\includegraphics[height=7.3cm]{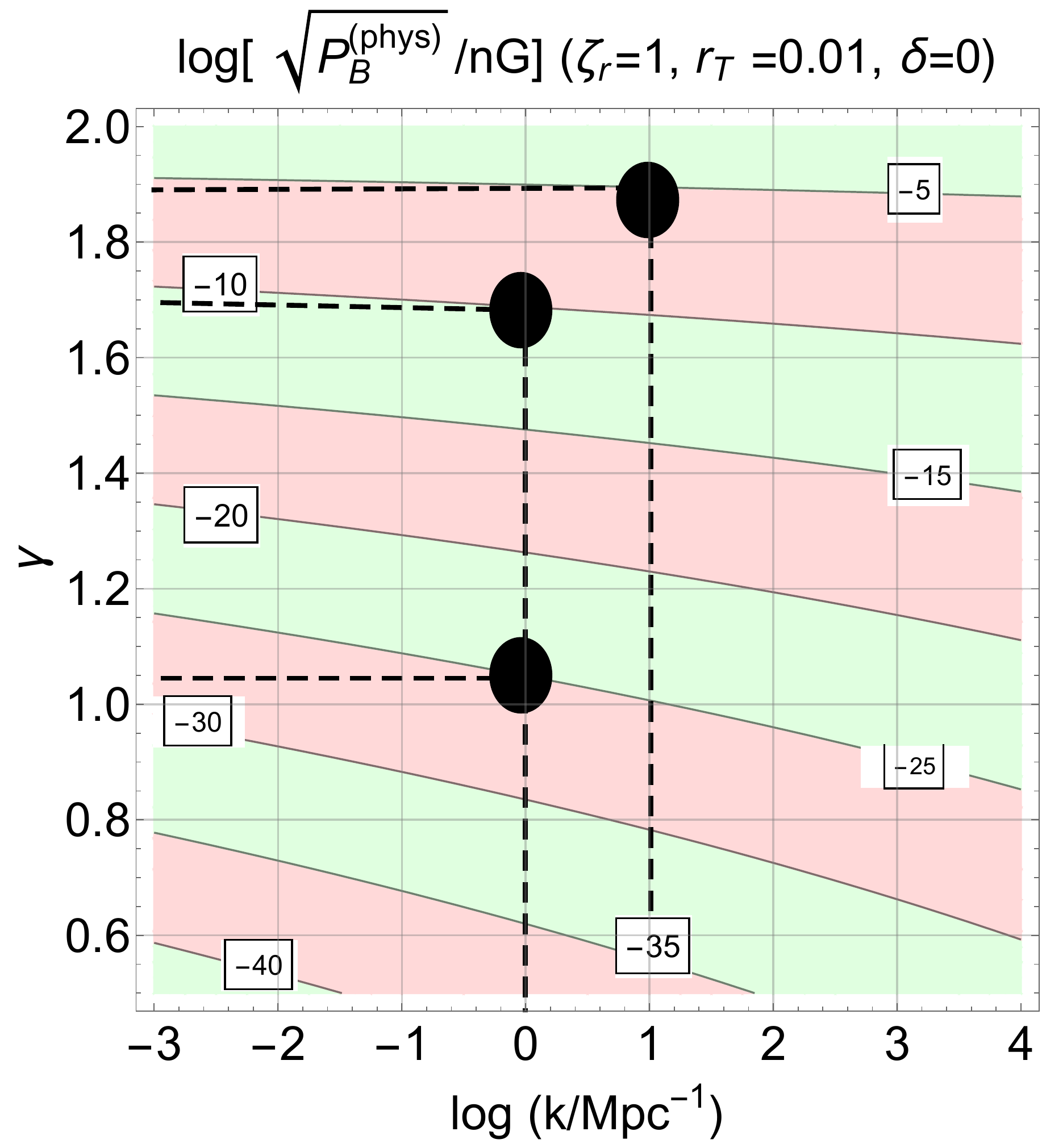}
\caption[a]{We illustrate the common logarithm of the 
power spectrum in the plane $(\gamma, \, r_{T})$. We remind that 
$r_{T}$ denotes, within the present notations, the tensor to scalar ratio. }
\label{FFF2}      
\end{figure}
Most of the work in the context of the dynamo 
theory focuses on reproducing the correct features of the 
magnetic field of our galaxy. The dynamo term may  be responsible for the origin of the magnetic field of the galaxy. The galaxy has a typical rotation period of 
$3 \times 10^{8}$ yrs and comparing this figure with the typical age of the galaxy, ${\cal O}(10^{10} {\rm yrs})$, it can be appreciated that the galaxy performed about $30$ rotations since the time 
of the protogalactic collapse. The achievable amplification produced by the 
dynamo instability can be at most of $10^{13}$, i.e. $e^{30}$. Thus, if 
the present value of the galactic magnetic field is ${\mathcal O}(\mu\mathrm{G})$, its value 
right after the gravitational collapse of the protogalaxy might have 
been as small as ${\mathcal O}(10^{-10})$ nG over a typical scale of $30$--$100$ kpc.

The compressional amplification and the dynamical action are typically combined 
together so that the magnetogenesis requirements 
 roughly demand that the magnetic fields at the time of the gravitational collapse of the protogalaxy should be approximately larger than a (minimal) power spectrum which can be estimated between ${\mathcal O}(10^{-32})\,\mathrm{nG}^2$ and ${\mathcal O}(10^{-22})\, \mathrm{nG}^2$:
\begin{equation}
\log{\biggl(\frac{\sqrt{P^{(phys)}_{B}}}{\mathrm{nG}} \biggr)} \geq - \xi, \qquad 11 < \xi < 16.
\label{mg3}
\end{equation}
The least demanding requirement  of Eq. (\ref{mg3}) (i.e. $\sqrt{P^{(phys)}_{B}} \geq 10^{-16}\,\mathrm{nG}$) follows by assuming 
that, after compressional amplification, every rotation of the galaxy increases the initial magnetic field of one $e$-fold. According to some this requirement is not completely since it takes more than one $e$-fold  to increase the value of the magnetic field by one order of magnitude  and this is the rationale for the most demanding condition of Eq. (\ref{mg3}), i.e. $\sqrt{P^{(phys)}_{B}} \geq 10^{-11}\,\mathrm{nG}$.
\begin{table}
\begin{center}
\vskip 0.5truecm
\begin{tabular}{| c || c || c ||  c || c |}
\hline
\quad\quad & $\gamma =1.8, \,\,\, \delta =0.01$ & $\gamma =1.9, \,\,\, \delta =0.01$ &  $\gamma=2,, \,\,\, \delta =0.01$\\ \hline
$k = 1\, \mathrm{Mpc}^{-1}$ &   $ \sqrt{P_{B}^{(phys)}} = 10^{-7.3} \, \mathrm{nG}$ & $\sqrt{P_{B}^{(phys)}} = 10^{-4.9} \, \mathrm{nG}$ &  $\sqrt{P_{B}^{(phys)}}=10^{-2.30} \,\mathrm{nG}$ \\ \hline
$k =  0. 1\, \mathrm{Mpc}^{-1}$ & $ \sqrt{P_{B}^{(phys)}} =  10^{-7.25} \, \mathrm{nG}$ & $\sqrt{P_{B}^{(phys)}} =  10^{-4.78} \, \mathrm{nG}$ &  $\sqrt{P_{B}^{(phys)}}=10^{-2.29} \, \mathrm{nG}$  \\ \hline
$k =  0. 01\, \mathrm{Mpc}^{-1}$ & $ \sqrt{P_{B}^{(phys)}} = 10^{-7.44} \, \mathrm{nG}$ & $\sqrt{P_{B}^{(phys)}} =  10^{-4.87} \, \mathrm{nG}$ &  $\sqrt{P_{B}^{(phys)}}= 10^{-2.28} \, \mathrm{nG}$  \\ \hline
\hline
\end{tabular}
\caption{Numerical values of the magnetic power spectrum at different scales and in the framework 
of the $(\gamma,\delta)$ transition where the gauge coupling is perturbative throughout all the stages 
of its evolution.}
\label{Table1}
\end{center}
\end{table}

\subsection{Charting the parameter space}
\label{subs5}
Let us now consider a simplified estimate where $\zeta_{r} \to 1$ and $\alpha \to 0$. This is is the situation 
of the concordance paradigm and for the typical values of the parameters 
given above the results for $\sqrt{P_{B}^{(phys)}(k,\tau_{0})}$ are reported in Tab. \ref{Table1} for 
different values of $\gamma$ and $k$.
Since the non-screened vector modes of the hypercharge
field project on the electromagnetic fields through the cosine of the Weinberg angle the estimates of Tab. \ref{Table1} follow from Eq. (\ref{PH19}) 
after multiplying the obtained result by $\cos{\theta_{W}}$ (recall, in this respect, that $\sin^2{\theta_W} \simeq 0.22$).  The illustrative orders of magnitude of Tab. \ref{Table1} will now be complemented 
with more detailed estimates allowing for a simultaneous variation 
of the different parameters in the light of the typical figures required by the 
magnetogenesis considerations.   

We shall distinguish the results obtained in the smooth 
 (i.e. $\delta\ll \gamma$) and the sudden (i.e.  $\delta \to 0$) approximations.
In practice this will be achieved by illustrating all the subsequent numerical results for $\delta =0.1$ and 
the case $\delta =0$.
In Fig. \ref{FFF1} the physical power spectrum is illustrated in 
the $(k,\, \gamma)$ plane. The black blobs in both 
plots indicate three reasonable regions of the parameter space where 
all the pertinent phenomenological constraints are satisfied. The 
numbers on the various curves denote, as explained on top of the plot,
the common logarithms of $\sqrt{P_{B}^{(phys)}}$ (expressed in $\mathrm{nG}$) 
along that curve. By looking at the intercept on the $\gamma$ axis we see 
that the phenomenologically reasonable values of $\gamma$ correspond to spectra that are blue or, at most, quasi-flat but always slightly increasing with $k$ (as already suggested by the results of Tab. \ref{Table1}). 

The results of Fig. \ref{FFF1} are complemented by Fig. \ref{FFF2}  where 
the power spectra are illustrated in the  $(\gamma, \, r_{T})$ plane
for fixed wavenumber $k = \mathrm{Mpc}^{-1}$. The black blobs 
correspond this time to the regions where $r_{T} = {\mathcal O}(0.01)$: for instance in Ref. \cite{RT3} the combination 
of different data sets implies $r_{T} <0.07$ while in Ref. \cite{RT4} the range of values $r_{T} <0.064$ is quoted.
By looking simultaneously at Figs. \ref{FFF1} and \ref{FFF2} 
we can appreciate that the viable range of $\gamma$ roughly extends between $1.3$ and $2$.

\subsection{Post-inflationary phases preceding the radiation epoch and reheating dynamics}
\label{subs6}
\begin{figure}[!ht]
\centering
\includegraphics[height=7.3cm]{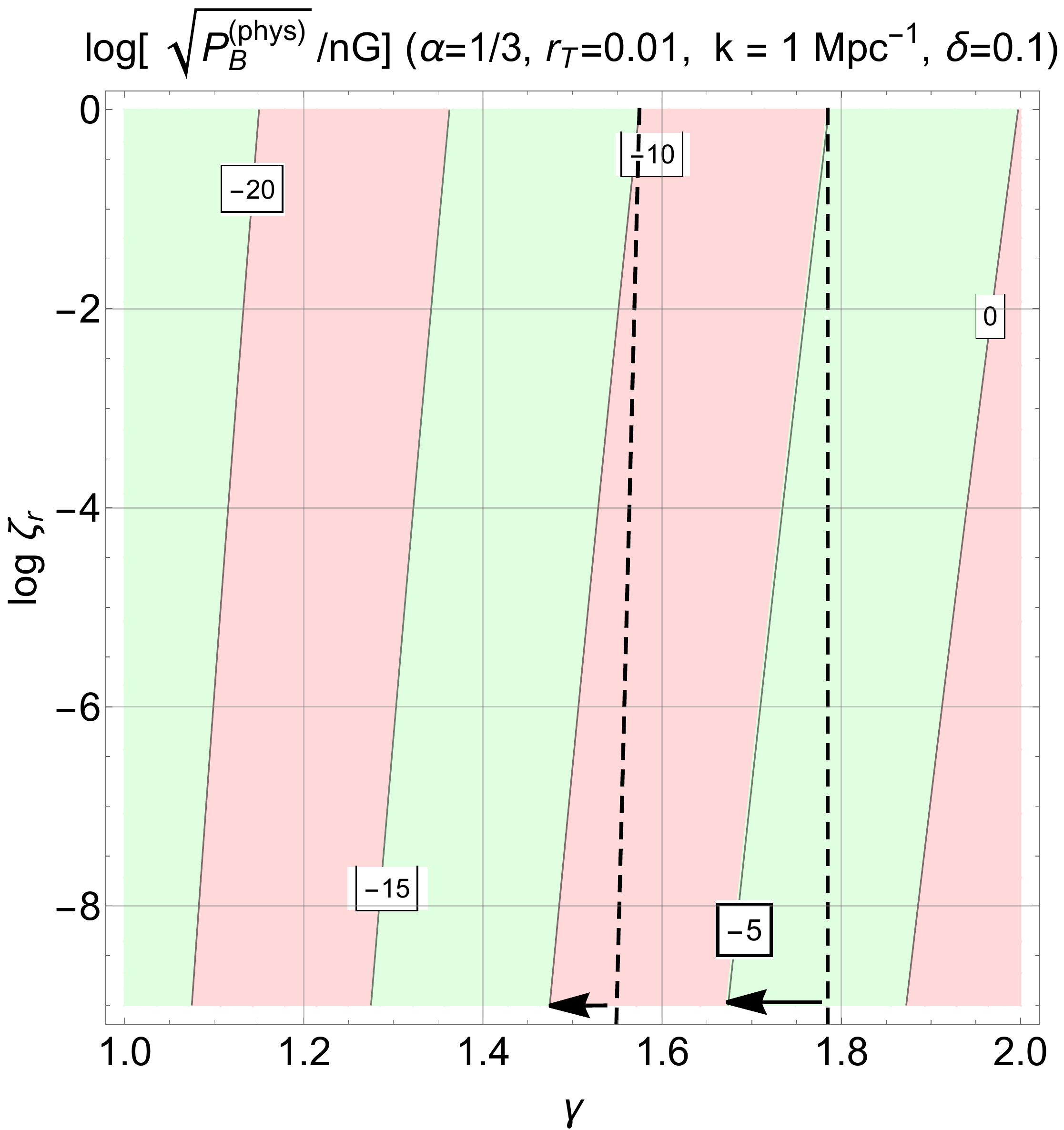}
\includegraphics[height=7.3cm]{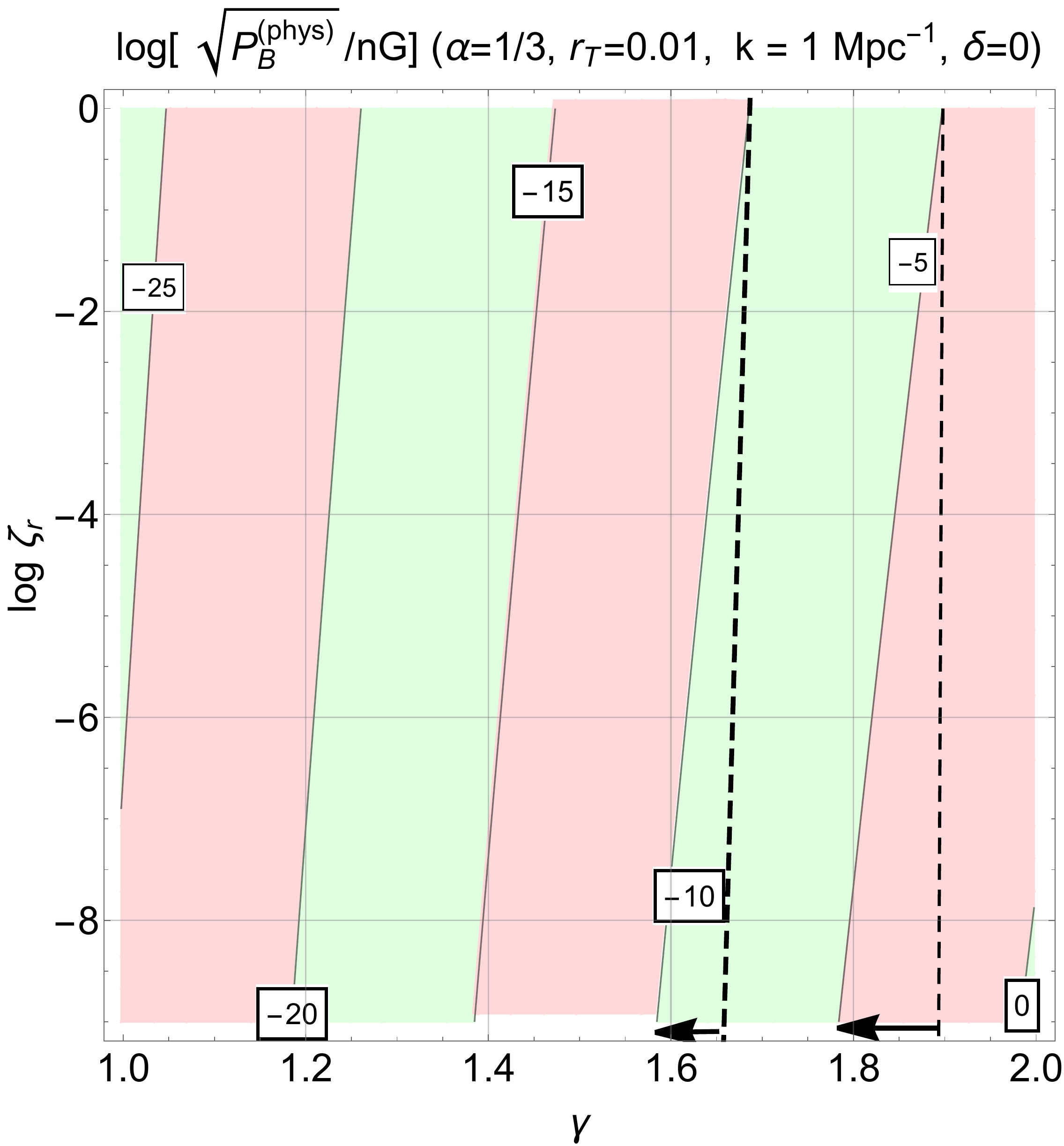}
\caption[a]{We illustrate the parameter space in the $(\gamma, \zeta_{r})$ plane for a typical expansion rate 
$\alpha =1/3$ corresponding to a post-inflationary stiff phase.}
\label{FFF3}      
\end{figure}

The results obtained so far also follow the continuity of the extrinsic curvature 
across the inflationary boundary. This approximation is customarily dubbed instant reheating and it is 
often used when setting bounds on quantities like the number of inflationary 
$e$-folds and the tensor to scalar ratio (see e.g. \cite{RT3,RT4}). The explosive 
particle decay as well as other non-perturbative effects related to the preheating dynamics probably bring the 
situation closer to the case of an instant reheating (see e.g. \cite{linde1,linde2}).
However, if the reheating is perturbative as originally suggested in the new inflationary scenario \cite{farhi1} 
the effect of the duration of the reheating phase may be relevant (see also \cite{GW3}). 

\begin{figure}[!ht]
\centering
\includegraphics[height=7.3cm]{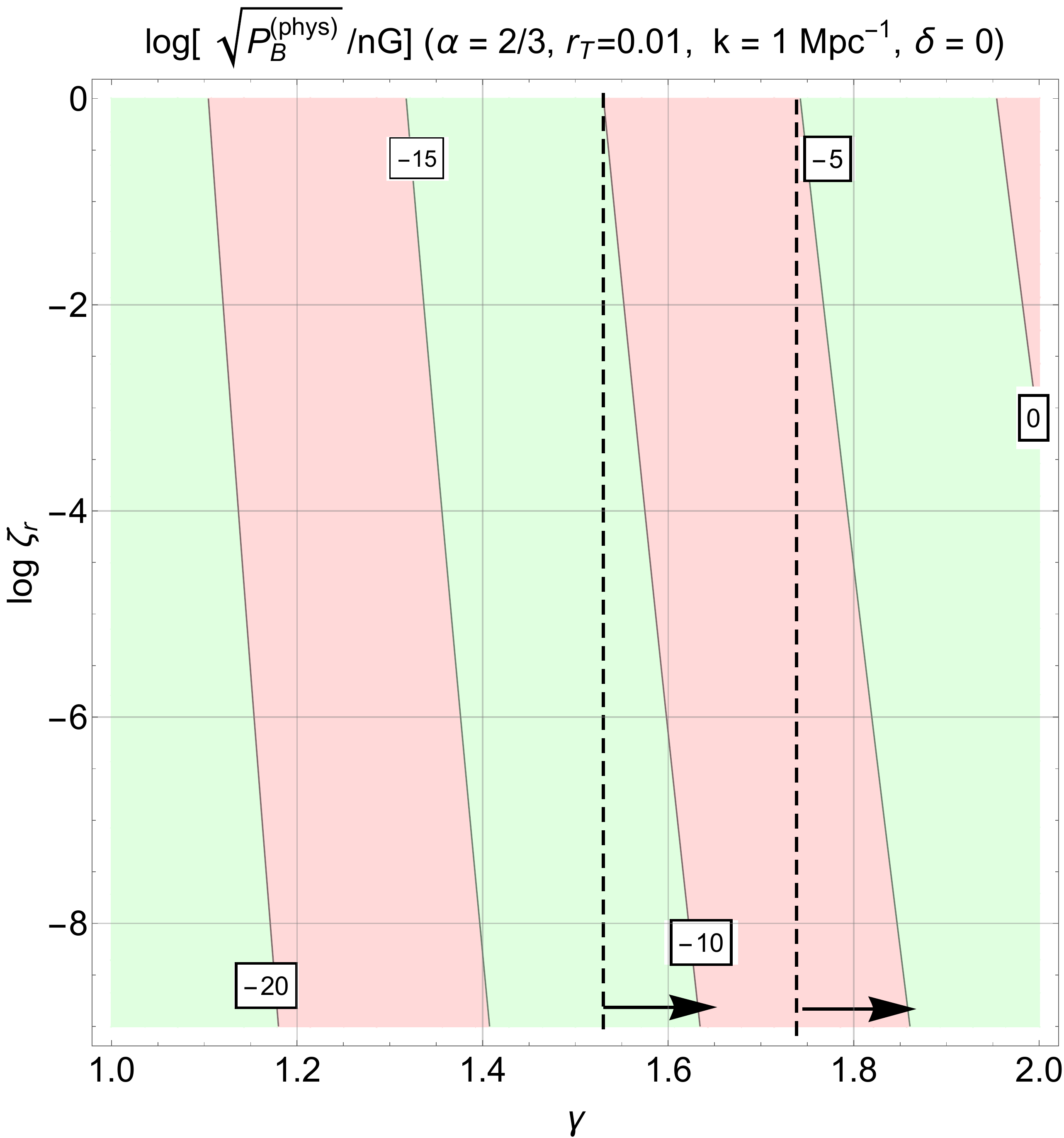}
\includegraphics[height=7.3cm]{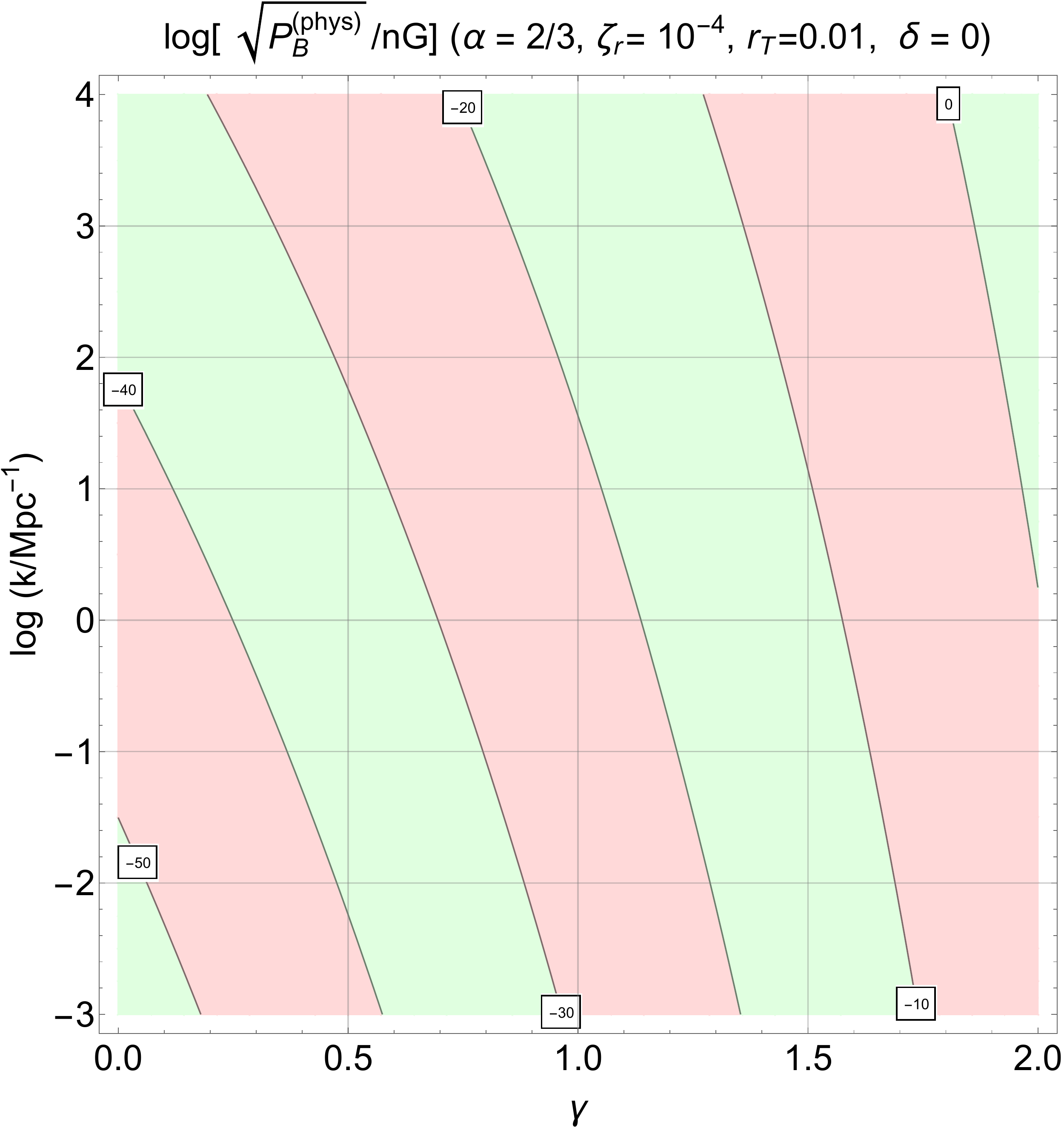}
\caption[a]{In the plot at the left the parameter space is illustrated in the $(\gamma, \zeta_{r})$ plane while in the plot at the right 
the parameter space is instead illustrated in the $(\gamma, k)$ plane. In both plots the typical expansion rate is faster than radiation (i.e. 
$\alpha =2/3$) as it happens, for instance, in the context of the reheating dynamics.}
\label{FFF3a}      
\end{figure}

\begin{figure}[!ht]
\centering
\includegraphics[height=7.3cm]{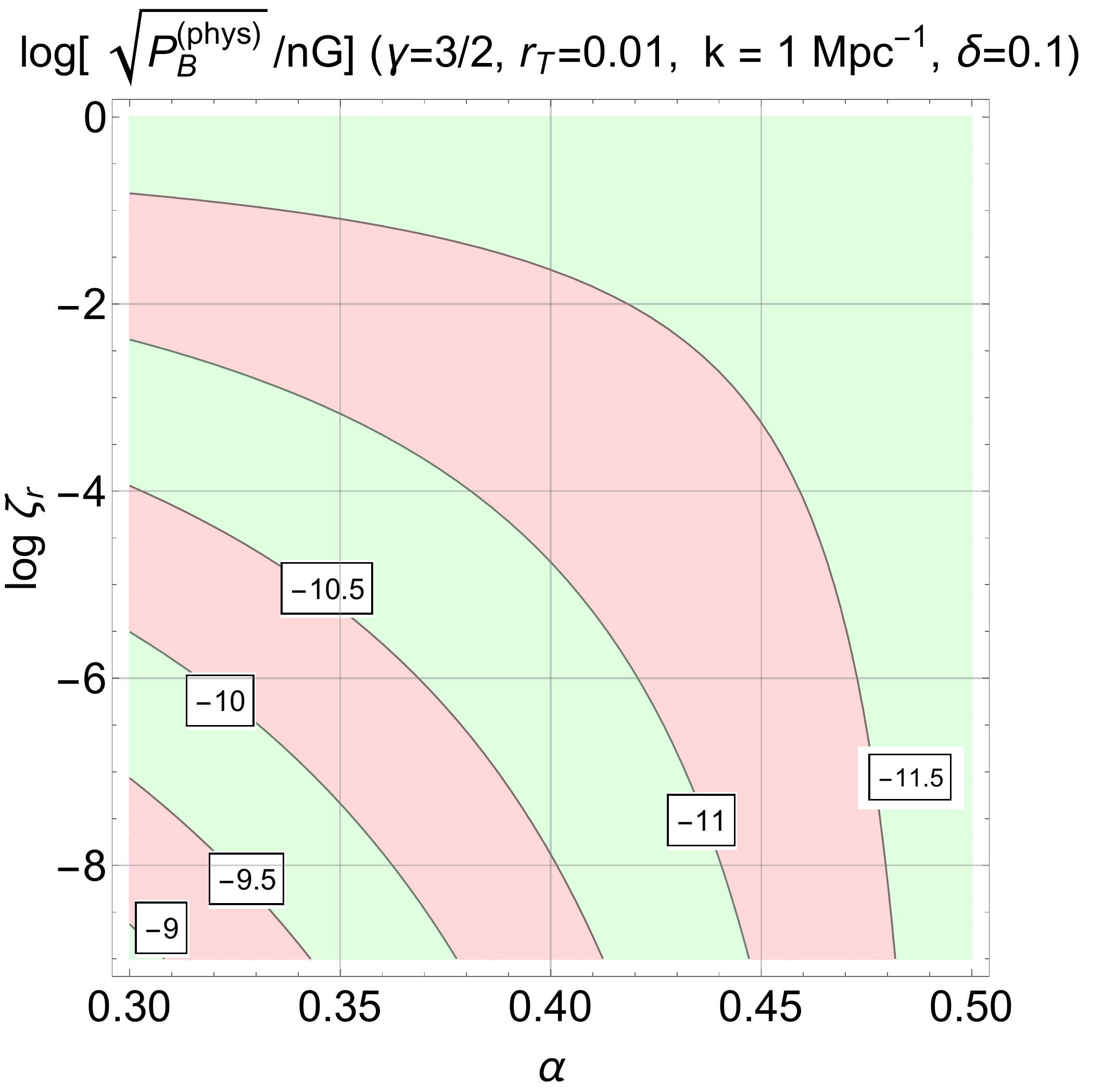}
\includegraphics[height=7.3cm]{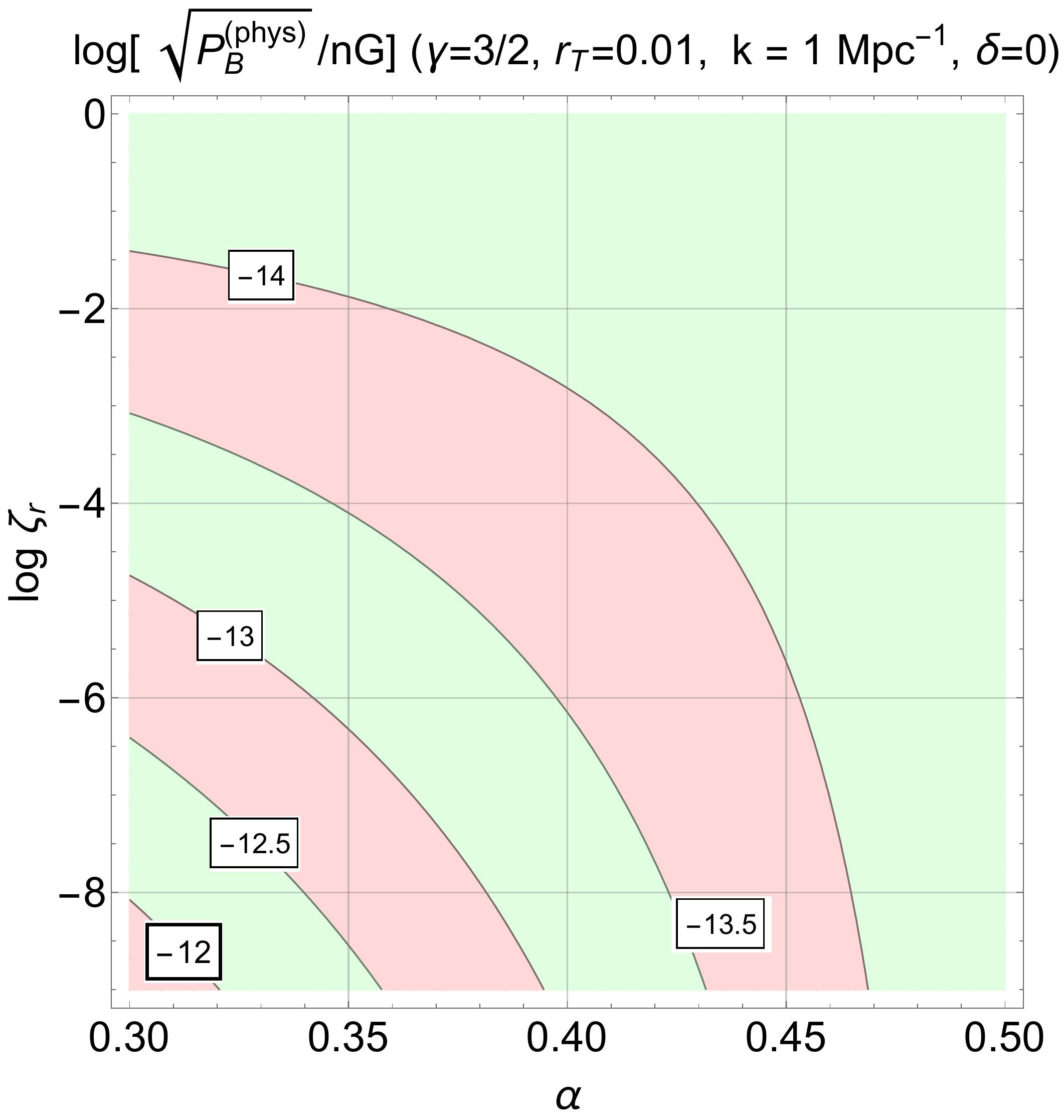}
\caption[a]{The allowed region of the parameter space is illustrated $(\alpha, \, \zeta_{r})$ plane for a fixed value of $\gamma$.}
\label{FFF4}      
\end{figure}

In Figs. \ref{FFF3}, \ref{FFF3a} and \ref{FFF4} we illustrated the situation where 
the dominance of radiation is {\em preceded} by a post-inflationary phase extending down to the 
relatively small values of the curvature scale. We remind from Eqs. (\ref{PH3})--(\ref{PH4}) that 
 $\zeta_{r} = H_{r}/H_{1}$  and $H_{r}$ denotes the Hubble rate 
 at the onset of the radiation dominance. Since $\alpha$ measures the expansion 
 rate between $H_{1}$ and $H_{r}$ the parameter space of the model can also be investigated 
 by varying $\alpha$ and $\zeta_{r}$. For instance in Fig. \ref{FFF3} the 
 value of $\alpha$ has been fixed (i.e. $\alpha \to 1/3$ as indicated in each of the two plots)
 and the contours of constant magnetic spectrum\footnote{We remind that the values of the magnetic power 
 spectrum reported on each of the curves 
remain {\em the same all along the curve} since this is the meaning of a contour plot. }
 have been investigated in the ($\gamma$, $\zeta_{r}$) plane. It is now relevant to consider with care the dashed lines in Fig. \ref{FFF3}: we see that  the same amplitude of the magnetic power spectrum
associated to a certain value of $\gamma$ when $\zeta_{r} \to 1$ (i.e. in the absence of an intermediate phase) 
can be obtained for a comparatively {\em lower value} of $\gamma$ if the intermediate phase is present 
(i.e. $\zeta_{r} < 1$).  This is the meaning of the arrows appearing in  Fig. \ref{FFF3}: for $\alpha=1/3$ (and for a 
given $\zeta_{r}< 1$) the $\gamma$ required to obtain a certain value of the power spectrum is comparatively smaller than in the case $\zeta_{r} =1$. This conclusion depends on the value of $\alpha$: in Fig. \ref{FFF3} 
the case $\alpha=1/3$ is meant to represent a post-inflationary phase dominated by 
a stiff fluid expanding at a rate {\em slower than radiation}. 

In the case of an extended reheating dynamics the expansion rate prior to the dominance of radiation is expected to be faster than in the stiff case. In Fig. \ref{FFF3a} we then considered 
the possibility of a post-inflationary phase preceding the ordinary radiation phase 
where the expansion rate is effectively faster than radiation. 
By looking at the dashed lines in Fig. \ref{FFF3a} we see that the situation 
differs from the one illustrated in Fig. \ref{FFF3}: for $\alpha = 2/3$ (and for a 
given $\zeta_{r}< 1$) the $\gamma$ required to obtain a certain value of the power spectrum is comparatively 
larger than in the case $\zeta_{r} =1$.  The same conclusion holds for different expansion rates 
faster than radiation.
Along this  perspective we can therefore consider the plane $(\alpha, \zeta_{r})$ for a fixed 
value of $\gamma$ (e.g. $\gamma\to3/2$, as in Fig. \ref{FFF4}).   
As  $\zeta_{r}$ diminishes (i.e. the post-inflationary phase gets longer) the power spectrum gets {\em larger}. 
In Fig. \ref{FFF4} we considered the situation where the expansion rate during the intermediate phase was slower 
than radiation and we have also purposely chosen $\gamma =3/2$. For larger values of $\gamma$ the amplitude of the power spectra will be larger but the basic trend of Fig. \ref{FFF4} will remain the same.
All in all we can therefore say that the inclusion of a post-inflationary phase preceding the ordinary radiation-dominated stage has a threefold effect depending on the expansion rate:
\begin{itemize}
\item{} if the post-inflationary phase preceding radiation expands faster than radiation the amplitude of the power spectrum 
gets mildly reduced for a fixed value of $\gamma$; 
this means that to achieve the same amplitude we should have a 
comparatively larger value of $\gamma$;
\item{} if the post-inflationary phase expands at a rate slower than radiation (as it happens in the case of stiff post-inflationary 
phases) the power spectrum is mildly enhanced, i.e. to achieve the same spectral amplitude we should have a comparatively 
smaller value of $\gamma$;
\item{} in both cases the length of the post-inflationary phase is crucial: for a long stiff post-inflationary phase the amplitude 
of the magnetic power spectrum {\em gets larger}; conversely for a delayed reheating expanding faster than radiation the 
power spectrum gets reduced.
\end{itemize}
We finally mention that the effects of a post-inflationary phase have been qualified as {\em mild} in the previous list 
of items. By this we simply mean that for an excursion of $\zeta_{r}$ of about $9$ orderers of magnitude 
the spectral index gets renormalised by a factor of the order of $0.1$. So this preliminary analysis 
seems to confirm that these are not large effects. 
\begin{figure}[!ht]
\centering
\includegraphics[height=7.4cm]{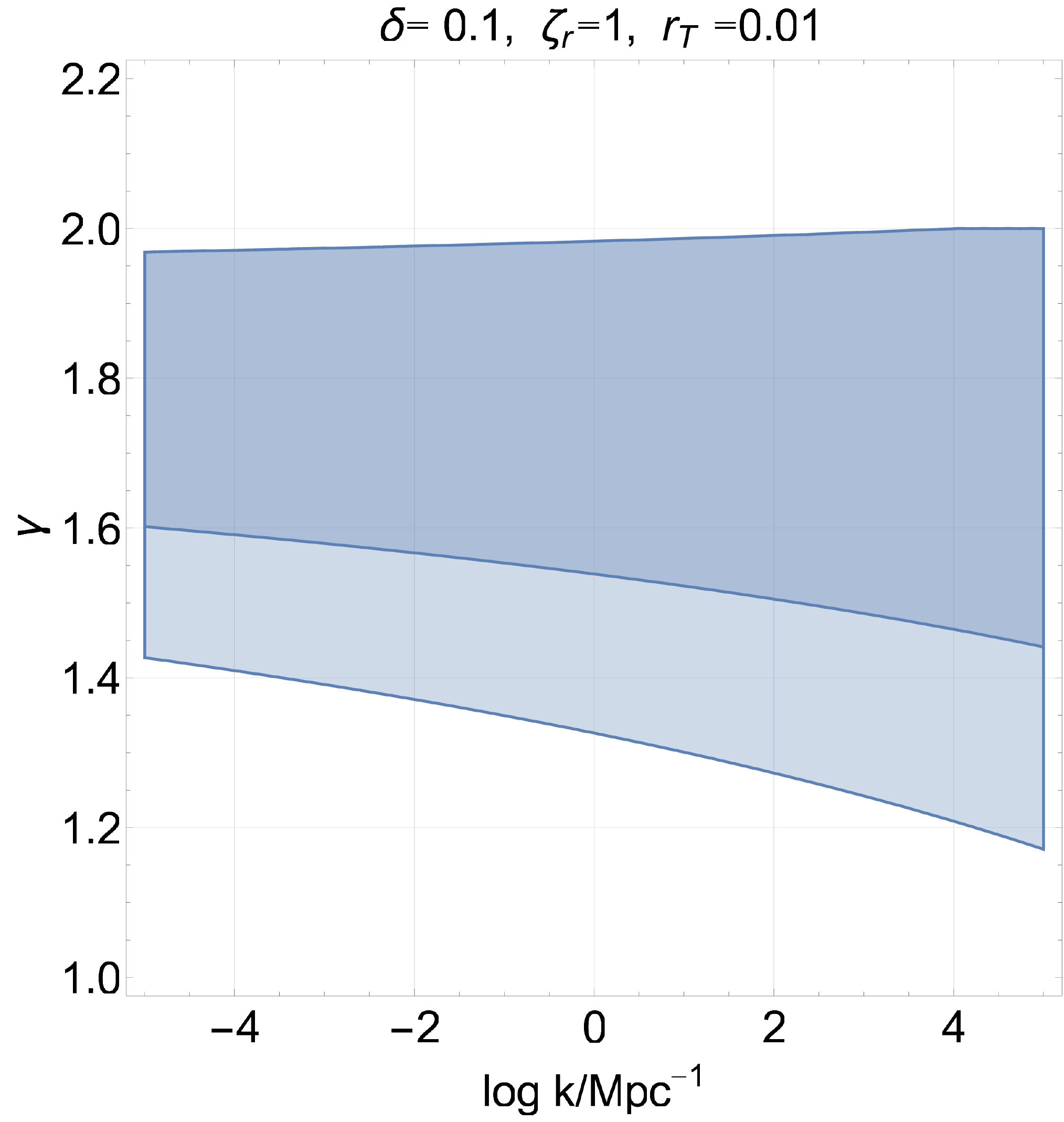}
\includegraphics[height=7.4cm]{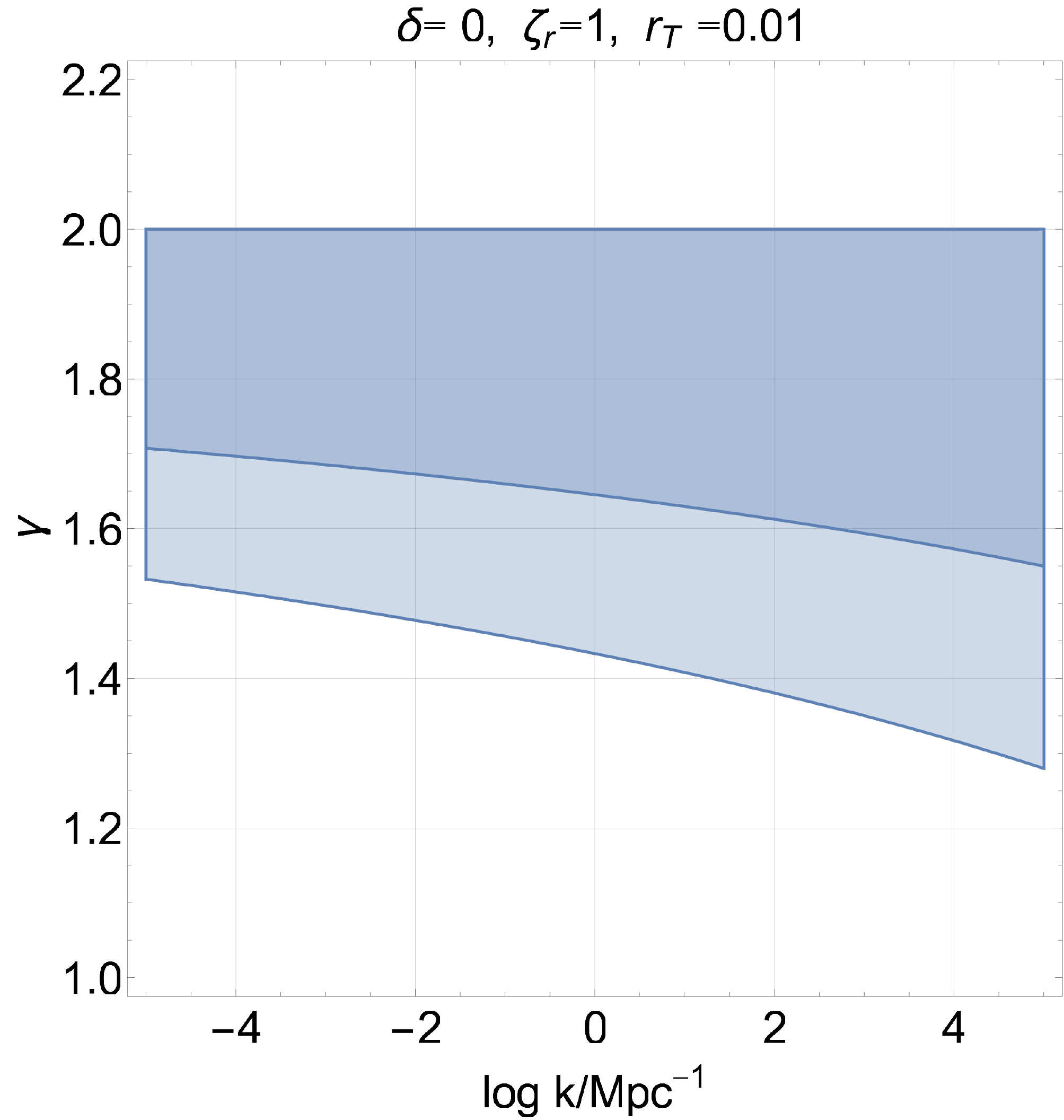}
\caption[a]{The allowed region of the parameter space is illustrated with a 
shaded area for fixed $\delta$ and in the $(\gamma,\,k)$ plane. The different shading corresponds to the two extreme values of $\xi$ appearing in Eq. (\ref{mg3}).}
\label{FFF5}      
\end{figure}
\subsection{Final assessment of the obtained results and the meaning of the parameter space}
\label{subs7}
 Since this analysis has not been done before it is appropriate, as we conclude, to summarize 
the main results. The first aspect we stress is that has been be explicitly used to deduce the gauge power spectra both during inflation and at late times. The symmetry can therefore be used to verify 
and constrain the explicit expressions of the power spectra. 

If the gauge coupling increases during a quasi-de Sitter stage of expansion only the hyperelectric spectrum can be flat for typical wavelengths larger than the effective horizon during inflation [see e.g. Eq. (\ref{TWO11}) and discussions therein]. The duality symmetry implies therefore that 
only the hypermagnetic spectrum can be flat [see Eq. (\ref{THREE9})] during a quasi-de Sitter 
stage of expansion and for typical wavelengths larger than the effective horizon.
These results, even if not explicitly related to the duality symmetry, were known also in Refs. \cite{DT1}
and suggested the idea that to have a successful magnetogenesis scenario we should 
necessarily consider a dynamical situation where, in the language of this paper, the gauge coupling decreases.
Only in this case we could obtain a quasi-flat magnetic spectrum during the quasi-de Sitter phase. 
Even if it has been later on recognised that non only the inflaton but also some other spectator fields could effectively play the role of a dynamical gauge coupling \cite{DT2} the conclusion remained: only if the gauge coupling decreases a flat hypermagnetic spectrum can be obtained during inflation. 

The first observation is therefore that, within the present approach the twofold conclusion mentioned in the previous paragraph, follows directly from the dual treatment of the problem and of the related spectra; see, in particular the discussions after  Eqs. (\ref{TWO9})--(\ref{TWO10}) and also after Eqs. (\ref{THREE6})--(\ref{THREE7}). For a number of years, it has been assumed that only when the gauge coupling decreases the magnetogenesis constraints could be satisfied \cite{DT3}.  This possibility has however some drawbacks that have been accurately discussed here  (see subsections \ref{subs241} and \ref{subs242}): if the hypermagnetic spectrum is flat during inflation strong coupling is expected when the gauge coupling starts its evolution. 

The main idea of this paper has been to consider a more complete scenario where the 
gauge coupling first increases (or decreases) and then flattens out at late times as illustrated in Fig. \ref{FFF0a}
and \ref{FFF0b}. This approach, heavily based on the continuity of the mode functions and of the 
extrinsic curvature also considers specifically the transient regime where the gauge coupling relaxes 
and it does it in a computable manner. In the past the common approximation was that the gauge 
coupling (and hence $\sqrt{\lambda}$) abruptly freezes after inflation: we argued that this approach 
is not fully consistent a priori since it implies that the relevant pump fields (like $\sqrt{\lambda}^{\prime\prime}$)
are singular across the inflationary boundary.  If the gauge coupling flattens out in a smooth 
manner the present results showed that the late-time hypermagnetic power spectra outside the horizon in the radiation epoch are determined by the hyperelectric fields at the end of inflation whereas the opposite is true in the case of decreasing coupling. The obtained results then suggest that a slightly blue hyperelectric spectrum during inflation may lead to a quasi-flat hypermagnetic spectrum prior to matter radiation equality and before the relevant wavelengths reenter the effective horizon. Once more these results are  a consequence 
of duality and of a very accurate analysis of the transition regime between the early inflationary stage and 
the subsequent evolution, as illustrated in Figs. \ref{FFF0a} and \ref{FFF0b}. The obtained magnetic fields 
have been confronted with the magnetogenesis requirements and we can then 
 conclude that large-scale magnetic fields can be generated during a quasi-de Sitter 
stage of expansion while the gauge coupling remains perturbative throughout all the stages 
of the dynamical evolution.

\begin{figure}[!ht]
\centering
\includegraphics[height=7.4cm]{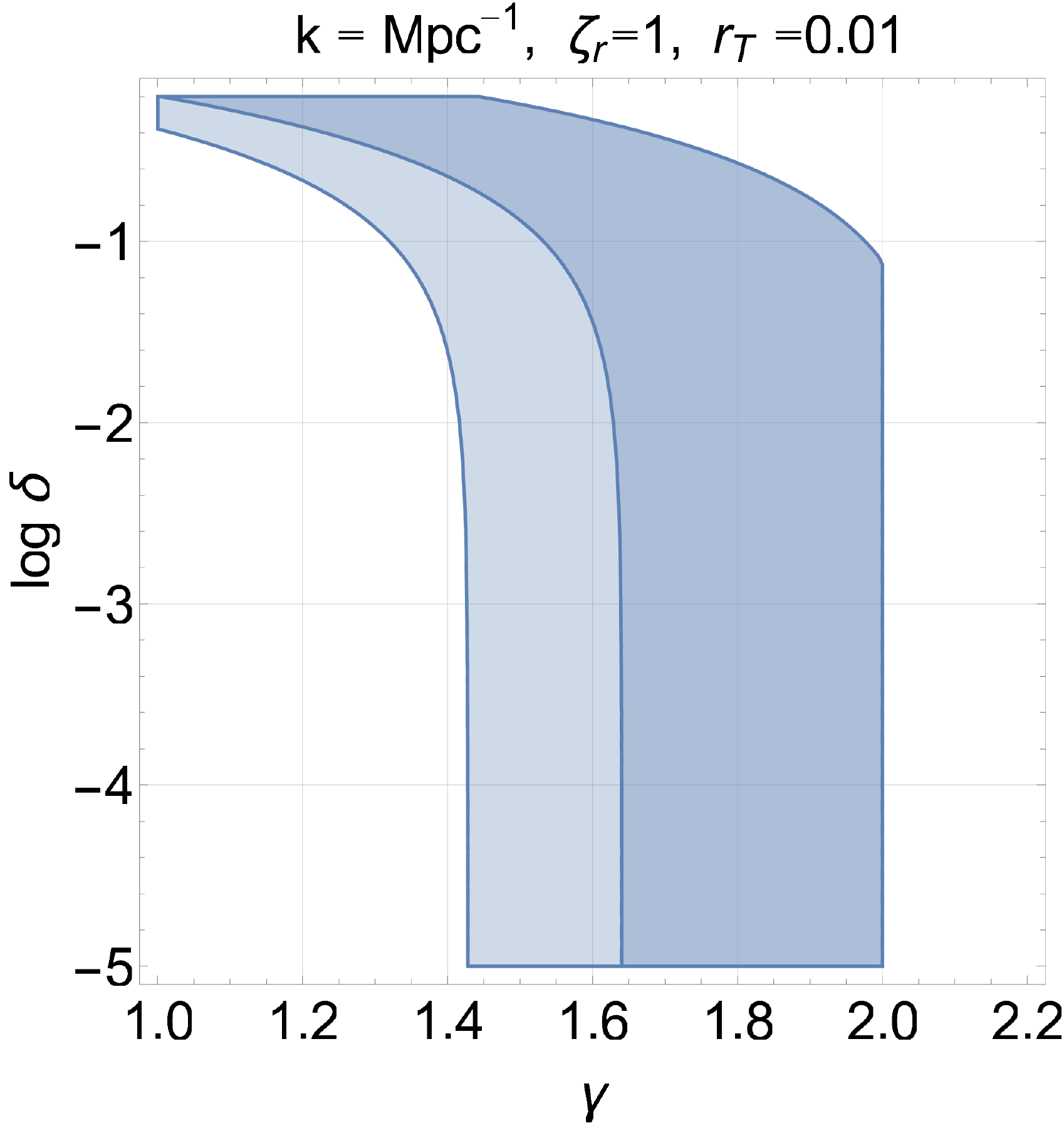}
\includegraphics[height=7.4cm]{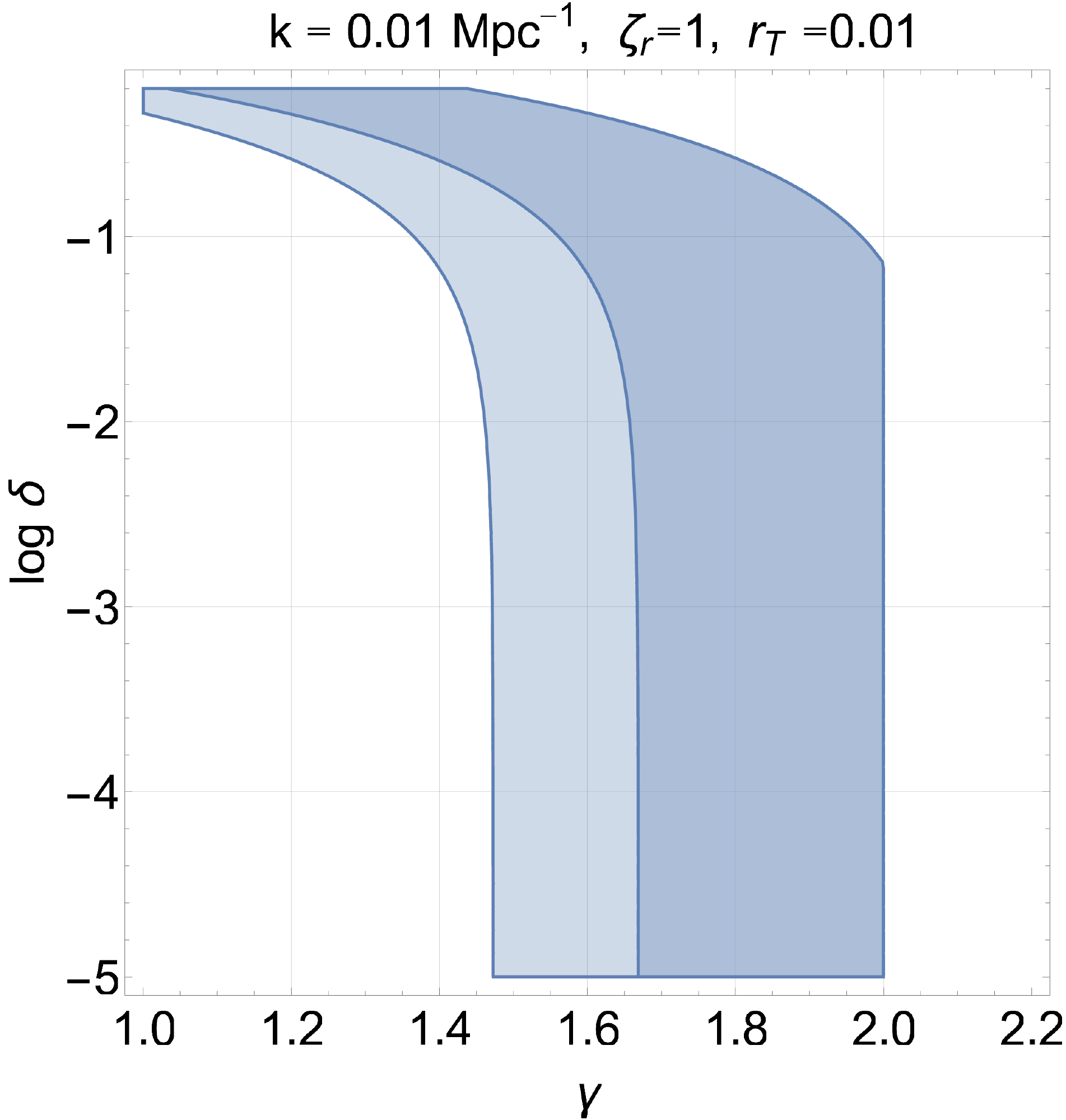}
\caption[a]{The allowed region of the parameter space is illustrated with a 
shaded area for fixed $k$ and in the $(\gamma,\,\delta)$ plane. As in Fig. \ref{FFF5} the different shadings correspond to the two extreme values of $\xi$ appearing in Eq. (\ref{mg3}).}
\label{FFF6}      
\end{figure}

The results of this analysis have been summarized in Figs. \ref{FFF5} and \ref{FFF6}.
In Fig. \ref{FFF5} the shaded areas denote the region where the spectral energy density 
is subcritical both during and after inflation while the magnetogenesis 
and the Cosmic Microwave Background (CMB) constraints 
are all satisfied. While the discussion 
of the CMB constraints would probably require a separate analysis, some general results 
could be used and these are essentially the ones reproduced in Fig. \ref{FFF3}: they amount to 
requiring that the physical power spectrum after equality but before decoupling 
is smaller than $10^{-2}$ nG for typical wavenumbers comparable with the pivot 
scale $k_{p} = 0.002\, \mathrm{Mpc}^{-1}$ at which the scalar and tensor power spectra 
are customarily assigned.  The proper analysis of the CMB effects associated with 
the magnetic random fields started already after the first releases of the WMAP data; through the years 
various direct constraints have been derived both from the temperature and from the polarization anisotropies 
(see, for instance, \cite{CMBM} for a recent review).
The main difference between the two plots of Fig. \ref{FFF5} 
stems from the magnetogenesis requirements: the larger area in the right plot corresponds to the 
situation where each galactic rotation amplifies the initial magnetic field value by one $e$-fold
while the smaller area follows by requiring that the physical power spectrum exceed $10^{-22}\,
\mathrm{nG}^2$ (see Eq. (\ref{mg3}) and discussion therein).
In Fig. \ref{FFF6} we consider the same constraints of Fig. \ref{FFF5} but in the $(\gamma, \delta)$ plane.
The physical region obviously corresponds to the case $\delta \ll 1$. 

All in all we can conclude that magnetogenesis is phenomenologically viable in the perturbative regime 
provided the gauge coupling increases during a quasi-de Sitter stage and then flattens out after the 
end of the inflationary phase. The sweet spot of the parameter space (i.e. where all reasonable constraints are met) corresponds to a slightly increasing (blue) spectrum where ${\mathcal O}(1.6)\leq \gamma < 2$. If the 
 magnetogenesis requirement are relaxed $\gamma$ can be as small as ${\mathcal O}(1.4)$. 
 
 It does not make much sense to compare the present findings with the parameter space of the models 
 where the gauge coupling decreases since, in those cases, a strongly coupled regime is expected at the 
 onset of inflation. We instead conclude with a discussion of the various parameters of the scenario. 
 These parameters are: 
 \begin{itemize}
 \item{} $\gamma$ and $\delta$ (or $\widetilde{\gamma}$ and $\widetilde{\delta}$ in the dual situation) describing the dynamics of the gauge coupling according to Figs. 
 \ref{FFF0a} and \ref{FFF0b};
 \item{} the maximal inflationary scale which is however determined by the tensor to scalar ratio $r_{T}$ which we 
 took smaller than about $0.64$ \cite{RT3,RT4};
 \item{} the possible presence of a post-inflationary phase whose length has been parametrized by $\zeta_{r}$;
 \item{} the expansion rate during the latter phase.
 \end{itemize}
 To these parameters we could eventually add $e_{1}= \sqrt{4\pi/\lambda_{1}}$, i.e. the value of the 
 gauge coupling at the end of inflation but this is implicitly included in the initial profile 
 of the gauge couplings of Figs. \ref{FFF1} and \ref{FFF2}.
Depending on how we count we therefore have overall $5$ or $6$ parameters. If we would consider 
 the case of instantaneous reheating the number of parameters could be reduced by $2$.
 We are not interested here in the minimisation of the free parameters since the purpose of this paper is just to illustrate a potentially novel approach to the problem: more parameters might be included 
 to account for more detailed physical aspects of the scenario. 
It is actually debatable if the minimisation of the parameters should be the ultimate purpose of a scenario and the sole criterion to assess its viability: an increase in the parameters typically entails a more accurate 
description of certain phenomena and a good example is actually the effect of reheating on magnetogenesis, as discussed above. Other examples could be made but it is not our purpose 
here to debate the heuristic value of this type of arguments. 

\newpage

\renewcommand{\theequation}{6.\arabic{equation}}
\setcounter{equation}{0}
\section{Concluding considerations}
\label{sec6}
Duality implies that the hypermagnetic power spectra parametrically amplified from quantum fluctuations during a quasi-de Sitter 
stage of expansion are scale-invariant  (or slightly blue) if the gauge coupling decreases while an increase of the gauge coupling is only compatible with a flat  hyperelectric spectrum for wavelengths larger than the effective horizon at the corresponding epoch. The same duality symmetry demands that the late-time gauge spectra do not always coincide with the results obtained at the end of inflation: the late-time hypermagnetic spectra follow directly from the hypermagnetic mode functions 
at the end of inflation whenever the gauge coupling decreases. Conversely 
if the gauge coupling increases the late-time hypermagnetic spectra are determined by the 
hyperelectric mode functions at the end of the inflationary phase. On a technical ground these 
results follow from the specific analysis of an appropriate transition matrix whose elements have well defined transformation properties under the duality symmetry and control the form of the late-time spectra.

Form a semiclassical viewpoint the mechanism analyzed here is the gauge analog of the Sakharov 
oscillations where travelling waves are transformed into  standing waves with well defined phases that depend  
on the dynamics  of the underlying geometry. The standing oscillations associated with the hyperelectric 
and with the hypermagnetic mode functions follow from the dynamics of the gauge couplings.
Only when the gauge coupling decreases it is therefore reasonable to compute the post-inflationary physical spectra by a simple redshift of the comoving result during inflation. Conversely such a procedure would lead to an incorrect result in the case of an increasing gauge coupling where the late time hypermagnetic spectrum is determined by the hyperelectric spectrum at the end of inflation. 

In summary, all the magnetogenesis constraints can be successfully satisfied when the gauge coupling remains perturbative throughout all the stages of its evolution. More precisely a slightly blue hyperelectric spectrum during inflation may lead to a quasi-flat hypermagnetic spectrum at late times. The induced large-scale magnetic fields turn out to be of the order of a few thousands of a nG over typical length scales between  few Mpc and $100$ Mpc. The magnetogenesis requirements are therefore satisfied together with all the backreaction constraints both during and after inflation. For the sweet spots of the parameter space the only further amplification required to seed the galactic magnetic fields is the one associated with the compressional amplification.

\section*{Acknowledgments} 
It is a pleasure to thank T. Basaglia and S. Rohr of the CERN Scientific Information Service for their kind help throughout various stages of this investigation.

\newpage

\begin{appendix}

\renewcommand{\theequation}{A.\arabic{equation}}
\setcounter{equation}{0}
\section{Transition matrix and increasing gauge coupling}
\label{APPA}
In this first part of this appendix we shall consider a gauge coupling that sharply increases during the inflationary phase and then flattens out in the post-inflationary 
era. The profile of $\sqrt{\lambda}$ corresponds, in this case, to Eqs. (\ref{TWO1})--(\ref{FIVE2}) with $0\leq \delta \ll \gamma$ (see also Fig. \ref{FFF0a} and discussion therein). Since the parametrization of $\sqrt{\lambda}$ is  continuous and differentiable  the evolution of the mode functions during the post-inflationary stage 
(i.e. for $\tau \geq - \tau_{1}$) follows from the solution of Eqs. (\ref{ONE8}) and (\ref{ONE8w}). For $ \tau \geq - \tau_{1}$ 
the mode functions $f_{k}(\tau)$ and $g_{k}(\tau)$ are given in terms of certain 
coefficients $A_{\pm}(k,\tau_{1})$:
\begin{eqnarray}
f_{k}(\tau) &=& \frac{\sqrt{ k y}}{\sqrt{2 k}} \biggl[ A_{-}(k,\tau_{1})\,N_{\nu}\,H_{\nu}^{(1)}(k y) +  A_{+}(k,\tau_{1})\,N_{\nu}^{*}\,H_{\nu}^{(2)}(k y)\biggr],
\label{FIVE4}\\
g_{k}(\tau) &=& \sqrt{\frac{k}{2}} \sqrt{ k y} \biggl[A_{-}(k,\tau_{1})\,N_{\nu}\,H_{\nu-1}^{(1)}(k y) +  A_{+}(k,\tau_{1})\,N_{\nu}^{*}\,H_{\nu-1}^{(2)}(k y)\biggr],
\label{FIVE5}
\end{eqnarray}
where the following auxiliary quantities have been introduced:
\begin{eqnarray}
N_{\nu} &=& \sqrt{\frac{\pi}{2}} e^{i \pi( 2 \nu + 1)/4}, \qquad y(\tau) = \tau + \tau_{1} (1 + q), 
\nonumber\\
q &=& q(\delta,\gamma) =  \frac{\delta}{\gamma}, \qquad \nu = \biggl(\delta + \frac{1}{2}\biggr).
\label{FIVE6}
\end{eqnarray}
The separate continuity of  Eqs.  (\ref{TWO2})--(\ref{TWO3}) and (\ref{FIVE4})--(\ref{FIVE5}) across $\tau = - \tau_{1}$, determines the explicit forms of $A_{\pm}(k,\tau_{1})$ in the case\footnote{The inflationary mode functions depend on the range of $\gamma$ (see, in in this respect,
Eqs. (\ref{TWO2}) and (\ref{TWO3})--(\ref{TWO4})). Therefore that Eq. (\ref{FIVE8}) only holds for $\gamma>1/2$ while Eq. (\ref{FIVE10}) holds for $0<\gamma < 1/2$.} $\gamma > 1/2$:
\begin{eqnarray}
&&A_{-}(k, \tau_{1}) = \frac{ i \pi N_{\mu}}{4 N_{\nu}} \sqrt{q} x_{1} \biggl[ H_{\mu}^{(1)}(x_{1})  H_{\nu -1 }^{(2)}(q\,x_{1}) - H_{\mu + 1}^{(1)}(x_{1})  H_{\nu}^{(2)}(q\,x_{1}) \biggr],
\nonumber\\
&&A_{+}(k, \tau_{1}) = \frac{ i \pi N_{\mu}}{4 N_{\nu}^{\ast}} \sqrt{q} x_{1} \biggl[ H_{\mu+ 1}^{(1)}(x_{1})  H_{\nu }^{(1)}(q\,x_{1}) - H_{\mu}^{(1)}(x_{1})  H_{\nu-1}^{(1)}(q\,x_{1}) \biggr],\qquad \gamma > 1/2,
\label{FIVE8}
\end{eqnarray}
where  $x_{1}= k\tau_{1}$. For $ 0 < \gamma < 1/2$  the expressions of $A_{\pm}(k,\tau_{1})$ are instead obtained 
by matching Eqs. (\ref{TWO2}) and (\ref{TWO4}) with Eqs. (\ref{FIVE4})--(\ref{FIVE5}):
\begin{eqnarray}
&& A_{-}(k, \tau_{1}) = \frac{ i \pi N_{\mu}}{4 N_{\nu}} \sqrt{q} x_{1} \biggl[ H_{\mu}^{(1)}(x_{1})  H_{\nu -1 }^{(2)}(q\,x_{1}) + H_{\mu - 1}^{(1)}(x_{1})  H_{\nu}^{(2)}(q\,x_{1}) \biggr],
\nonumber\\
&& A_{+}(k, \tau_{1}) = -\frac{ i \pi N_{\mu}}{4 N_{\nu}^{\ast}} \sqrt{q} x_{1} \biggl[ H_{\mu- 1}^{(1)}(x_{1})  H_{\nu }^{(1)}(q\,x_{1}) + H_{\mu}^{(1)}(x_{1})  H_{\nu-1}^{(1)}(q\,x_{1}) \biggr], \quad 0< \gamma < 1/2.
\label{FIVE10}
\end{eqnarray}
Since the electric and the magnetic mode functions must obey the Wronskian normalization of Eq. (\ref{ONE8w}), $A_{\pm}(k,\tau_{1})$ must satisfy the condition $\bigl|A_{+}(k,\tau_{1})\bigr|^2 - \bigl|A_{-}(k,\tau_{1})\bigr|^2 =1$. 
If the mixing coefficients $A_{\pm}(k,\tau_{1})$  are redefined as $\overline{A}_{+} = N_{\nu} A_{+}$ and $\overline{A}_{-} = N_{\nu}^{*} \widetilde{A}_{-}$ the general expressions of Eqs. (\ref{FIVE4}) and (\ref{FIVE5}) become
\begin{eqnarray}
f_{k}(\tau) &=& \frac{1}{\sqrt{2 k}} \sqrt{ k y} \biggl[\biggl(\overline{A}_{+} + \overline{A}_{-}\biggr) J_{\nu}(k y) + i \biggl(\overline{A}_{-} - \overline{A}_{+}\biggr) Y_{\nu}(k y)\biggr],
\label{FIVE13a}\\
g_{k}(\tau) &=& \sqrt{\frac{k}{2}}\sqrt{ k y} \biggl[\biggl(\overline{A}_{+} + \overline{A}_{-}\biggr) J_{\nu-1}(k y) + i \biggl(\overline{A}_{-} - \overline{A}_{+}\biggr) Y_{\nu-1}(k y)\biggr].
\label{FIVE13b}
\end{eqnarray}
Equations (\ref{FIVE13a}) and (\ref{FIVE13b}) can now be referred to the electric and magnetic mode functions of Eqs. (\ref{TWO2}) and (\ref{TWO3})--(\ref{TWO4}) {\em evaluated at the end of inflation} i.e. 
\begin{eqnarray}
\overline{f}_{k} &=&  \frac{N_{\mu}}{\sqrt{2 k}} \sqrt{x_{1}} H_{\mu}^{(1)}(x_{1}), \qquad \overline{g}_{k} =  \sqrt{\frac{k}{2}} N_{\mu} \sqrt{x_{1}} H_{\mu+1}^{(1)}(x_{1}), \qquad \gamma > 1/2,
\nonumber\\
\overline{f}_{k} &=&  \frac{N_{\mu}}{\sqrt{2 k}} \sqrt{x_{1}} H_{\mu}^{(1)}(x_{1}), \qquad \overline{g}_{k} =  - \sqrt{\frac{k}{2}} N_{\mu} \sqrt{x_{1}} H_{\mu- 1}^{(1)}(x_{1}), \qquad 0< \gamma < 1/2,
\label{FIVE12}
\end{eqnarray}
where, as in the bulk o the paper, $\overline{f}_{k} = f_{k}(-\tau_{1})$ and $\overline{g}_{k} = g_{k}(-\tau_{1})$. Since $x_{1} = k \tau_{1} <1$ 
for all the modes of the spectrum the ratio of the mode functions of Eq. (\ref{FIVE12}) can always be evaluated in the small argument limit and it is:
\begin{equation}
\biggl|\frac{k \overline{f}_{k}}{\overline{g}_{k}} \biggr|= \biggl|\frac{H_{|\gamma-1/2|}^{(1)}(x_{1})}{H_{\gamma+1/2}^{(1)}(x_{1})}\biggr| \to \frac{\Gamma(|\gamma-1/2|)}{\Gamma(\gamma +1/2)}\, \biggl(\frac{x_{1}}{2}\biggr)^{\gamma +1/2 - |\gamma -1/2|}.
\label{FIVE12aa}
\end{equation}
The values of $f_{k}(\tau)$ and $g_{k}(\tau)$ for $\tau\geq -\tau_{1}$ will then be expressible in terms of $\overline{f}_{k}$ and $\overline{g}_{k}$  in the following manner
\begin{equation}
\left(\matrix{ f_{k}(\tau) &\cr
g_{k}(\tau)/k&\cr}\right) = \left(\matrix{ A_{f\, f}(k, \tau, \tau_{1})
& A_{f\,g}(k,\tau, \tau_{1})&\cr
A_{g\,f}(k,\tau, \tau_{1}) &A_{g\,g}(k,\tau, \tau_{1})&\cr}\right) \left(\matrix{ \overline{f}_{k} &\cr
\overline{g}_{k}/k&\cr}\right).
\label{FIVE13def2a}
\end{equation}
The entries of the matrix appearing at the right hand side of Eq. (\ref{FIVE13def2a}) have been given in Eq. (\ref{FIVE13});
they follow directly from Eqs. (\ref{FIVE13a})--(\ref{FIVE13b}) once the various coefficients are made explicit
in terms of Eqs. (\ref{FIVE8}) and (\ref{FIVE10}).
\renewcommand{\theequation}{B.\arabic{equation}}
\setcounter{equation}{0}
\section{Transition matrix and decreasing gauge coupling}
\label{APPB}
When the gauge coupling decreases 
 Eqs. (\ref{THREE1}) and (\ref{SIX2}) describe the evolution of $\sqrt{\lambda}$ and of $\sqrt{\lambda}^{\,\prime}$ interpolating 
 between the inflationary stage and the subsequent radiation epoch (see also Fig. \ref{FFF0b}). In full analogy with Eqs. (\ref{FIVE4}) and (\ref{FIVE5}) the mode functions for $\tau \geq - \tau_{1}$  are determined in terms 
 of a set of new coefficients $\widetilde{A}_{\pm}(k,\tau_{1})$:
\begin{eqnarray}
f_{k}(\tau) &=& \frac{\sqrt{ k y}}{\sqrt{2 k}} \biggl[ \widetilde{A}_{-}(k,\tau_{1})\,N_{\widetilde{\,\nu\,}}\,H_{\widetilde{\,\nu\,}}^{(1)}(k y) +  \widetilde{A}_{+}(k,\tau_{1})\,N_{\widetilde{\,\nu\,}}^{*}\,H_{\widetilde{\,\nu\,}}^{(2)}(k y)\biggr],
\label{SIX4}\\
g_{k}(\tau) &=& -\sqrt{\frac{k}{2}} \sqrt{ k y} \biggl[\widetilde{A}_{-}(k,\tau_{1})\,N_{\widetilde{\,\nu\,}}\,H_{\widetilde{\,\nu\,}+1}^{(1)}(k y) +  \widetilde{A}_{+}(k,\tau_{1})\,N_{\widetilde{\,\nu\,}}^{*}\,H_{\widetilde{\,\nu\,} +1}^{(2)}(k y)\biggr],\qquad \widetilde{\,\delta\,} > \frac{1}{2},
\label{SIX5}\\
g_{k}(\tau) &=& \sqrt{\frac{k}{2}} \sqrt{ k y} \biggl[\widetilde{A}_{-}(k,\tau_{1})\,N_{\widetilde{\,\nu\,}}\,H_{\widetilde{\,\nu\,}-1}^{(1)}(k y) +  \widetilde{A}_{+}(k,\tau_{1})\,N_{\widetilde{\,\nu\,}}^{*}\,H_{\widetilde{\,\nu\,}-1}^{(2)}(k y)\biggr],\qquad 0<\widetilde{\,\delta\,} < 1/2,
\label{SIX6}
\end{eqnarray}
where, in analogy with Eq. (\ref{FIVE6}), the following auxiliary quantities have been introduced:
\begin{eqnarray}
N_{\widetilde{\,\nu\,}} &=& \sqrt{\frac{\pi}{2}} e^{i \pi( 2 \widetilde{\,\nu\,} + 1)/4}, \qquad y(\tau) = \tau + \tau_{1} ( 1 + q), 
\nonumber\\
q &=& q(\widetilde{\,\delta\,}, \widetilde{\,\gamma\,}) = \frac{\widetilde{\,\delta\,}}{\widetilde{\,\gamma\,}},\qquad \widetilde{\,\nu\,} = \biggl|\widetilde{\,\delta\,} - \frac{1}{2}\biggr|.
\label{SIX7}
\end{eqnarray}
The expression of  $\widetilde{\,\nu\,}$ now depends on the range of $\widetilde{\,\delta\,}$. More precisely  for $\widetilde{\,\delta\,}>1/2$ the mixing coefficients are:
\begin{eqnarray}
&& \widetilde{\,A\,}_{-}(k, \tau_{1}) = - \frac{ i \pi N_{\widetilde{\mu}}}{4 N_{\widetilde{\,\nu\,}}} \sqrt{q} x_{1} \biggl[ H_{\widetilde{\mu}}^{(1)}(x_{1})  H_{\widetilde{\,\nu\,} +1 }^{(2)}(q\,x_{1}) - H_{\widetilde{\mu} - 1}^{(1)}(x_{1})  H_{\widetilde{\,\nu\,}}^{(2)}(q\,x_{1}) \biggr],
\nonumber\\
&& \widetilde{\,A\,}_{+}(k, \tau_{1}) = -\frac{ i \pi N_{\widetilde{\mu}}}{4 N_{\widetilde{\,\nu\,}}^{\ast}} \sqrt{q} x_{1} \biggl[ H_{\widetilde{\mu}- 1}^{(1)}(x_{1})  H_{\widetilde{\,\nu\,} }^{(1)}(q\,x_{1}) - H_{\widetilde{\mu}}^{(1)}(x_{1})  H_{\widetilde{\,\nu\,}+1}^{(1)}(q\,x_{1}) \biggr],\qquad \widetilde{\,\delta\,} > 1/2,
\label{SIX9}
\end{eqnarray}
where, as in Eqs. (\ref{FIVE8})--(\ref{FIVE10}), $x_{1}= k\tau_{1}$. In the range $0< \widetilde{\,\delta\,} < 1/2$ we have instead that $\widetilde{\,A\,}_{\pm}(k, \tau_{1})$ are:
\begin{eqnarray}
&& \widetilde{\,A\,}_{-}(k, \tau_{1}) = \frac{ i \pi N_{\widetilde{\mu}}}{4 N_{\widetilde{\,\nu\,}}} \sqrt{q} x_{1} \biggl[ H_{\widetilde{\mu}}^{(1)}(x_{1})  H_{\widetilde{\,\nu\,} -1 }^{(2)}(q\,x_{1}) + H_{\widetilde{\mu} - 1}^{(1)}(x_{1})  H_{\widetilde{\,\nu\,}}^{(2)}(q\,x_{1}) \biggr],
\nonumber\\
&& \widetilde{\,A\,}_{+}(k, \tau_{1}) = -\frac{ i \pi N_{\widetilde{\mu}}}{4 N_{\widetilde{\,\nu\,}}^{\ast}} \sqrt{q} x_{1} \biggl[ H_{\widetilde{\mu}- 1}^{(1)}(x_{1})  H_{\widetilde{\,\nu\,} }^{(1)}(q\,x_{1}) + H_{\widetilde{\mu}}^{(1)}(x_{1})  H_{\widetilde{\,\nu\,}-1}^{(1)}(q\,x_{1}) \biggr], \quad 0\leq \widetilde{\,\delta\,} < 1/2.
\label{SIX11}
\end{eqnarray}
As in the case of the results of appendix \ref{APPA}, because of the Wronskian normalization of Eq. (\ref{ONE8w}), the mixing coefficients of Eqs. 
(\ref{SIX9})--(\ref{SIX11}) must satisfy $\bigl|\widetilde{A}_{+}(k,\tau_{1})\bigr|^2 - \bigl|\widetilde{A}_{-}(k,\tau_{1})\bigr|^2 =1$.  We can therefore summarize the situation by saying that 
while $A_{\pm}(k,\tau_{1})$ of Eqs. (\ref{FIVE8})--(\ref{FIVE10}) depend upon the range of $\gamma$, 
the $\widetilde{\,A\,}_{\pm}(k, \tau_{1})$ of Eqs. (\ref{SIX9})--(\ref{SIX11}) depend on the range of $\widetilde{\,\delta\,}$.
Inserting the expressions of the coefficients (\ref{SIX9}) and (\ref{SIX11}) back into Eqs. (\ref{SIX4}), (\ref{SIX5}) and (\ref{SIX6}) the values of $f_{k}(\tau)$ and $g_{k}(\tau)$ for different 
values of $\widetilde{\,\delta\,}$ can then be referred to the inflationary mode functions at $\tau = - \tau_{1}$
directly obtainable from Eqs. (\ref{THREE2})--(\ref{THREE3}):
\begin{equation}
\overline{f}_{k}= \frac{N_{\widetilde{\mu}}}{\sqrt{2 k}} \, \sqrt{x_{1}} \, H_{\widetilde{\mu}}^{(1)}(x_{1}),\qquad 
\overline{g}_{k} = -N_{\widetilde{\mu}} \,\sqrt{\frac{k}{2}} \,  \sqrt{x_{1}} \, H_{\widetilde{\mu}-1}^{(1)}(x_{1}),
\label{SIX14a}
\end{equation}
where $\widetilde{\mu} = \widetilde{\,\gamma\,} +1/2$. As in the case of Eq. (\ref{FIVE12aa})  the ratio of the mode functions of Eq. (\ref{SIX14a}) can always be evaluated in the small argument limit and and the result is: 
\begin{equation}
\biggl|\frac{k \overline{f}_{k}}{\overline{g}_{k}} \biggr|= \biggl|\frac{H_{\widetilde{\,\gamma\,}+1/2}^{(1)}(x_{1})}{H_{|\widetilde{\,\gamma\,}-1/2|}^{(1)}(x_{1})}\biggr| \to \frac{\Gamma(\widetilde{\,\gamma\,} +1/2)}{\Gamma(|\widetilde{\,\gamma\,} -1/2|)}\, \biggl(\frac{x_{1}}{2}\biggr)^{|\widetilde{\,\gamma\,} -1/2| - \widetilde{\,\gamma\,} -1/2}.
\label{SIX14c}
\end{equation}
The magnetic and the electric mode functions for $\tau > - \tau_{1}$ can therefore be related to $\overline{f}_{k}$ and $\overline{g}_{k}$ 
of Eqs. (\ref{SIX14a}) as:
\begin{equation}
\left(\matrix{ f_{k}(\tau) &\cr
g_{k}(\tau)/k&\cr}\right) = \left(\matrix{ \widetilde{\,A\,}_{f\, f}(k, \tau, \tau_{1})
& \widetilde{\,A\,}_{f\,g}(k,\tau, \tau_{1})&\cr
\widetilde{\,A\,}_{g\,f}(k,\tau, \tau_{1}) &\widetilde{\,A\,}_{g\,g}(k,\tau, \tau_{1})&\cr}\right) \left(\matrix{ \overline{f}_{k} &\cr
\overline{g}_{k}/k&\cr}\right),
\label{SIX14b}
\end{equation}
where the explicit form of the various coefficients appearing at the right hand side of Eq. (\ref{SIX14b}) 
has been reported in Eq. (\ref{SIX11a})  which is the analog\footnote{Equations (\ref{SIX11a}) and (\ref{FIVE13}) do not coincide since $\nu$ and $\widetilde{\,\nu\,}$ are different. Under duality, the elements of the transition matrices have a well defined transformation law given in Eq. (\ref{DUAL2}) (see also appendix \ref{APPC} hereunder). Along a similar perspective,  
unlike the case of Eqs. (\ref{TWO3}) and (\ref{TWO4}), the expression of $g_{k}(\tau)$ is unique for all the range of $\widetilde{\,\gamma\,} > 0$; this occurrence follows from the difference between $\mu$ and $\widetilde{\mu}$. } of Eq. (\ref{FIVE13}).

\renewcommand{\theequation}{C.\arabic{equation}}
\setcounter{equation}{0}
\section{Transition matrix and duality transformations}
\label{APPC}
The transition matrices of Eqs.  (\ref{FIVE13def2a}) and  (\ref{SIX11a}) 
are related by duality transformations. As it can be explicitly verified from Eqs. (\ref{TWO1})--(\ref{FIVE2}) and (\ref{THREE1})--(\ref{SIX2}) a duality transformation implies that $(\gamma,\,\delta)$ are transformed into $(\widetilde{\,\gamma\,},\,\widetilde{\,\delta\,})$:
\begin{equation}
\sqrt{\lambda} \to 1/\sqrt{\lambda}\,\qquad \Rightarrow \qquad \gamma \to \widetilde{\,\gamma\,} , \qquad \delta \to \widetilde{\,\delta\,}.
\label{APC1}
\end{equation}
It has been shown [see Eq. (\ref{FOUR1}) and discussion therein] that during inflation the transformation 
$\gamma \to \widetilde{\,\gamma\,}$ exchanges electric into magnetic power 
spectra and vice versa. We shall now demonstrate that after inflation the 
corresponding duality transformation exchanges the elements of the transition matrices (\ref{FIVE13def2a}) and  (\ref{SIX11a}).  Let us first start by noting that when $\delta \to \widetilde{\,\delta\,}$ the corresponding Bessel indices 
$\nu$ and $\widetilde{\,\nu\,}$ are related as:
\begin{equation}
\delta \to \widetilde{\,\delta\,} \qquad \Rightarrow \qquad \nu \to 1 - \widetilde{\,\nu\,}.
\label{APC2}
\end{equation}
Equation (\ref{APC2}) is a  consequence of the definitions of $\nu(\delta) = \delta+1/2$ and $\widetilde{\,\nu\,}(\widetilde{\,\delta\,}) = | \widetilde{\,\delta\,} -1/2|$ given, respectively, in Eqs. (\ref{FIVE13def1}) and (\ref{SIX11b}).
By now applying the transformation Eq. (\ref{APC2}) to the matrix elements of ${\mathcal M}(\delta, x_{1}, x)$ and 
$\widetilde{\,{\mathcal M}\,}(\widetilde{\,\delta\,}, x_{1}, x)$  the following transformations are easily deduced:
\begin{eqnarray}
&& \, A_{f\,f} \to \widetilde{\,A\,}_{g\,g}, \qquad A_{g\,g} \to \widetilde{\,A\,}_{f\,f},
\label{APC3}\\
&& \, A_{f\,g} \to -  \widetilde{\,A\,}_{g\,f}, \qquad \, A_{g\,f} \to -  \widetilde{\,A\,}_{f\,g}. 
\label{APC4}
\end{eqnarray}
Equations (\ref{APC3}) and (\ref{APC4}) follow from the explicit expression of each matrix 
element. Consider, for instance,  $A_{f\,f}(\delta, x_{1}, x)$ which we write, for immediate 
convenience, in terms of Hankel functions of first and second kinds: 
\begin{equation}
A_{f\,f}(\delta, x_{1}, x) = \frac{i}{2} \sqrt{ q x_{1}} \, \sqrt{k y} \biggl[ H_{\nu}^{(1)}(k\, y)\, H_{\nu -1}^{(2)}( q x_{1}) - 
H_{\nu -1}^{(1)}( q x_{1}) \,H_{\nu}^{(2)}(k\, y)\biggr].
\label{APC5}
\end{equation}
By now performing the transformation $\gamma \to \widetilde{\,\gamma\,}$ and $\delta \to \widetilde{\,\delta\,}$ (i.e. $\nu \to 1- \widetilde{\,\nu\,}$) we obtain, from Eq. (\ref{APC5}), 
\begin{equation}
A_{f\,f}(\widetilde{\,\delta\,}, x_{1}, x) = \frac{i}{2} \sqrt{ q x_{1}} \, \sqrt{k y} \biggl[ H_{1- \widetilde{\,\nu\,}}^{(1)}(k\,y)\, H_{- \widetilde{\,\nu\,} }^{(2)}( q x_{1}) - 
H_{- \widetilde{\,\nu\,}}^{(1)}( q x_{1}) \,H_{1- \widetilde{\,\nu\,}}^{(2)}(k\, y)\biggr].
\label{APC6}
\end{equation}
It is understood that while doing the transformation also the arguments of $q$ and $y$ will change 
according to  Eqs. (\ref{FIVE13def1}) and (\ref{SIX11b}):
\begin{equation}
q(\delta,\gamma) = \delta/\gamma \to \widetilde{\,\delta\,}/\widetilde{\,\gamma\,} = q (\widetilde{\,\delta\,}, \widetilde{\,\gamma\,}),
\label{APC7}
\end{equation}
and similarly for $y$. We now recall that the Hankel functions of generic index $\mu$ and generic argument $z$ 
obey \cite{abr1,abr2}
\begin{equation}
H_{- \mu}^{(1)}(z) = e^{i \pi \mu} H_{\mu}^{(1)}(z), \qquad H_{- \mu}^{(2)}(z) = e^{-i \pi \mu} H_{\mu}^{(2)}(z).
\label{APC8}
\end{equation}
Inserting Eq. (\ref{APC8}) into Eq. (\ref{APC6}) we therefore obtain 
\begin{equation}
A_{f\,f}(\widetilde{\,\delta\,}, x_{1}, x) = - \frac{i}{2} \sqrt{ q x_{1}} \, \sqrt{k y} \biggl[ H_{ \widetilde{\,\nu\,}- 1}^{(1)}(k\,y)\, H_{\widetilde{\,\nu\,} }^{(2)}( q x_{1}) - 
H_{ \widetilde{\,\nu\,}}^{(1)}( q x_{1}) \,H_{\widetilde{\,\nu\,} - 1}^{(2)}(k\, y)\biggr],
\label{APC9}
\end{equation}
which coincides with $\widetilde{\,A\,}_{g\, g}(\widetilde{\,\delta\,}, x_{1}, x)$, as anticipated in Eq. (\ref{APC3}).
With a similar procedure also the remaining transformation rules of Eqs. (\ref{APC3}) and (\ref{APC4}) 
can be easily deduced.

\end{appendix}


\newpage

\begin{thebibliography}{99}
\itemsep 2pt

\bibitem{lich} A. Lichnerowicz, {\it Magnetohydrodynamics: waves and shock waves in curved space-time}, 
(Kluwer academic publisher, Dordrecht, 1994).

\bibitem{duality1}  S.~Deser and C.~Teitelboim,  Phys.\ Rev.\  D {\bf 13}, 1592 (1976).

\bibitem{duality2} S. Deser,  J. Phys. A {\bf 15}, 1053 (1982).

\bibitem{parker1} L.~Parker,  Phys.\ Rev.\ Lett.\  {\bf 21},  562 (1968).

\bibitem{birrell} N. D. Birrell and P. C. W. Davies, {\it Quantum fields in curved space}, 
(Cambridge University Press, Cambridge, UK, 1982).

\bibitem{parker2} L. Parker and D. Toms, {\it Quantum Field Theory in Curved Spacetime: Quantized Fields and Gravity}, (Cambridge University Press, Cambridge, UK, 2009).

\bibitem{ONEa} E. Fermi,  Phys. Rev. {\bf 75}, 1169 (1949).

\bibitem{ONEb}  H. Alfv\'en, Phys. Rev. {\bf 75}, 1732 (1949); R. D. Richtmyer, E. Teller,  Phys. Rev. {\bf 75}, 1729 (1949).

\bibitem{SIX1} F. Hoyle, Proc. of Solvay Conference {\it ``La structure et l'evolution de 
l'Univers''}, (ed. by R. Stoop, Brussels) p. 59 (1958). 

\bibitem{SIX2} Ya. B. Zeldovich,  Sov. Phys. JETP {\bf 21}, 656 (1965) [Zh. Eksp. Teor. Fiz., {\bf 48} 986 (1965)].

\bibitem{SIX3} K. S. Thorne, Astrophys. J. {\bf 148}, 51 (1967).

\bibitem{SIX3a} M. P. Ryan and L. C. Shepley {\it Homogeneous Relativistic Cosmologies} (Princeton, NJ, Princeton
University Press, 1975).

\bibitem{SIX4}  E. R. Harrison, Phys. Rev. Lett. {\bf 18}, 1011 (1967).

\bibitem{SIX5}  E. R. Harrison, Phys. Rev. {\bf 167}, 1170 (1968).

\bibitem{mgenesis} M. Giovannini,  Phys. Rev. D {\bf 62}, 123505 (2000).

\bibitem{b1}  H. Alfv\'en and C.-G. F\"althammer, {\it Cosmical Electrodynamics}, 2nd edn., (Clarendon press, Oxford, 1963).

\bibitem{b2} E. N. Parker, {\it Cosmical Magnetic Fields} (Clarendon Press, Oxford, 1979).

\bibitem{b3} Ya. B. Zeldovich, A. A. Ruzmaikin, D.D. Sokoloff  {\it Magnetic Fields in Astrophysics} (Gordon  Breach Science, New York, 
1983).

\bibitem{rev1} K. Enqvist, Int.\ J.\ Mod.\ Phys.\  D  {\bf 7}, 331 (1998).

\bibitem{rev2} M. Giovannini, Int. J. Mod. Phys. D {\bf 13},  391 (2004); A. Kandus, K. Kunze, C. Tsagas,  arXiv 1007.3891.

\bibitem{rev3} J. L. Han,  Annu. Rev. Astron. Astrophys. {\bf 55},  111 (2017).

\bibitem{wein1} S.~Weinberg, {\it ``Cosmology''}, (Oxford, Oxford University Press, 2008).

\bibitem{DT1} B. Ratra,  Astrophys.\, J.\, Lett.  {\bf 391}, L1 (1992);
 M.~Gasperini, M.~Giovannini, and G.~Veneziano, Phys. Rev. Lett. {\bf 75}, 3796 (1995);
 M.~Giovannini,  Phys.\ Rev.\  D {\bf 64}, 061301 (2001).
 
\bibitem{DT2} K.~Bamba and M.~Sasaki,  JCAP {\bf 02}, 030 (2007); K. Bamba JCAP {\bf 10}, 015 (2007);
 M. Giovannini, Phys.\ Lett.\  B {\bf 659}, 661 (2008).

\bibitem{DT3} K. Bamba, Phys. Rev. D {\bf 75} 083516 (2007); J. Martin and J. Yokoyama, JCAP {\bf 0801}, 025 (2008);
M. Giovannini, Lect.\ Notes Phys.\  {\bf 737}, 863 (2008); V.~Demozzi, V.~Mukhanov and H.~Rubinstein, JCAP \textbf{08}, 025 (2009).

\bibitem{DT4}   I. Brown, Astrophys. J. {\bf 733}, 83 (2011); T.~Fujita and S.~Mukohyama,  JCAP {\bf 1210}, 034 (2012); T.~Kahniashvili, A.~Brandenburg, L.~Campanelli, B.~Ratra and A.~G.~Tevzadze, Phys.\ Rev.\ D {\bf 86}, 103005 (2012); R.~Z.~Ferreira and J.~Ganc, JCAP {\bf 1504}, 029 (2015); L.~Campanelli, Phys. Rev. D {\bf 93}, 063501 (2016); C. Tsagas, arXiv 1412.4806; arXiv 1508.06604.

 \bibitem{c0} R. D. Peccei and H. R. Quinn, Phys. Rev. Lett. {\bf 38}, 1440  (1977); Phys. Rev. D {\bf 16}, 1791 (1977).

\bibitem{c0a} J. Kim, Phys. Rep. {\bf 150}, 1 (1987); H.-Y. Cheng, {\em ibid}., {\bf 158}, 1 (1988); G. G. Raffelt, Phys. Rep. {\bf 198}, 1 (1990);  Lect.\ Notes Phys.\  {\bf 741}, 51 (2008).

\bibitem{c1}  S. Carroll, G. Field and R. Jackiw, Phys. Rev. D {\bf 41},  1231 (1990);  W. D. Garretson, G. Field and S. Carroll, Phys. Rev. D {\bf 46}, 5346 (1992).

\bibitem{c1a} G. Field and S. Carroll Phys.Rev.D {\bf 62}, 103008 (2000); M.~Giovannini,
  Phys.\ Rev.\ D {\bf 61}, 063502 (2000); Phys.\ Rev.\ D {\bf 61}, 063004 (2000); K.~Bamba, Phys.\ Rev.\ D {\bf 74}, 123504 (2006); K.~Bamba, C.~Q.~Geng and S.~H.~Ho,
  Phys.\ Lett.\ B {\bf 664}, 154 (2008).

\bibitem{c2}  L.~Campanelli,  Int.\ J.\ Mod.\ Phys.\ D {\bf 18}, 1395 (2009); L.~Campanelli and M.~Giannotti,
Phys.\ Rev.\ D {\bf 72}, 123001 (2005);  Phys.\ Rev.\ Lett.\  {\bf 96}, 161302 (2006); M. Giovannini,  Phys.\ Rev.\ D {\bf 88}, 063536 (2013); K.~Bamba,
   Phys.\ Rev.\ D {\bf 91},  043509 (2015).

\bibitem{mm1}  M.~Giovannini and M.~E.~Shaposhnikov,  Phys.\ Rev.\ D {\bf 57}, 2186 (1998);  Phys.\ Rev.\ Lett.\  {\bf 80}, 22 (1998); M.~Giovannini,  Phys.\ Rev.\ D {\bf 61}, 063004 (2000).
 
\bibitem{mm2} M. Giovannini, Phys.\ Rev.\ D {\bf 61}, 063502 (2000); M.~Dvornikov and V.~B.~Semikoz,
  JCAP {\bf 1202}, 040 (2012); S.~Alexander, A.~Marciano and D.~Spergel,
  JCAP {\bf 1304}, 046 (2013); N.~D.~Barrie and A.~Kobakhidze,  JHEP {\bf 1409}, 163 (2014).

\bibitem{mm3} M.~Giovannini, Phys. Rev. D {\bf 92},  121301 (2015); Phys. Rev. D {\bf 93},  103518 (2016).

\bibitem{cme1} D.~E.~Kharzeev,  Prog.\ Part.\ Nucl.\ Phys.\  {\bf 75}, 133 (2014); M.~Giovannini,
  Phys.\ Rev.\ D {\bf 88}, 063536 (2013).
    
  \bibitem{cme2}  D. Kharzeev, L. McLerran and H. Warringa, Nucl. Phys. A {\bf 803}, 227 (2008); 
K.~Fukushima, D.~Kharzeev and H.~Warringa, Phys.\ Rev.\ D {\bf 78}, 074033 (2008);
D.~Kharzeev,  Annals Phys.\  {\bf 325}, 205 (2010); M. Giovannini, Phys. Rev. D {\bf 94}, 081301 (2016).

\bibitem{SAK1} A. D. Sakharov,  Sov. Phys. JETP {\bf 22}, 241 (1966) [Zh. Eksp. Teor. Fiz. {\bf 49}, 345 (1965)].

\bibitem{SAK2} P.~J.~E.~Peebles and J.~T.~Yu,  Astrophys.\ J.\  {\bf 162} 815 (1970).

\bibitem{SAK3} R.~A.~Sunyaev and Y.~B.~Zeldovich,  Astrophys.\ Space Sci.\  {\bf 7}, 3 (1970).

\bibitem{SAK4} P. Naselsky and I. Novikov, Astrophys. J. {\bf 413}, 14 (1993).
  
\bibitem{wein2} S.~Weinberg, Phys.\ Rev.\ D {\bf 67}, 123504 (2003).

\bibitem{abr1}  A. Erdelyi, W. Magnus, F. Oberhettinger, and F. R. Tricomi {\em Higher Trascendental Functions} (Mc Graw-Hill, New York, 1953).

\bibitem{abr2} M. Abramowitz and I. A. Stegun, {\em Handbook of Mathematical Functions} (Dover, New York, 1972).

\bibitem{poss} M. Giovannini,  J. Cosmol. Astropart. Phys. {\bf 04} 003 (2010); Phys. Rev. D {\bf 85}, 101301 (2012).

\bibitem{GW1} M.~Giovannini, Class. Quant. Grav. \textbf{26}, 045004 (2009); Prog. Part. Nucl. Phys. 112, 103774 (2020).

\bibitem{GW2} A.~Ito, J.~Soda and M.~Yamaguchi, [arXiv:2009.03611 [astro-ph.CO]].

\bibitem{GW3} K.~Bamba, E.~Elizalde, S.~D.~Odintsov and T.~Paul, [arXiv:2012.12742 [gr-qc]].

\bibitem{SCH1} T. Kobayashi and N. Afshordi, J. High Energy Phys. {\bf 10}, 166 (2014); C. Stahl, E. Strobel, and S. S. Xue, Phys. Rev. D {\bf 93}, 025004 (2016); R. Sharma and S. Singh, Phys. Rev. D {\bf 96}, 025012 (2017); E. Bavarsad, S. P. Kim, C. Stahl, and S. S. Xue, Phys. Rev. D {\bf 97}, 025017 (2018).

\bibitem{SCH2} M. Giovannini, Phys. Rev. D {\bf 97}, 061301(R) (2018).

\bibitem{RT3} P.~A.~R.~Ade {\it et al.} [BICEP2 and Keck Array Collaborations],  Phys.\ Rev.\ Lett.\  {\bf 116}, 031302 (2016).
  
\bibitem{RT4}  Y.~Akrami {\it et al.} [Planck Collaboration], Astron. Astrophys. {\bf 641}, A10 (2020).

\bibitem{linde1} L. Kofman, A. D. Linde, and A. A. Starobinsky, Phys. Rev. Lett. {\bf 73},  3195 (1994).

\bibitem{linde2} L. Kofman, A. D. Linde, and A. A. Starobinsky,  Phys.Rev. D {\bf 56}, 3258 (1997).

\bibitem{farhi1}  L. Abbott, E. Farhi, and M. B. Wise, Physics Letters B {\bf 117}, 29  (1982).

\bibitem{CMBM} M.~Giovannini, Class. Quant. Grav. {\bf 35}, 084003 (2018).

\end{thebibliography}
\end{document}